\documentclass[a4paper,fleqn,usenatbib]{mnras}

\usepackage[T1]{fontenc}
\usepackage{ae,aecompl}

\usepackage{graphicx}	
\usepackage{amsmath}	
\usepackage{amssymb}	
\usepackage{subfigure}

\newcommand{\msun}{M$_\odot$}
\newcommand{\ha}{H$\alpha$}
\newcommand{\mum}{$\mu$m}

\newcommand{\sigmah}{$\Sigma_{\textrm{H}_2}$}

\newcommand{\sigmasfr}{$\Sigma_{\textrm{SFR}}$}
\newcommand{\mrc}{RC\&K17}

\graphicspath{{./pictures/}}

\title[Environmental variations of YSC properties in M51]{The young star cluster population of M51 with LEGUS - II. Testing environmental dependencies}

\author[M. Messa et al.]{Matteo Messa,$^{1}$\thanks{E-mail: matteo.messa@astro.su.se}
A. Adamo$^{1}$
D. Calzetti,$^{2}$
M. Reina-Campos,$^{3}$
D. Colombo,$^{4}$
\newauthor
E. Schinnerer,$^{5}$
R. Chandar,$^{6}$
D.A. Dale,$^{7}$
D.A. Gouliermis,$^{8,5}$
K. Grasha,$^{2}$
\newauthor
E.K. Grebel,$^{3}$
B.G. Elmegreen,$^{9}$
M. Fumagalli,$^{10}$
K.E. Johnson,$^{11}$
J.M.D. Kruijssen,$^{3}$
\newauthor
G. \"{O}stlin,$^{1}$
F. Shabani,$^{3}$
L.J. Smith,$^{12}$
B.C. Whitmore$^{13}$
\\
$^{1}$Dept. of Astronomy, The Oskar Klein Centre, Stockholm University, Stockholm, Sweden\\
$^{2}$Dept. of Astronomy, University of Massachusetts -- Amherst, Amherst, MA 01003\\
$^{3}$Astronomisches Rechen-Institut, Zentrum f\"ur Astronomie der Universit\"at Heidelberg, M\"onchhofstr.\ 12--14, 69120 Heidelberg, Germany\\
$^{4}$Max Planck Institut f\"{u}r Radioastronomie, D-53010 Bonn, Germany\\
$^{5}$Max Planck Institute for Astronomy,  K\"{o}nigstuhl\,17, 69117 Heidelberg, Germany\\
$^{6}$Dept. of Physics and Astronomy, University of Toledo, Toledo, OH\\
$^{7}$Dept. of Physics and Astronomy, University of Wyoming, Laramie, WY\\
$^{8}$Zentrum f\"ur Astronomie der Universit\"at Heidelberg, Institut f\"ur Theoretische Astrophysik, Albert-Ueberle-Str.\,2, 69120 Heidelberg, Germany\\
$^{9}$IBM Research Division, T.J. Watson Research Center, Yorktown Hts., NY\\
$^{10}$Institute for Computational Cosmology and Centre for Extragalactic Astronomy, Durham University, Durham, United Kingdom\\
$^{11}$Dept. of Astronomy, University of Virginia, Charlottesville, VA\\
$^{12}$European Space Agency/Space Telescope Science Institute, Baltimore, MD\\
$^{13}$Space Telescope Science Institute, Baltimore, MD\\
}
\date{Accepted 2018 February 28. Received 2018 February 14; in original form 2017 December 13}

\pubyear{2017}

\begin{document}
\label{firstpage}
\pagerange{\pageref{firstpage}--\pageref{lastpage}}
\maketitle

\begin{abstract}
It has recently been established that the properties of young star clusters (YSCs) can vary as a function of the galactic environment in which they are found.
We use the cluster catalogue produced by the Legacy Extragalactic UV Survey (LEGUS) collaboration to investigate cluster properties in the spiral galaxy M51. 
We analyse the cluster population as a function of galactocentric distance and in arm and inter-arm regions.
The cluster mass function exhibits a similar shape at all radial bins, described by a power law with a slope close to $-2$ and an exponential truncation around $10^5$ \msun. 
While the mass functions of the YSCs in the spiral arm and inter-arm regions have similar truncation masses, the inter-arm region mass function has a significantly steeper slope than the one in the arm region; a trend that is also observed in the giant molecular cloud mass function and predicted by simulations.
The age distribution of clusters is dependent on the region considered, and is consistent with rapid disruption only in dense regions, while little disruption is observed at large galactocentric distances and in the inter-arm region.
The fraction of stars forming in clusters does not show radial variations, despite the drop in the $H_2$ surface density measured as function of galactocentric distance.
We suggest that the higher disruption rate observed in the inner part of the galaxy is likely at the origin of the observed flat cluster formation efficiency radial profile.
\end{abstract}


\newpage
\section{Introduction}	         

Star formation is believed to be a hierarchical  process both in space and time. 
At the density peaks of the hierarchy star clusters may form, stellar systems that remain gravitationally bound for hundreds of Myr (see, e.g., \citealp{elmegreen2008,portegieszwart2010,elmegreen2011,kruijssen2012}). Due to their long lifetimes, these systems can be used as probes of the star formation process in   galaxies. 
Until now, the effort of studying star formation has focused either on the star/cluster scale only or on the galaxy scale only, without a real connection. The Legacy 
Extragalactic UV Survey (LEGUS) aims to fill the gap between these two scales, via the observations of 50 nearby galaxies in broadband imaging from the near-$UV$ to the $I$ band \citep{legus1}. Particularly important for cluster studies is the inclusion of two blue broadband filters of the Wide-Field Camera 3 (WFC3), namely F275W and F336W, which provide the information necessary for an accurate age analysis of the young clusters. A clear example of the power given by the availability of a set of filters ranging from the $UV$ to the $I$ band is the analysis of the super star clusters in the galaxy NGC 5253 by \citet{calzetti2015}. 
Observational studies of star formation in LEGUS are also supported by simulations, which are nowadays able to study entire galaxies but achieving the resolution of individual clusters \citep[e.g.,][]{dobbs2017}.

Evidences of star formation hierarchy in LEGUS galaxies are found by analyzing the UV-light structures of both spiral and dwarf galaxies \citep{elmegreen2014}. Other studies of the clustering of stars and clusters within LEGUS suggest that both tracers find the same underlying hierarchical structure (\citealp{gouliermis2015,gouliermis2017} and \citealp{grasha2017b}). The evolution in time of the hierarchical distribution of clusters is also tested by \citet{grasha2015} and \citet{grasha2017}, who analyze how the clustering strength of clusters changes with time, using a two-point correlation function and considering clusters of different ages for 6 LEGUS galaxies. The strength of the clustering is found to decrease with increasing age of the clusters, and disappears after $\sim 40-60$~Myr in all the studied cases.

One of the main goals of the LEGUS project is to link the properties of the star and cluster populations to the properties of the host galaxies, in order to understand how the galactic environment (e.g., the density of the ISM) affects the star formation process on various scales.
It was recently observed that the properties of clusters vary as a function of the distance from the centre of the galaxy M83. In the inner regions of M83, where the molecular gas has high density, the mass function is truncated at higher masses and the disruption has smaller timescales compared to the external regions \citep{silvavilla2014,adamo2015}. It was also observed that in environments with high gas density in M83, a higher fraction of the star formation happens inside clusters (measured through the cluster formation efficiency). 
Analyses of giant molecular cloud (GMC) properties in M83 showed similar radial variations, suggesting a close link between the gas clouds and the clusters forming from them, and these analyses also confirm that the galactic environment is capable of regulating the star formation process \citep{freeman2017}. 

Understanding how the environment can affect the cluster properties is however not straightforward: \citet{ryon2015}, for example, found that the sizes and shapes of clusters in M83 do not show radial dependence (differently from cluster truncation masses in the same galaxy), suggesting that some cluster properties may be more universal. In addition, a recent study of 5 dwarf galaxies of the LEGUS sample does not find any variation of the fraction of star formation happening in clusters (Hunter et al., in preparation), in contrast to the finding in M83. These findings highlight the necessity of expanding the number of galaxies (and of different environments) where cluster properties are studied.

Among the galaxies of the LEGUS sample, M51 stands out as an interesting case due to its large cluster population. In a previous work \citep[hereafter Paper~I]{paper1}, we analysed the cluster population of M51 as a whole.
The YSC mass function is well-described by a power law with an exponent of $\sim-2$, and compatible with an exponential truncation at $\sim10^5$ \msun. Similar results have been found in other spirals, e.g., M83 \citep{adamo2015}, NGC 1566 \citep{hollyhead2016}, and NGC 628 \citep{legus2}. A power law with an exponent of $\sim-2$ is expected if star formation takes place from a hierarchical medium \citep{elmegreen2010} while the presence of an exponential truncation suggest that galaxies may be limited in the formation process of high-mass compact structures.
The fraction of stars forming in bound clusters in M51 is $\sim20\%$, again in line with the values of similar galaxies in the nearby Universe. Finally, only moderate disruption seems to affect the clusters of ages between $\sim3-200$~Myr. It has been suggested that all these properties may depend on the galactic environment, and a comparison with the same cluster properties in different galaxies like M31 \citep{johnson2016,johnson2017} or the Antennae system \citep{whitmore1999,whitmore2007} seem to confirm such a relation. 
In this case, we should be able to spot differences in the cluster properties also as a function of the environment within the same galaxy, as observed for M83. 

In this work, we propose to test the presence of variations of cluster properties in different environments of M51.
We will use gas and SFR densities to probe our current understanding of cluster formation and evolution via simple theoretical models.
To complete the analysis, we compare our findings with the results of \citet{colombo2014a}, who studied the properties of GMCs in M51 dividing the sample in dynamical regions. \citet{huges2013} already suggested that the properties of the CO gas distribution for different M51 environments are strongly correlated with properties of GMC and YSC populations, and in particular with their mass functions.

The cluster catalogue of M51 produced with the LEGUS dataset is described in Paper~I. We nonetheless summarize the main properties of the dataset and the steps followed to produce the catalogue in Section~\ref{sec:data}. The division of the galaxy into subregions is described in Section~\ref{sec:environment}, while the analysis of the main cluster properties is in Section~\ref{sec:analysis}. In Section~\ref{sec:discussion}, we discuss the results of the cluster analysis and we apply a simple model to predict the mass properties of the clusters starting from gas data. Finally, the main conclusions of this work are summarised in Section~\ref{sec:conclusions}.

\section {Data and Cluster Catalogue Construction} 
\label{sec:data}
The general description of the LEGUS dataset and its standard reduction process are given in \citet{legus1}. The M51 dataset used in this study includes new UVIS/WFC3 imaging in the filters F275W ($UV$ band) and F336W ($U$ band), as well as archival ACS data in the filters F435W, F555W and F814W ($B$, $V$ and $I$ bands respectively).
The $U$ and $UV$ band data consist of 5 pointings, covering most of the spiral and companion galaxy NGC 5195. Exposures times are given in Paper~I for all filters. 
A detailed description of the procedure for producing the cluster catalog of M51 is given in Paper~I. The general procedure followed is a blind extraction of sources followed by a series of cuts aimed at removing spurious sources (e.g. stars, background galaxies) and by a morphological classification. Here we summarize the main steps. 

The $V$ band (filter F555W) is taken as the reference frame, where positions of cluster candidates are extracted using \texttt{SExtractor} \citep{sextractor}. On this extracted catalogue, a first cut is made based on the luminosity profile, keeping only sources with a concentration index (CI) bigger than 1.35\footnote{The concentration index is defined here as the difference in a source magnitude when measured in a 1 pixel radius aperture and in a 3 pixel radius aperture, i.e. $mag(r=1\ [px])-mag(r=3\ [px])$.}. A subsequent cut excludes the sources not detected photometrically in at least 2 contiguous bands (the reference $V$ band and either the $B$ or $I$ bands). The sources still remaining in the sample after these two cuts constitute the \textit{``automatic''} catalogue (according to the LEGUS nomenclature), which, being only automatically-selected, likely still includes some contaminating sources.
In order to reduce the contamination from stars and interlopers, a sub-catalogue is produced with additional cuts. We require the sources to be detected in at least 4 filters with a photometric error smaller than 0.30 mag and an absolute $V$-band magnitude brighter than $-6$ mag. This sub-catalog contains 10925 cluster candidates that were morphologically classified (a description of the morphological classes used in LEGUS is given in \citealp{legus2}). Almost 1/4 of those cluster candidates were visually classified, while the remaining ones were classified via a machine-learning (ML) code. The ML code used is described in a forthcoming paper (Grasha et al., submitted). 
As for the analyses of Paper~I, among the 10925 sources morphologically classified, we consider in the following analyses only the 2839 cluster candidates of classes 1 and 2, namely those that appear single-peaked, compact, and uniform in colour.

All sources detected in at least 4 filters were analysed via SED-fitting algorithms and values of age, mass and extinction were retrieved for each. Simple stellar population (SSP) models considering Padova-AGB evolutionary tracks with solar metallicity, the Milky Way extinction curve \citep{cardelli1989}, and a uniformly sampled \citet{kroupa2001} stellar initial mass function were used (see \citealp{ashworth2017} for a generalization to a variable IMF). Nebular continuum emission is also taken into account in the fit. The details of the SED-fitting techniques are described in \citet{legus2}.

The photometric completeness of the final cluster sample (2839 sources) and the comparison to older cluster catalogues of M51 were explored in Paper~I. The completeness is discussed also in Appendix~\ref{sec:a1}, where we derive the completeness value for a mass-limited sample inside the sub-regions defined in Section~\ref{sec:radial_division}.

\section{Galaxy Environment} 
\label{sec:environment}

\subsection{Environmental division of the catalogue} 
\label{sec:radial_division}
In the analysis of the cluster population, we exclude clusters found in the central part of the galaxy. This is due to the high level of incompleteness in the cluster detection near the centre, as already discussed in Paper I.
Excluding the region within 35'' (1.3 kpc at an assumed distance of 7.66~Mpc, \citealp{distancem51}) from the centre, we divide the area of the galaxy into 4 radial annuli. These are defined in order to contain the same number of clusters with $\rm M>5000$ \msun\ and ages younger than 200~Myr. No lower age cut has been applied. The cuts on age and mass are used to define a mass-limited sample, as in Paper~I. This choice allows us to have a sample not limited by luminosity.
In addition to these 4 bins, we also consider a central annulus corresponding to the molecular ring (MR) region defined in \citet{colombo2014a}, ranging from 23'' to 35'' (0.85 to 1.30 kpc) from the centre. In the MR the number of clusters is smaller compared to the other annuli and the completeness is worse (see Appendix~\ref{sec:a1}). However, this region is important for testing the cluster properties in a dense central region of the galaxy.

\begin{figure*}
\centering
\subfigure[Annuli of equal number of clusters]{\includegraphics[width=0.32\textwidth]{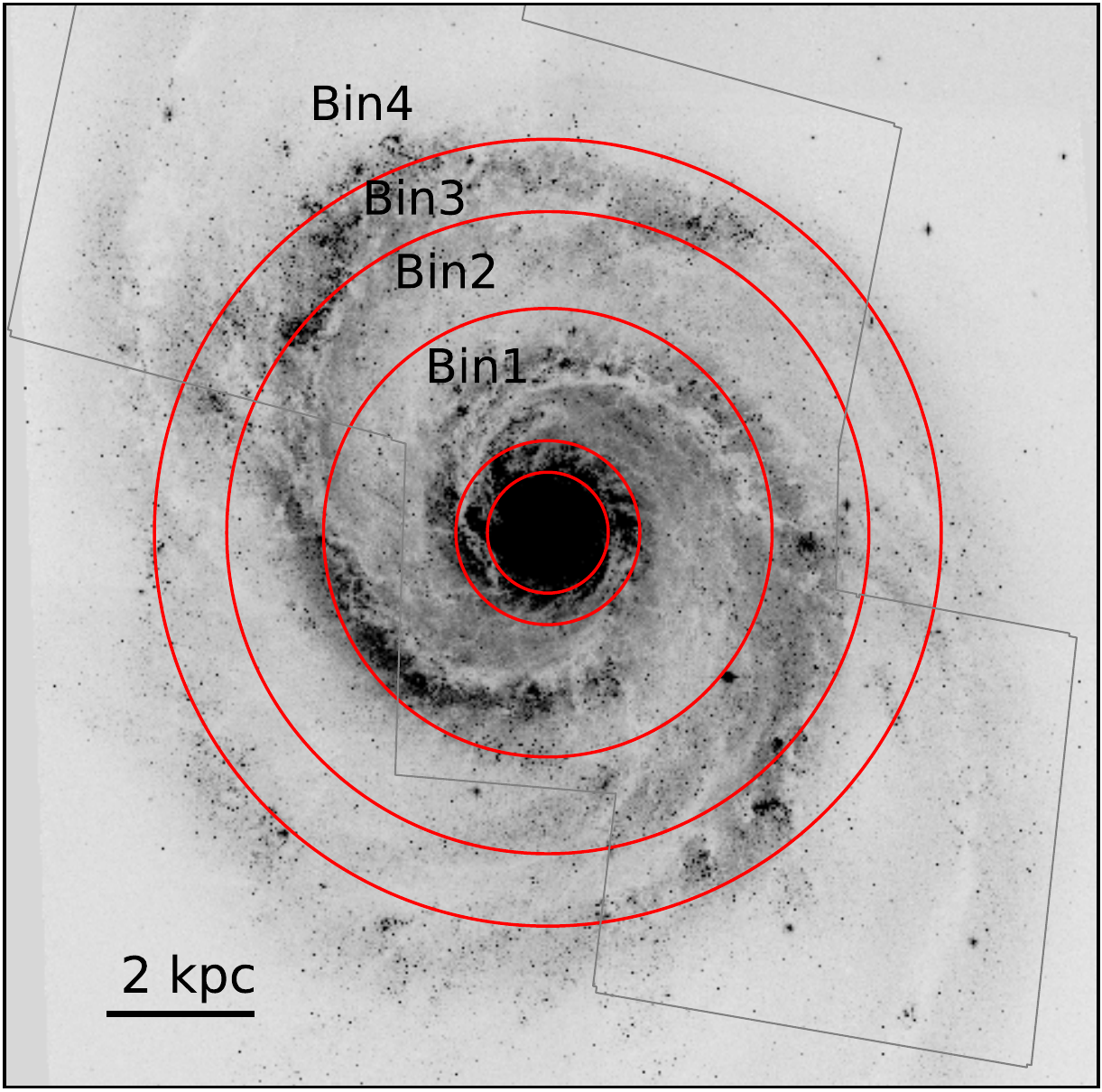}}
\subfigure[Annuli of equal area]{\includegraphics[width=0.32\textwidth]{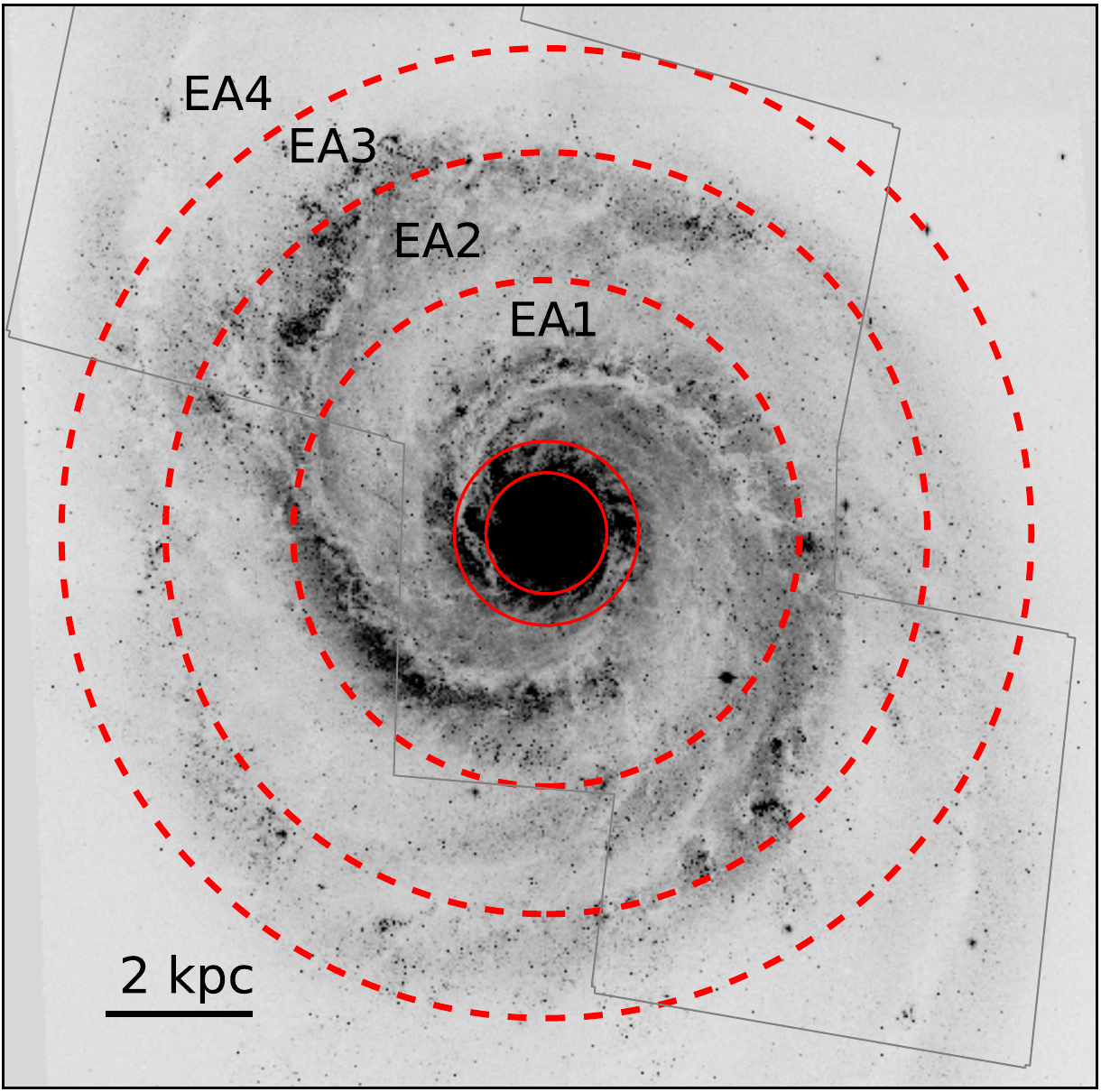}}
\subfigure[Arm and inter-arm division]{\includegraphics[width=0.32\textwidth]{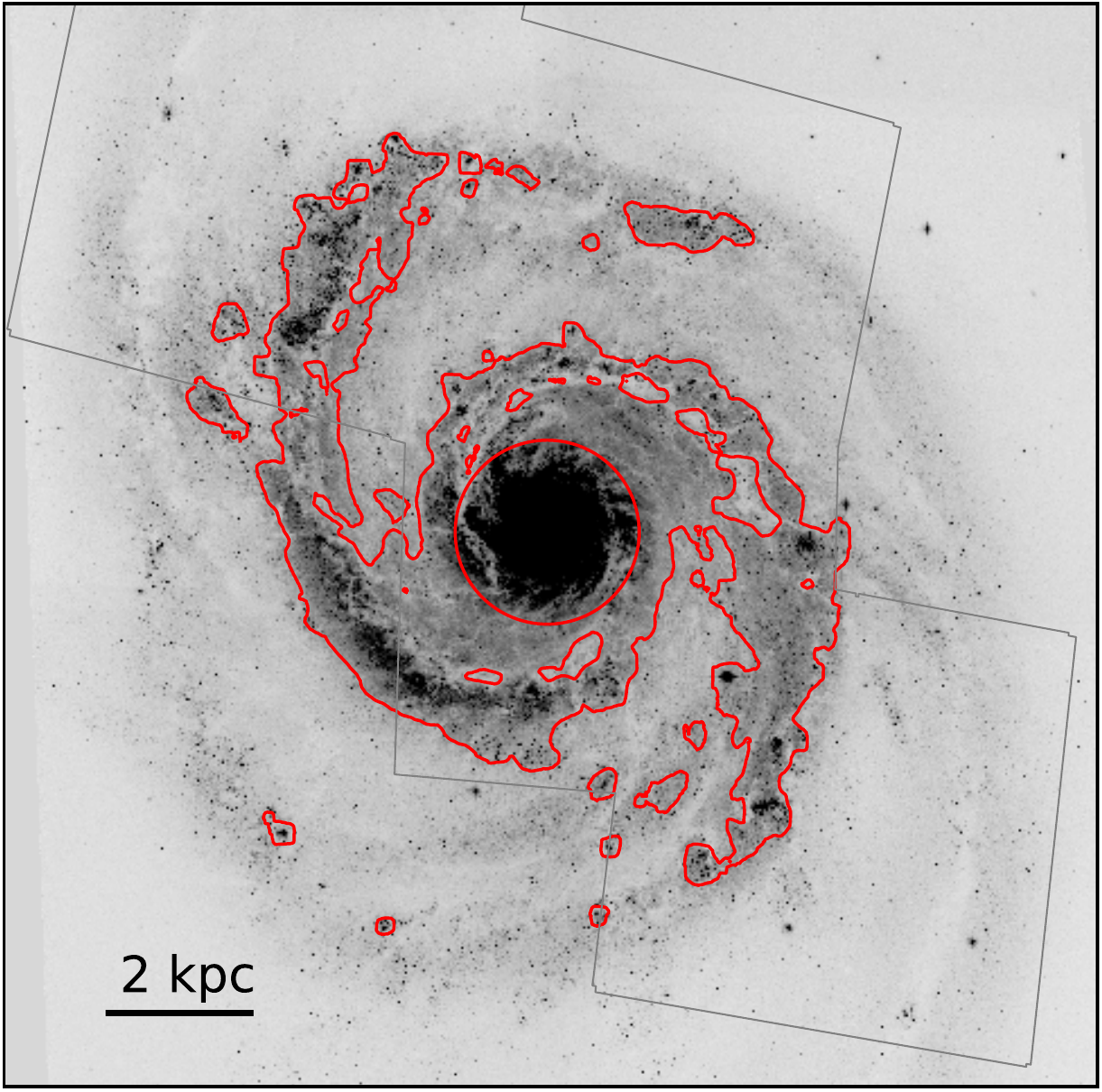}}
\caption{M51 in the V band, showing the division in radial annuli (red circles in the left and middle panels). The right panel shows the arm/inter-arm division. The two innermost circles in the panels enclose the molecular ring (MR) region.}
\label{fig:regions}
\end{figure*}

In order to test how the choice of radial binning affects the analyses, the cluster sample was also divided into 4 radial annuli of equal area. The radial division is graphically shown in Fig.~\ref{fig:regions}. Radii separating the bins are listed in Tab.~\ref{tab:properties} along with the number of clusters in each bin and other physical quantities used in the text.
\begin{table*}
\centering
\caption{Properties of the cluster population and of the galaxy environment. The part of the galaxy covered by UVIS observations is taken as a whole (Entire), divided into Spiral-Arm (SA) and Inter-Arm (IA) subregions according to the $V$ band brightness (see Section \ref{sec:radial_division}) and divided into radial regions according to the criteria in columns 2 and 3. Column 4 displays the number of class 1 and 2 clusters, and column 5 the number of those with $M>5000$ \msun\ and age $<200$~Myr. In column 6 the values of the average \sigmah are shown, derived from data of the PAWS \citep{schinnerer13}. Columns 7 to 10 contain SFR and $\Sigma_{\textrm{SFR}}$ derived from: A) FUV+24$\mu$m and B) H$\alpha$+24$\mu$m. The division of the galaxy in sub-regions is illustrated in Fig.~\ref{fig:regions}, radial profiles of \sigmah\ and \sigmasfr\ are illustrated in Fig.~\ref{fig:h2sfr}.}
\begin{tabular}{cccccccccc}
\hline
\multicolumn{1}{c}{Name} 		& \multicolumn{2}{c}{Interval} 	& \multicolumn{2}{c}{\# of clusters} 	& \multicolumn{1}{c}{$\langle$\sigmah$\rangle$} & \multicolumn{1}{c}{SFR$_A$} & \multicolumn{1}{c}{$\langle\Sigma_{\textrm{SFR,A}}\rangle$} & \multicolumn{1}{c}{SFR$_B$} & \multicolumn{1}{c}{$\langle\Sigma_{\textrm{SFR,B}}\rangle$} \\
\multicolumn{1}{c}{} 		& \multicolumn{1}{c}{arcsec} & \multicolumn{1}{c}{kpc} 	& \multicolumn{1}{c}{total} & \multicolumn{1}{c}{selection}	& \multicolumn{1}{c}{(\msun/pc$^2$)} & \multicolumn{1}{c}{$\left(\textrm{M}_{\odot}/\textrm{yr}\right)$} & \multicolumn{1}{c}{$\left(\textrm{M}_{\odot}/\textrm{yr kpc}^{-2}\right)$}  & \multicolumn{1}{c}{$\left(\textrm{M}_{\odot}/\textrm{yr}\right)$} & \multicolumn{1}{c}{$\left(\textrm{M}_{\odot}/\textrm{yr kpc}^{-2}\right)$} \\
\multicolumn{1}{c}{(1)} 		& \multicolumn{1}{c}{(2)}  & \multicolumn{1}{c}{(3)} 	& \multicolumn{1}{c}{(4)} 	& \multicolumn{1}{c}{(5)} & \multicolumn{1}{c}{(6)}& \multicolumn{1}{c}{(7)} & \multicolumn{1}{c}{(8)} & \multicolumn{1}{c}{(9)} & \multicolumn{1}{c}{(10)} \\
\hline
\hline
Entire 	& $-$ 			& $-$		& 2839	& 1625	& 30.4 	& 2.098  	& 0.0160 	& 1.803  	& 0.0138		\\
nocentr	& $>35.0$			& $>1.30$		& 2653	& 1471	& 25.3	& 1.636  	& 0.0139 	& 1.437  	& 0.0122		\\
SA 		& $-$ 			& $-$		& 1100	& 668	& 55.3 	& $-$  	& $-$ 	& $-$  	& $-$ 		\\
IA 		& $-$ 			& $-$		& 1553	& 803	& 16.5 	& $-$  	& $-$ 	& $-$  	& $-$ 		\\
\hline
MR  		& $23.0-35.0$ 		& $0.85-1.30$	& 122	& 106 	& 133.6	& 0.220  	& 0.0732 	& 0.176  	& 0.0584		\\ 
Bin 1		& $35.0-85.3$ 		& $1.30-3.17$	& 626	& 367	& 58.8	& 0.524  	& 0.0222 	& 0.404  	& 0.0171		\\
Bin 2		& $85.3-122.1$ 	& $3.17-4.54$	& 683	& 367	& 30.7	& 0.396  	& 0.0178 	& 0.314  	& 0.0141		\\
Bin 3		& $122.1-149.7$ 	& $4.54-5.56$	& 640	& 368	& 28.0	& 0.385  	& 0.0227 	& 0.373  	& 0.0220		\\
Bin 4		& $>149.7$ 		& $>5.56$		& 704	& 369	& 8.7		& 0.331  	& 0.0060 	& 0.346  	& 0.0063		\\
\hline	
EA 1		& $35.0-96.3$ 		& $1.30-3.58$	& 830	& 468	& 50.9	& 0.595  	& 0.0203 	& 0.448  	& 0.0153		\\
EA 2		& $96.3-144.9$ 	& $3.58-5.38$	& 994	& 564	& 32.0	& 0.641 	& 0.0218 	& 0.578  	& 0.0197		\\
EA 3		& $144.9-184.5$ 	& $5.38-6.85$	& 592	& 310	& 11.4	& 0.286  	& 0.0097 	& 0.293  	& 0.0100		\\
EA 4		& $>184.5$ 		& $>6.85$		& 237	& 129	& 7.0		& 0.114  	& 0.0039 	& 0.117  	& 0.0040		\\
\hline
\end{tabular}
\label{tab:properties}
\end{table*}

From Fig.~\ref{fig:regions} is clear that, for both divisions (equal number and equal area), each annulus consists of part of a spiral-arm and part of an inter-arm environment, but with different fractions. In order to understand the effect of the spiral arms on the resulting cluster populations, we divide also the cluster sample into arm and inter-arm environments. 
The arm is defined based on $V$ band brightness: from the F555W mosaic, smoothed with a boxcar average of 200 pixels, we consider all pixels with mag $<$ 28.231 (surface brightness of $\sim1.9\times10^{-20}$ erg/cm$^2$/\AA ) as part of the spiral-arm (SA).
The remaining area of the galaxy is considered inter-arm (IA). This division is very similar to the one used in \citet{haas2008} to separate the galaxy into regions of different backgrounds and allows us to make a direct comparison with their results. In addition, this cut on the magnitude also gives similar numbers of clusters in the SA and IA environments (see Tab.~\ref{tab:properties}).
A contour of the spiral arm region is shown in Fig.~\ref{fig:regions}. Once again, the central area of the galaxy, within a radius of 35'' from the centre, is excluded from the SA and IA regions. 

\begin{figure*}
\centering
\subfigure{\includegraphics[width=0.33\textwidth]{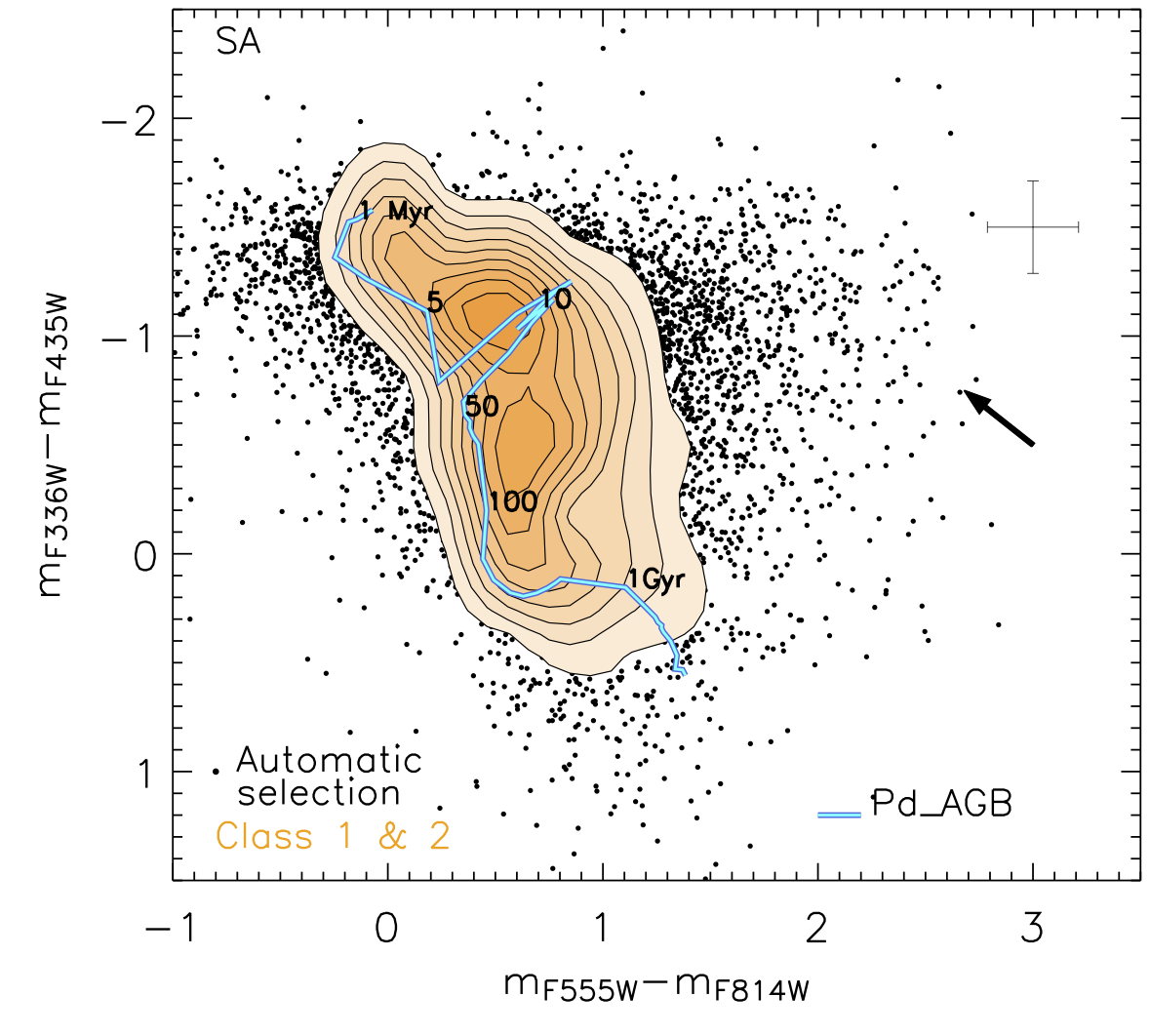}}
\subfigure{\includegraphics[width=0.33\textwidth]{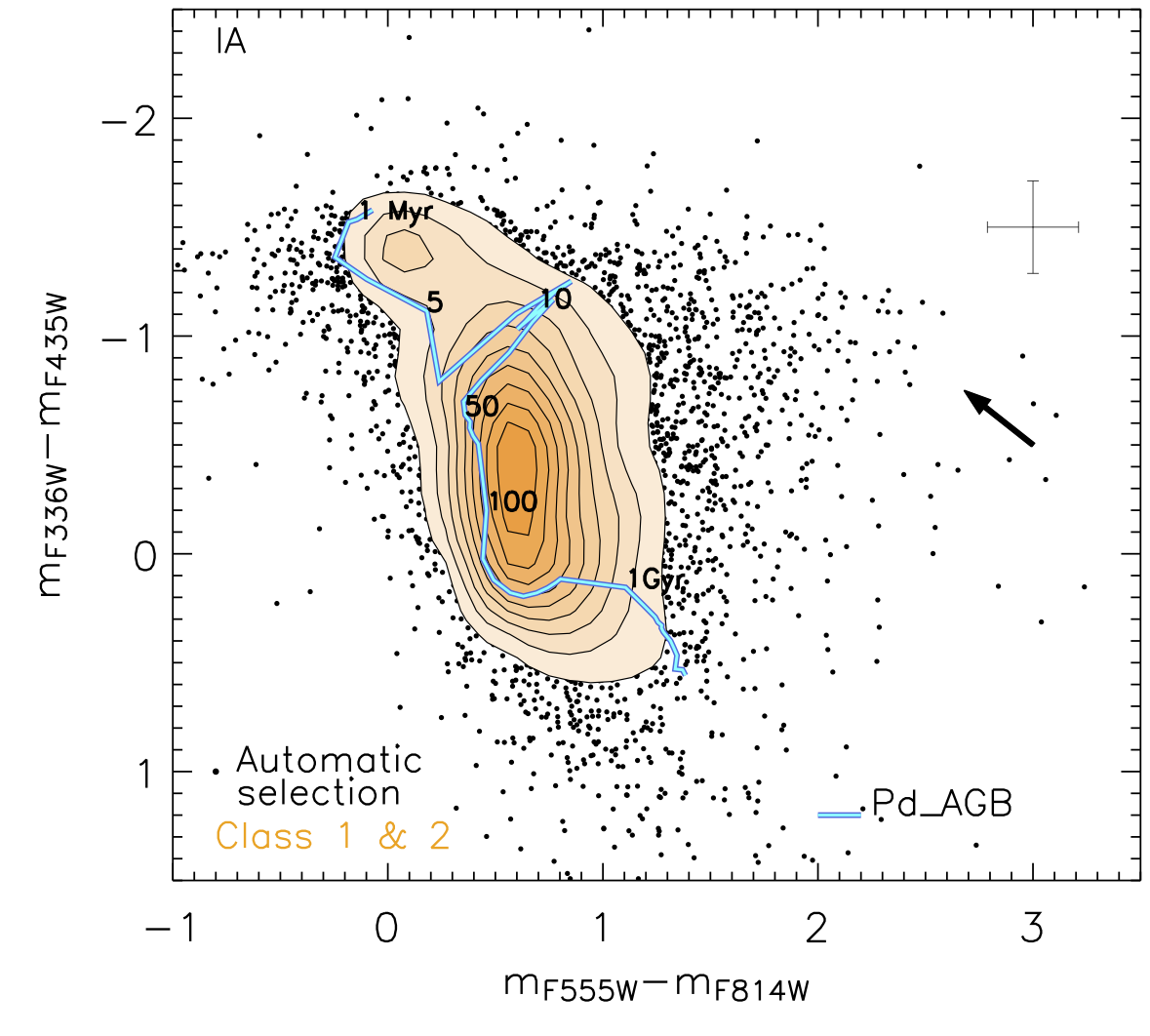}}
\subfigure{\includegraphics[width=0.33\textwidth]{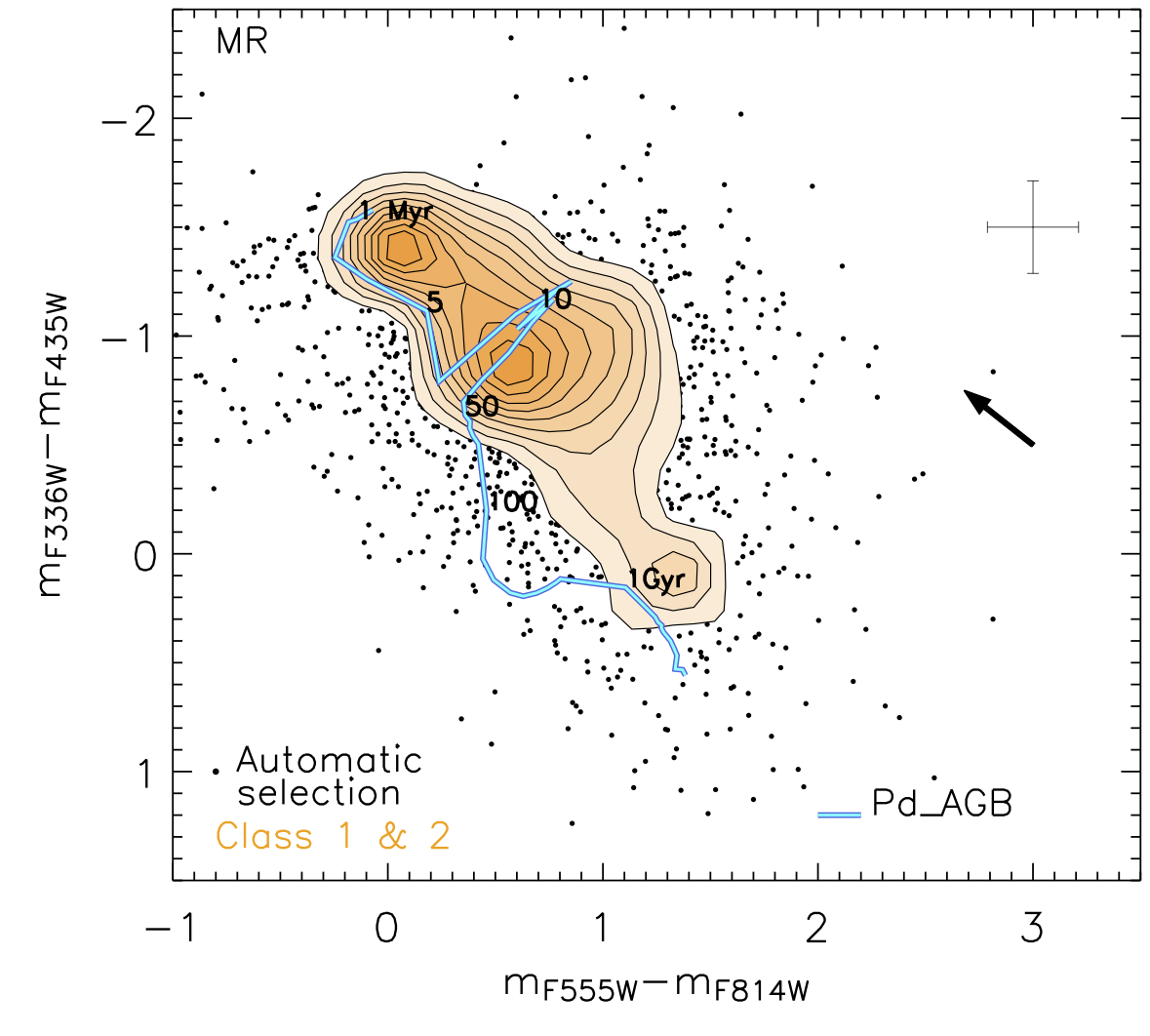}}
\subfigure{\includegraphics[width=0.4\textwidth]{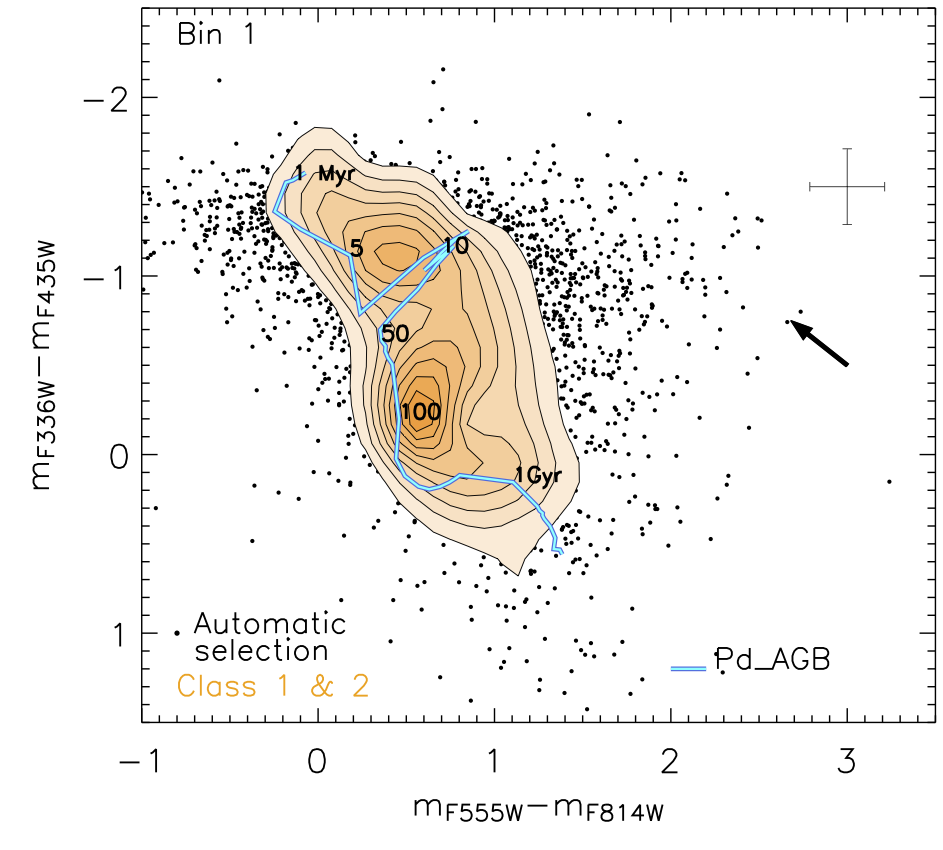}}
\subfigure{\includegraphics[width=0.4\textwidth]{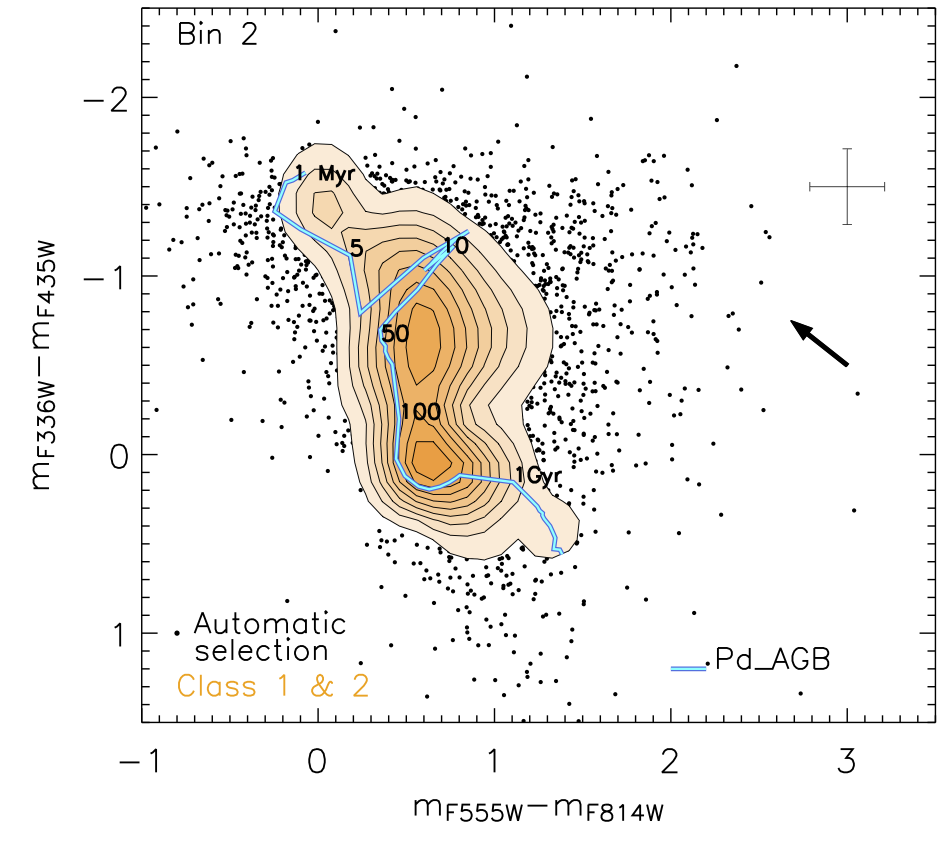}}
\subfigure{\includegraphics[width=0.4\textwidth]{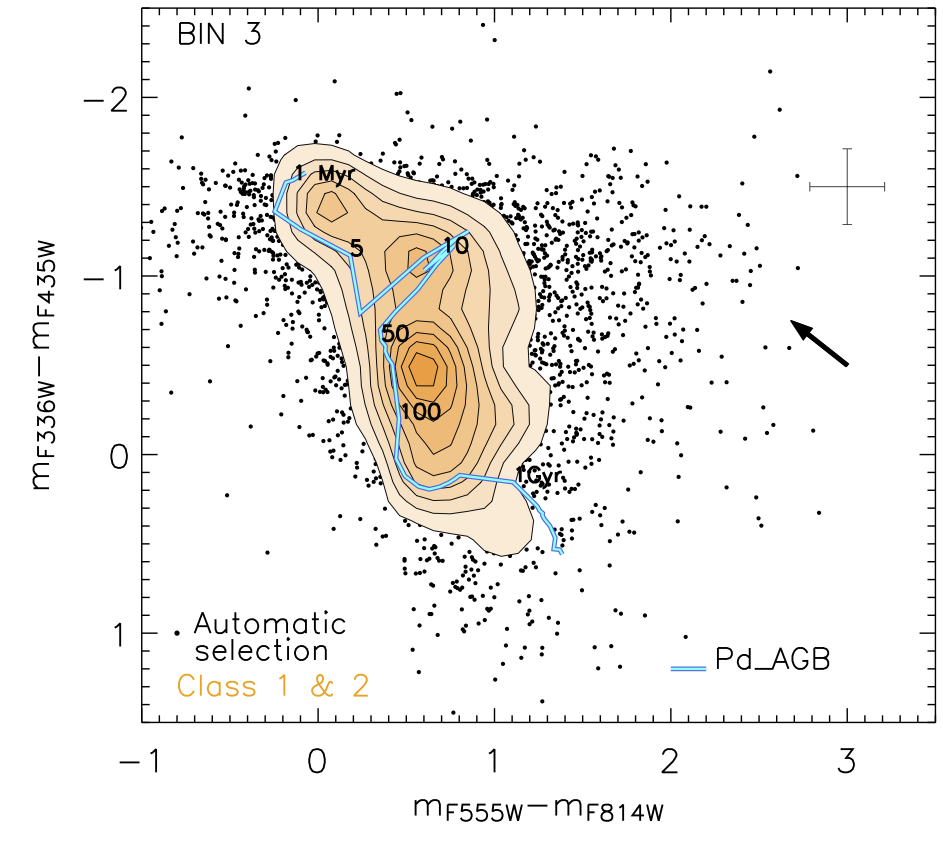}}
\subfigure{\includegraphics[width=0.4\textwidth]{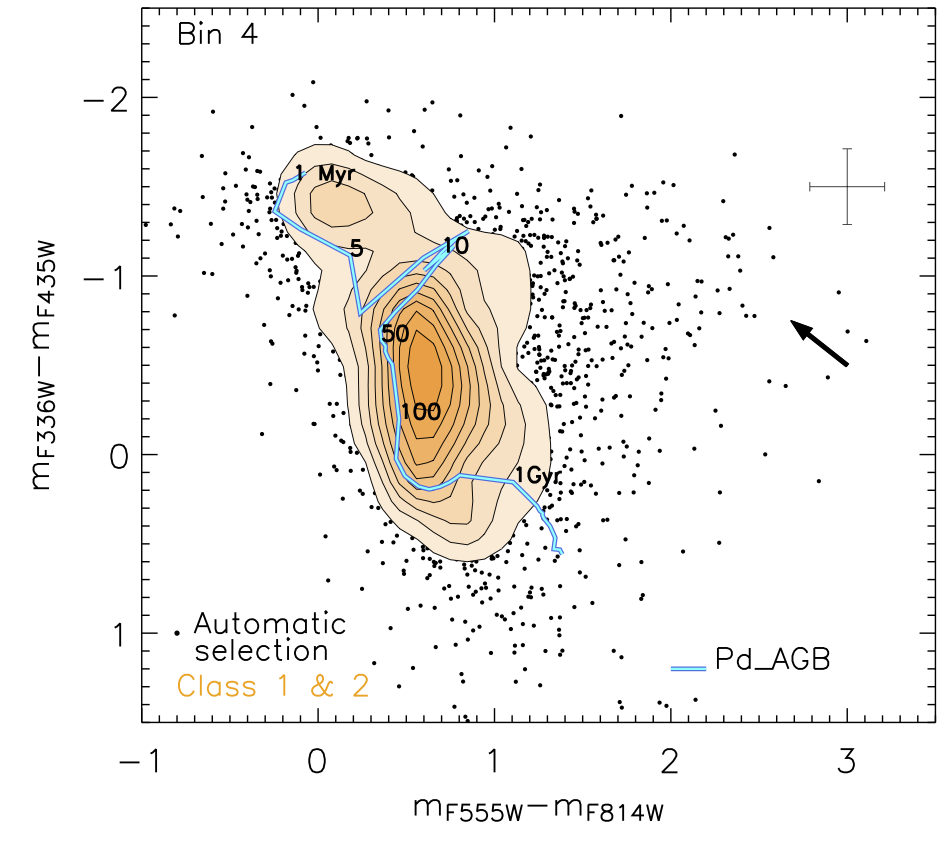}}
\caption{Colour-colour diagrams of the cluster sub-samples. The axes are $V-I$ (x-axis) and $U-B$ (y-axis) colours. Black points are all sources that passed the automatic selection, while the orange shaded area shows the colour distribution for class 1 and 2 clusters. Uncertainties on the colours (associated with typical photometric uncertainties of 0.15 mag) are shown in the top-right corner of each panel. The black solid arrows show how clusters move if corrected for a reddening of E(B-V)$=0.2$. The blue lines show simple stellar population (SSP) evolutionary tracks from Padova-AGB models covering ages from 1~Myr to 14~Gyr.}
\label{fig:ccd_bins}
\end{figure*}

The colour-colour diagrams of the population divided into regions are in Fig.~\ref{fig:ccd_bins}. 
The evolutionary track obtained from the \texttt{Yggdrasil} evolutionary models \citep{yggdrasil} is overplotted. Most of the clusters have ages between 10~Myr and 1~Gyr, with noticeable differences between regions. Bin 1 and Bin 3 seem to host, on average, younger populations compared to Bin 2 and 4. The molecular ring has a colour distribution that is clearly very different from all the others, a sign that this region could be biased by incompleteness against old and red clusters. Comparing the arm and inter-arm environments, we note that clusters in the spiral arm are on average younger than clusters in the inter-arm.

\begin{figure}
\centering
\includegraphics[width=\columnwidth]{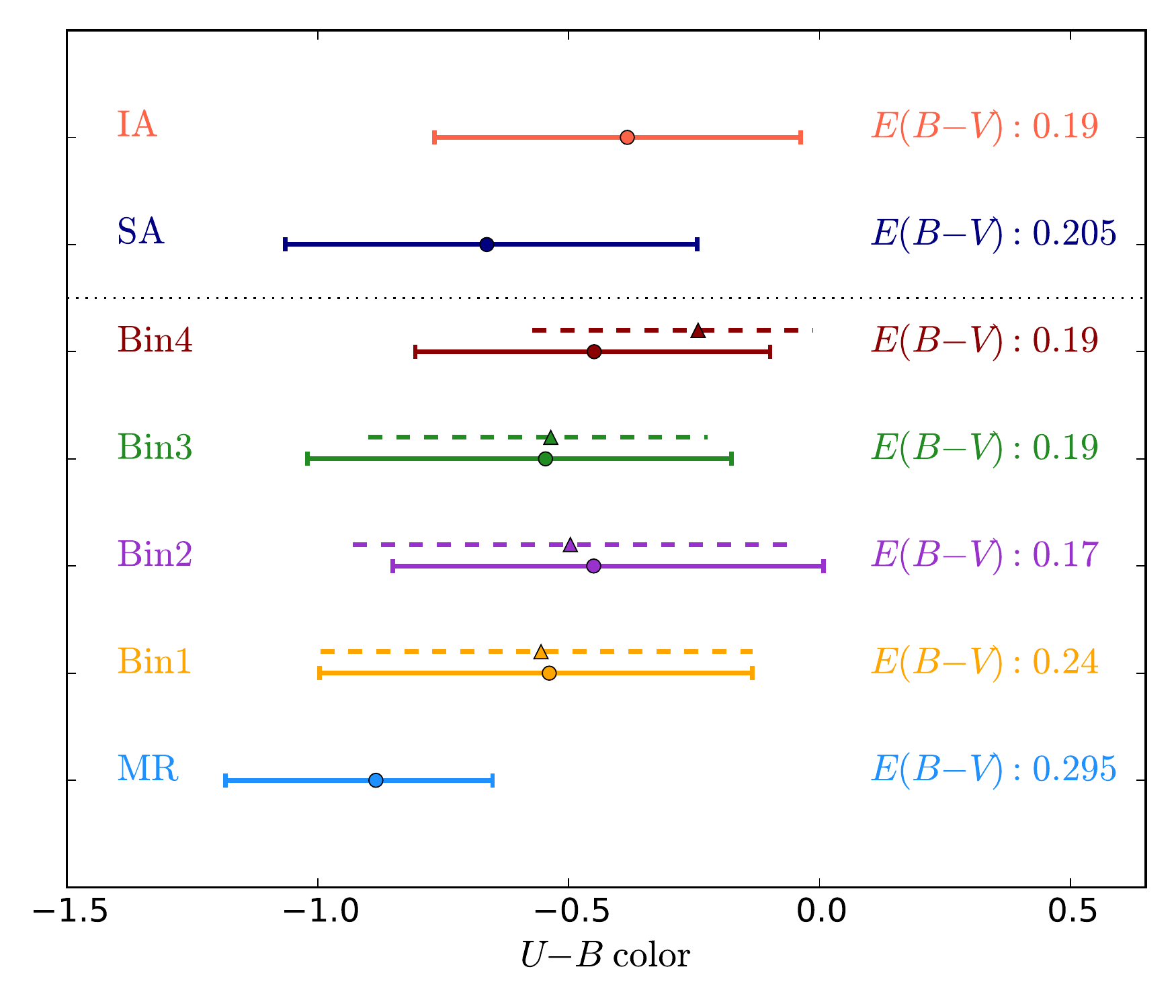}
\caption{$U-B$ colour, sensitive to age, of the sub-samples. Median value, 1st and 3rd quartiles are collected for each sample (solid lines). Dashed lines are used for equal area bins. Median $E(B-V)$ values (in mag units) for each of the sub-regions are displayed on the right of each distribution.}
\label{fig:ub_bins}
\end{figure}

The median, 1st and 3rd quartiles of the $U-B$ colours in each region are shown in Fig.~\ref{fig:ub_bins}. The colours for equal-area bins are also reported, showing very little difference from the equal-number bins. 
Trends similar to what was previously observed are recovered. Clusters in Bin 2 and 4 are on average older than in Bin 1 and 3. The $U-B$ colour distribution of the MR is very different from the other radial bins, possibly also due to the higher extinction in this region. However, extinction alone cannot explain the lack of sources populating the 100~Myr region in Fig.~\ref{fig:ccd_bins}.
The main difference is recovered again when comparing clusters in SA and IA environment, which have an average colour difference of $\sim0.3$ mag.
Median values of $E(B-V)$ derived from the SED fitting are also displayed in Fig.~\ref{fig:ub_bins}: as expected the MR is the region with the highest extinction. The median extinction values do not vary much in the other bins and therefore extinction alone can not explain the differences observed in the distribution of $U-B$ colours.


\subsection{Map of the gas surface density} 
\label{sec:co_data}
\begin{figure}
\centering
\includegraphics[width=\columnwidth]{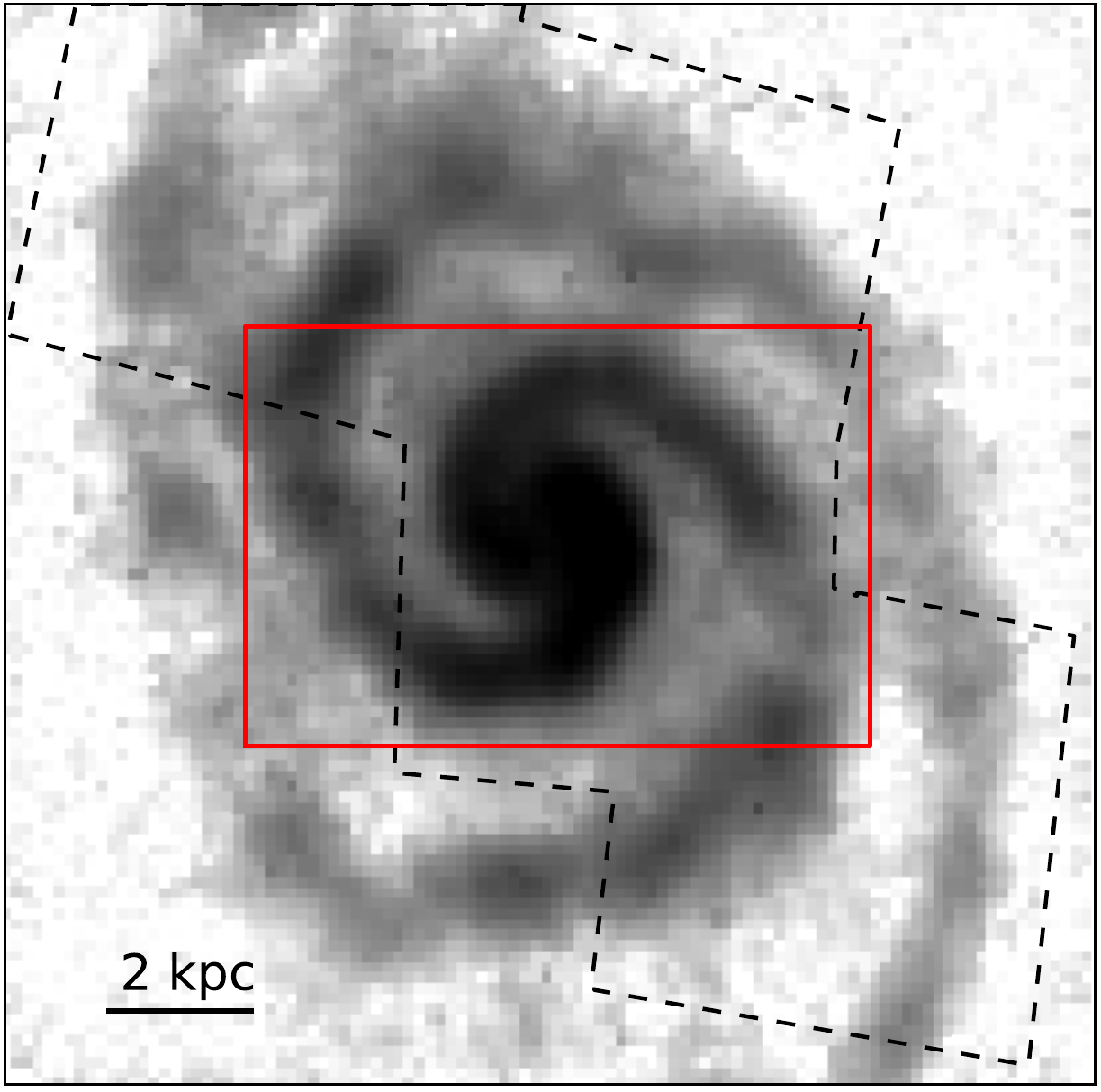}
\caption{CO(1-0) map of M51 from the PAWS survey \citep{schinnerer13}, single-dish observations ($22''.5$ angular resolution). The area covered by the high-resolution interferometry ($1''.1$ angular resolution) is given by the red box at the centre. The footprint of the UVIS observations (black dotted line) is also overplotted. The extracted \sigmah\ radial profile is plotted in Fig.~\ref{fig:h2sfr}}
\label{fig:co}
\end{figure}

\begin{figure}
\centering
\includegraphics[width=\columnwidth]{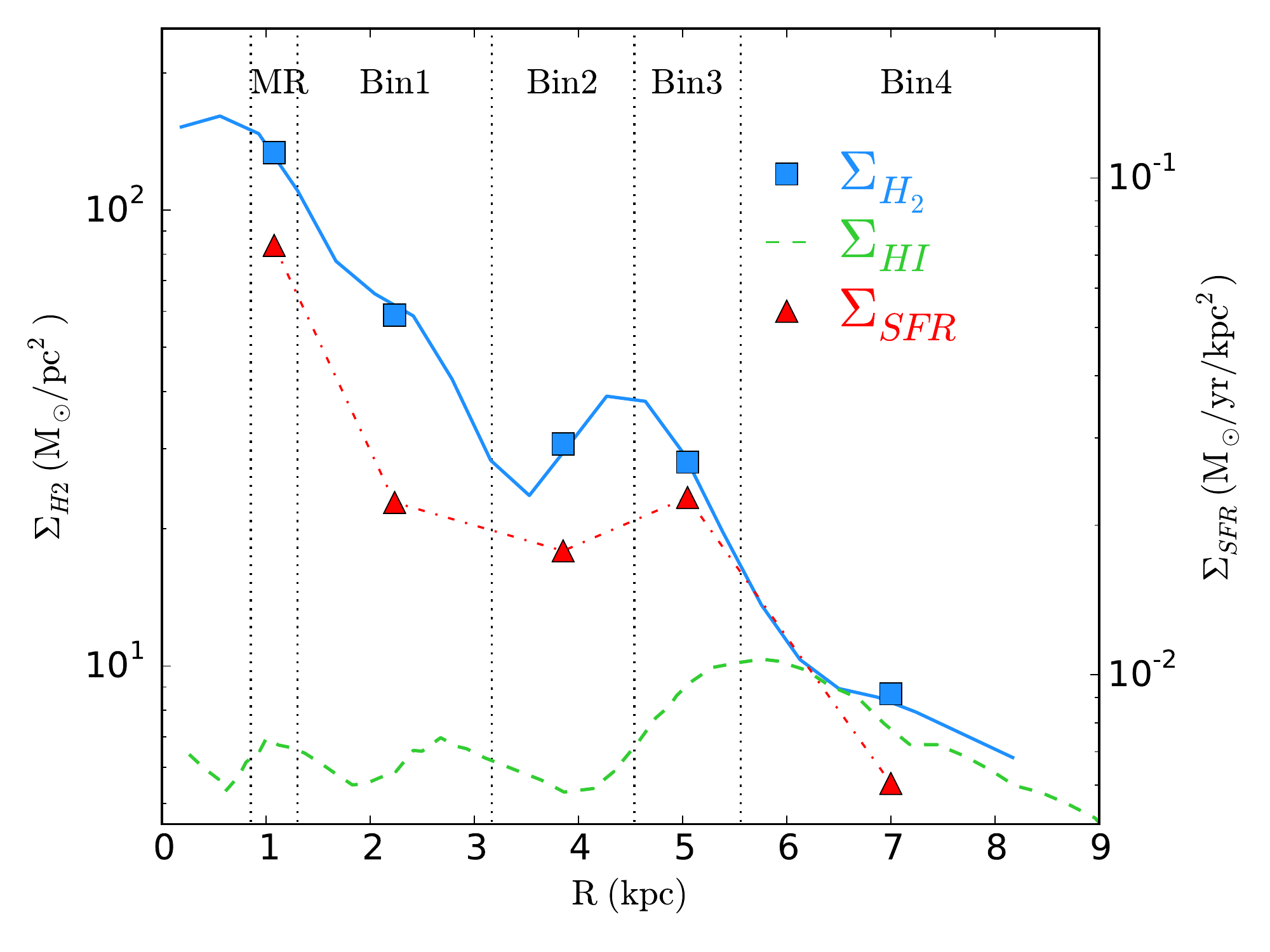}
\caption{Radial profile of the \sigmah (blue line) with the averages in each radial annulus (blue squares, values from Tab.~\ref{tab:properties}) Both the profile and the averages are calculated considering only the area of the galaxy covered by the UVIS observations. Average \sigmasfr\ in the annuli, derived from $FUV+24$ \mum\ (see Tab.~\ref{tab:properties}), are shown as red triangles. The surface density radial profile of HI, taken from \citet{schuster2007}, is shown as a green dashed line.
The black dotted vertical lines mark the edges of the radial regions considered.}
\label{fig:h2sfr}
\end{figure}

In order to investigate how the cluster formation and disruption processes are affected by the environment, we consider the properties of the molecular gas in M51. 
We used CO(1-0) single-dish mapping (angular resolution of 22.5 arcsec) covering the whole galaxy, in order to calculate average values for the surface density of the molecular gas (H$_2$). 
These data are made available via the PAWS project\footnote{http://www2.mpia-hd.mpg.de/PAWS/PAWS/Home.html} \citep{schinnerer13}. 
We refer to \citet{pety2013} for the details on data acquisition and reduction. 
Although the low resolution mapping suffers from beam dilution \citep[e.g.][]{leroy2013}, it enables us to recover the gas density also in the outer parts of the galaxy.  
The conversion from CO intensity to \sigmah\ is made via the conversion factor $X_{CO} = 2\times10^{20}$ cm$^{-2}$ K$^{-1}$ km$^{-1}$ s, used in \citet{schinnerer13}. 
Our cluster population is restricted to regions where we have UVIS coverage and for this reason we consider in the CO map only the region enclosed by the UVIS footprint (see Fig.~\ref{fig:co}). After masking the rest of the map, we measure how \sigmah\ varies radially (Fig.~\ref{fig:h2sfr}). Average values of \sigmah\ in each of the bins defined in Section~\ref{sec:radial_division} are given in Tab.~\ref{tab:properties}.
The H$_2$ surface density decreases monotonically moving outwards in the galaxy but between 4 and 5 kpc from the centre a second peak appears, in correspondence with the location of the outer part of the spiral arms.
When averaged over the radial bins considered, \sigmah\ is very high in the MR, but it rapidly decreases moving to outer bins. A similar decrease is found when the equal-area (EA) bins are considered.

A catalog with a list of the GMCs and their derived properties was produced by \citet{colombo2014a} (hereafter C14). It relies on the high-resolution map of the PAWS survey and therefore is limited to the area at the centre of the galaxy covered with interferometric observations (see Fig.~\ref{fig:co}). The area covered extends only to part of our Bin 2 region and the comparison of cluster properties with GMC properties (in Section~\ref{sec:gmc}) will be limited to those internal regions.

\subsection{Star formation rate in the galaxy} 
\label{sec:sfrmap}
The star formation rate (SFR) of M51 has been calculated using the far-$UV$ (FUV) emission from GALEX, corrected for the presence of dust via  the 24\mum\ emission from Spitzer/MIPS, using the recipe from \citet{hao2011}. The SFR has been normalized to a \citet{kroupa2001} IMF in the stellar mass interval $0.1-100$ \msun. A second value for the SFR has been derived using \ha\ emission instead of FUV. Also in this case, the 24\mum\ emission has been used to obtain an extinction-corrected SFR (using the recipe by \citealp{kennicutt2009}). Both methods provide mean SFR values over the last $\sim100$~Myr. The SFR values obtained for the whole galaxy and for the sub-regions defined in Section~\ref{sec:radial_division} are displayed in Tab.~\ref{tab:properties}. Differences between the two methods are in general below $\sim20\%$. 

An average SFR surface density, \sigmasfr, has also been derived for each region.
As for the surface density of H$_2$, \sigmasfr\ also decreases almost monotonically from the centre to the outskirts of the galaxy, with the exception of a small bump in correspondence to Bin 3 (Fig.~\ref{fig:h2sfr}). The molecular ring has a \sigmasfr\ that is much higher than in the radial bins, while the small value in Bin 4 suggests that little star formation happens in the outer part of the galaxy. The values of \sigmasfr\ are very similar in bins 1, 2 and 3, and a factor of 2 smaller in Bin 4.  

\section{Cluster analysis}
\label{sec:analysis}
\subsection{Luminosity Functions} 
\label{sec:lumfunc}
The luminosity function (LF) is an observed property of the cluster sample, obtained directly from photometry. 
Its shape is generally described by a power law, PL, of the type $dN/dL\propto L^{-\alpha}$ with an almost universal slope $\alpha\sim-2$ retrieved in a wide range of galaxies (see e.g. the reviews by \citealp{whitmore2003} and \citealp{larsen2006b}), including M51, where the slope has been found to vary from $-1.84$ in the F275W filter to $-2.04$ in the F814W filter (if the function is fitted by a single power law, see Paper~I). However, in Paper~I we show that the luminosity function of M51 is best fitted by a double power law, steeper at the bright end, revealing a dearth of bright sources, which is a sign of a similar behavior in the underlying mass function. The study of the average cluster ages at different luminosities and the comparison to Monte Carlo simulations confirmed that the luminosity function can be used to study the properties of the underlying mass function. In a similar way \citet{gieles2006} used the luminosity function of M51 to put constraint on the underlying mass function.

\begin{figure*}
\centering
\subfigure{\includegraphics[width=\columnwidth]{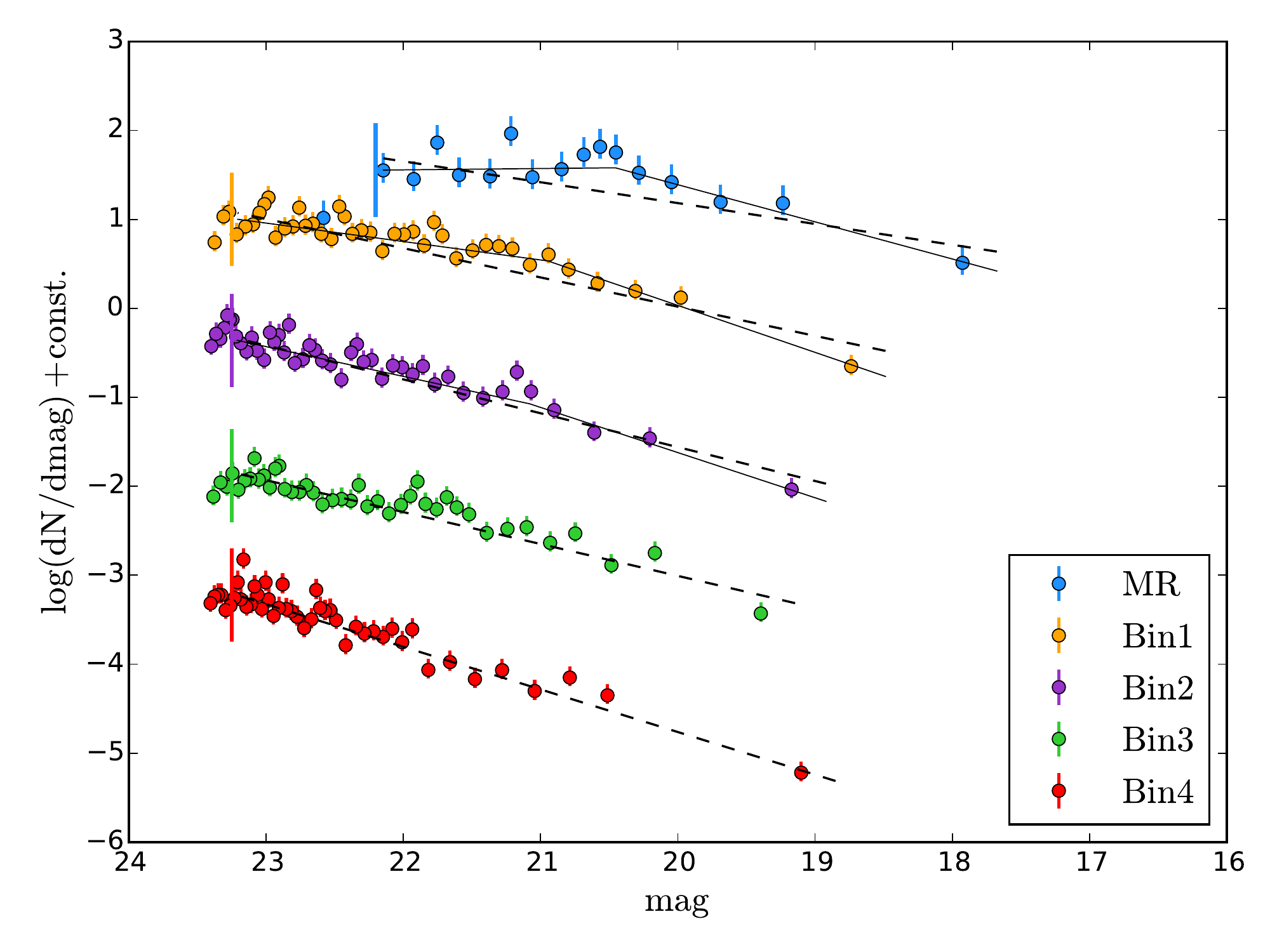}}
\subfigure{\includegraphics[width=\columnwidth]{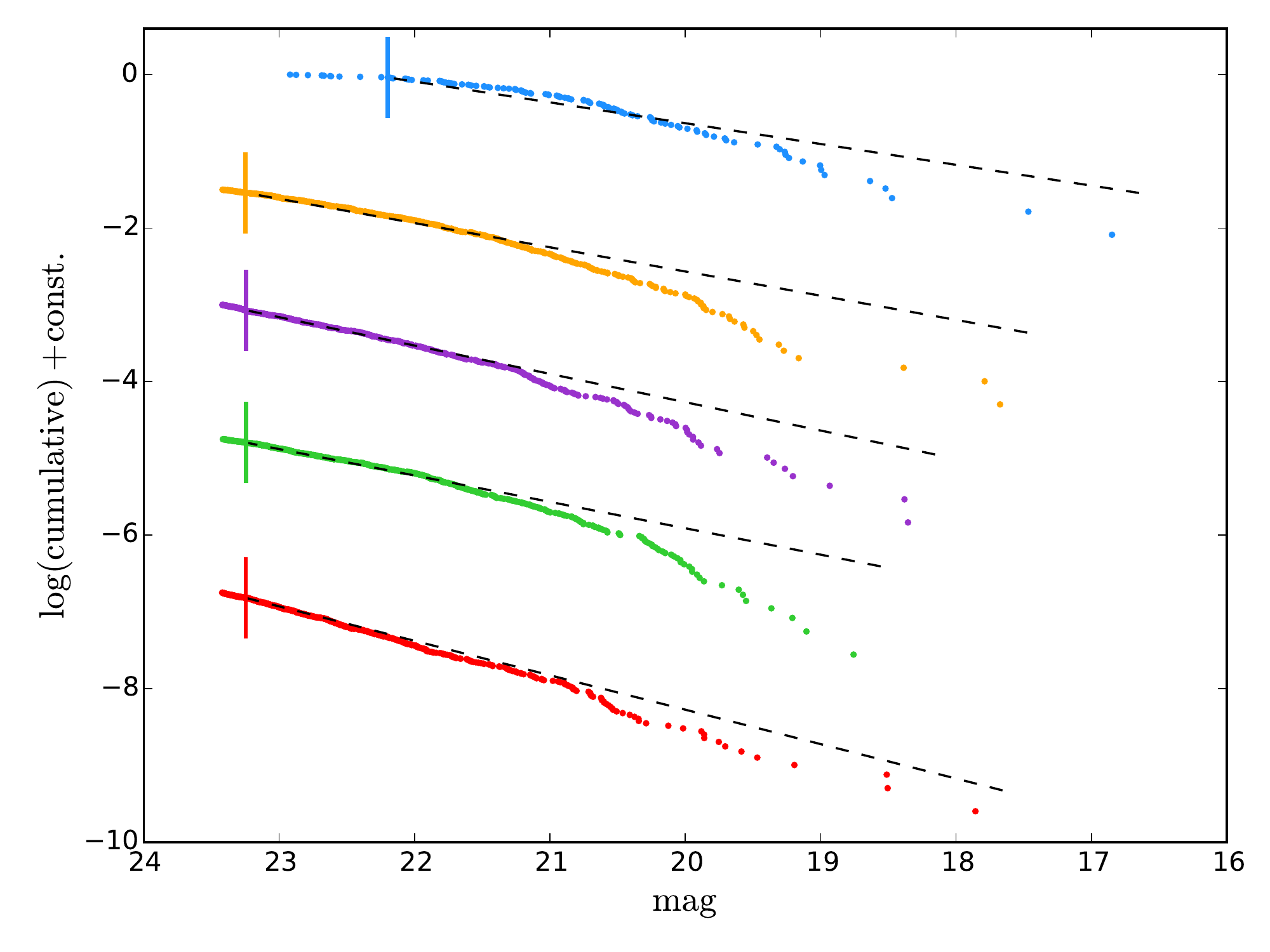}}
\caption{Luminosity functions in the binned form, with equal numbers of clusters per bin (left panel), and in the cumulative form (right panel). The black dashed lines are the best fits with single power laws, the black solid lines are the best fits with double power laws. A vertical line is plotted in correspondence to the magnitude cut used in each bin. Best fit values are listed in Tab.~\ref{tab:lumfit_bins}.}
\label{fig:lumfit_bins}
\end{figure*}

We now study the luminosity function in all filters in each of the radial bins. The function is studied both in a binned form, with luminosity bins containing an equal number of sources, according to \citet{maiz2005}, and in a cumulative form, following \citet{bastian2012}.
In Table~\ref{tab:lumfit_bins} we summarise the outcomes of the luminosity function analysis in the $V$ band (F555W filter). The analyses of the other filters show similar outcomes and are thus omitted. $V$ band luminosity functions are plotted in Fig.~\ref{fig:lumfit_bins}. All clusters down to the completeness limit were included in the fit. 
For the radial annuli (and the SA-IA environments) the same limit of 23.25 mag, as in Paper~I, is used as a lower limit. We remind the reader that this limit was derived looking at where the luminosity distribution was peaked at the faint end, deviating therefore from an expected power-law shape. The value is consistent with the luminosity completeness being set by the magnitude cut in the $V$ band applied in the process of defining the cluster catalogue, as also confirmed by the analysis in Appendix~\ref{sec:a1}. The MR has a brighter incompleteness, and we use as magnitude limit the value of 22.20 mag (Appendix~\ref{sec:a1}).

\begin{table*}
\centering
\caption{Results of the fit of the luminosity functions of the F555W filter in the sub-regions as described in Tab.~\ref{tab:properties}. The luminosity functions are plotted in Fig.~\ref{fig:lumfit_bins}. The magnitude cut used is $23.25$ mag, except for the MR, where the cut is $22.20$ mag.}
\begin{tabular}{ccccccccc}
\hline
\multicolumn{1}{c}{Bin}  	&\multicolumn{2}{l}{single power-law fit} 	& \multicolumn{4}{l}{double power-law fit} & \multicolumn{1}{l}{cumulative fit}\\
\multicolumn{1}{c}{} 	  	& \multicolumn{1}{c}{$\alpha$} & \multicolumn{1}{c}{$\chi^2_{red.}$}		& \multicolumn{1}{c}{$\alpha_1$}	& \multicolumn{1}{c}{$\textrm{mag}_{\textrm{break}}$} 	& \multicolumn{1}{c}{$\alpha_2$} & \multicolumn{1}{c}{$\chi^2_{red.}$} & \multicolumn{1}{c}{$\alpha$}	\\
\hline
\hline
SA  		& $1.75\ _{\pm 0.05}$	& 3.68 &	$1.22\ _{\pm 0.05}$ &	$20.07\ _{\pm  0.10}$ & 	$2.33\ _{\pm 0.08}$  		& 0.73 	& $2.03\ _{\pm 0.01}$ \\
IA  		& $2.30\ _{\pm 0.04}$	& 1.31 &	$2.02\ _{\pm 0.10}$ &	$21.93\ _{\pm  0.25}$ & 	$2.54\ _{\pm 0.10}$  		& 1.12 	& $2.51\ _{\pm 0.01}$ \\
\hline
MR		& $1.59\ _{\pm0.10}$		& 1.75	& $0.96\ _{\pm0.22}$	& $20.45\ _{\pm0.31}$	& $2.05\ _{\pm0.19}$	& 0.79	& $1.97\ _{\pm 0.03}$ \\
\hline
Bin 1		& $1.82\ _{\pm0.05}$	& 2.01	& $1.52\ _{\pm0.08}$	& $20.94\ _{\pm0.26}$	& $2.32\ _{\pm0.16}$	& 1.07	& $2.08\ _{\pm 0.01}$ \\
Bin 2		& $1.95\ _{\pm0.05}$	& 1.27	& $1.84\ _{\pm0.09}$	& $21.07\ _{\pm0.56}$	& $2.27\ _{\pm0.20}$	& 1.07	& $2.17\ _{\pm 0.02}$ \\
Bin 3		& $1.90\ _{\pm0.05}$	& 1.26	& $-$				& $-$				& $-$				&  $-$	& $2.16\ _{\pm 0.01}$ \\
Bin 4		& $2.20\ _{\pm0.06}$	& 1.28	& $-$				& $-$				& $-$				&  $-$	& $2.28\ _{\pm 0.02}$ \\
\hline
EA 1		& $1.90\ _{\pm0.04}$	& 1.77	& $1.51\ _{\pm0.07}$	& $21.16\ _{\pm0.18}$	& $2.37\ _{\pm0.12}$	& 0.86	& $2.12\ _{\pm 0.01}$ \\
EA 2		& $1.92\ _{\pm0.04}$	& 1.39	& $-$				& $-$				& $-$				& $-$	& $2.13\ _{\pm 0.01}$ \\
EA 3		& $2.11\ _{\pm0.05}$		& 1.16	& $1.94\ _{\pm0.17}$	& $21.95\ _{\pm0.68}$	& $2.23\ _{\pm0.12}$	& 1.16	& $2.29\ _{\pm 0.02}$ \\
EA 4		& $2.28\ _{\pm0.09}$	& 0.79	& $-$				& $-$				& $-$				& $-$	& $2.36\ _{\pm 0.04}$ \\
\hline
\end{tabular}
\label{tab:lumfit_bins}
\end{table*}

When the functions are fitted by single power laws, Bin 1 has the shallowest slope, $\alpha=1.82 \pm0.05$, while Bin 4 has the steepest, $\alpha=2.20 \pm0.06$. Bin 2 and 3 have slopes in between those two values. The functions are better fitted with double power law in Bin 1 and 2, but not in Bin 3 and 4. Bins of equal area show similar trends. In this case only bins EA 1 and EA 3 are better fitted with a double power law. 
The single power-law fit of the luminosity function in the MR region is very shallow ($\alpha=1.59 \pm0.10$). We note, however, that this is driven by the very flat part at magnitudes fainter than $20.5$ mag: after the break, the function has a slope comparable with the other bins ($\alpha=2.05 \pm0.19$). This result suggests that incompleteness could be affecting the MR region even at magnitudes brighter than 22.2 mag.
The plot of the cumulative luminosity function in Fig.~\ref{fig:lumfit_bins} (right panel) shows that, in all radial bins, there is a drop in the number of observed clusters at bright luminosities, compared to what would be expected from the best fit with a single power law.

\begin{figure}
\centering
\includegraphics[width=\columnwidth]{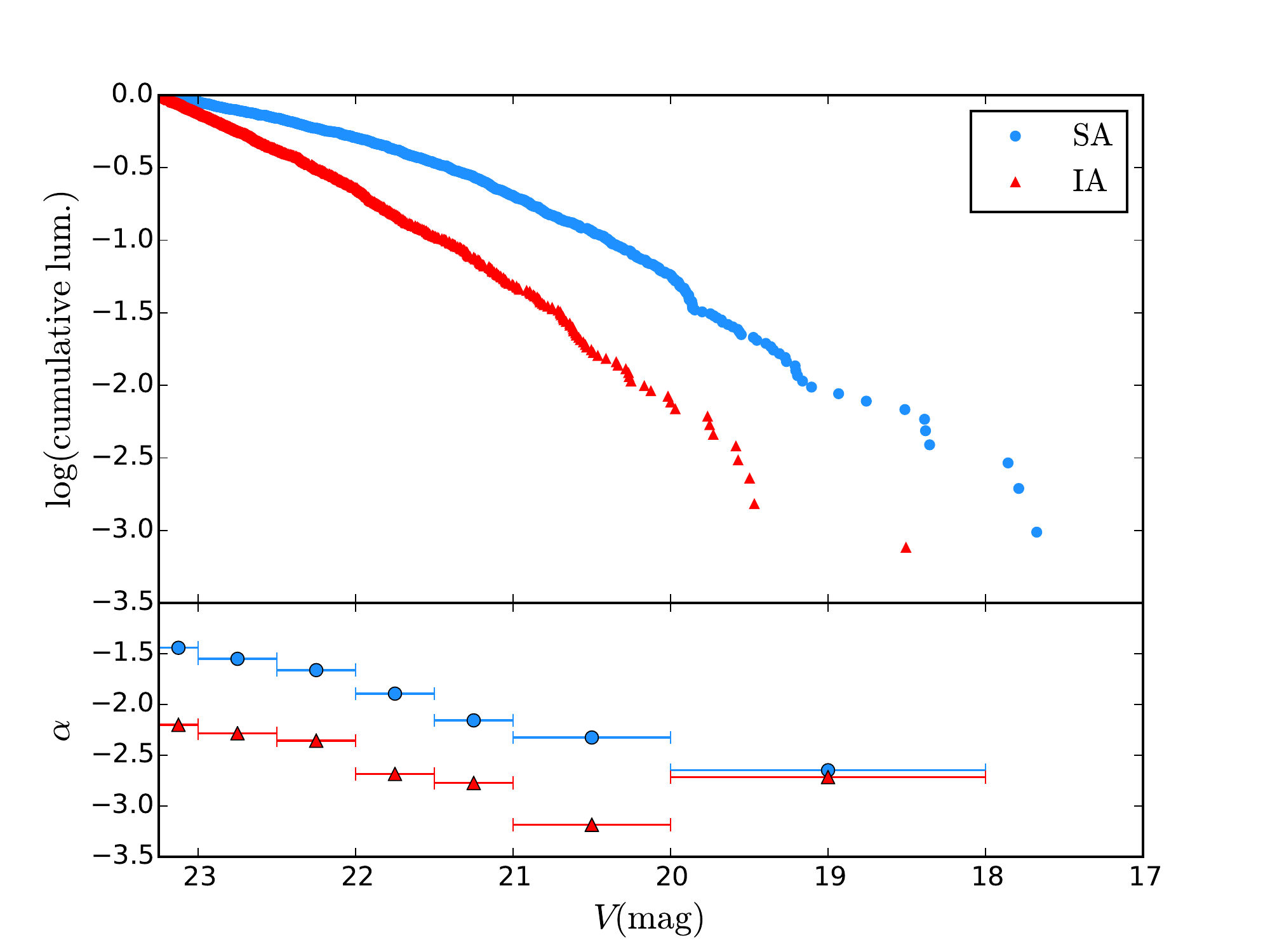}
\caption{Cumulative luminosity functions for the arm (SA, blue circles) and inter-arm (IA, red triangles) regions. The bottom panel shows the slopes of subparts of the functions (in bins of variable sizes, from 0.5 mag to 2 mag).}
\label{fig:lumfit_saia}
\end{figure}

Luminosity functions in the arm and inter-arm regions present significant differences. In Fig.~\ref{fig:lumfit_saia} the cumulative functions of arm and inter-arm clusters are compared. The slopes of the luminosity function at different luminosities (bottom panel) are calculated dividing the function in bins of 0.5 mag and fitting each bin with a power law (at high luminosities bins of 1 mag and 2 mag width have been considered to compensate for the low number of clusters). In the SA case the function is on average very shallow (best fit with a single power-law is $\alpha=1.75$) and it is clearly truncated, as the bright-magnitude sources fall off the slope observed at lower magnitudes. The improvement in the value of the recovered reduced $\chi^2$ in the double power-law fit confirms it. On the other hand, the function of the inter-arm region is steeper. It may present a truncation, since the slope steepens when moving to brighter magnitudes (bottom panel of Fig.~\ref{fig:lumfit_saia}), but can also be well described by a single power-law of slope $\alpha=2.30$. 

A similar trend for arm and inter-arm division was found already by \citet{haas2008}. The galaxy was divided into regions of different surface brightness, in a very similar way. The luminosity function of the bright regions of the galaxy was found to have a shallow low-luminosity end, and therefore also a more evident truncation. This shallow slope is not what is expected from a young population of clusters with an initial cluster mass function with a slope of $-2$. \citet{haas2008} invoked blending of the sources (which in the arms are frequently clustered) as the cause of turning low-luminosity sources into brighter ones, flattening the slope of the function. The higher background can also cause incompleteness for the low-luminosity sources, as it is the case in the MR region. From the analysis with the Monte Carlo populations in Paper~I we know that cluster disruption can also cause a flattening of the function. 
All these factors can have an impact on the shape of the luminosity function. We note, however, that the difference in slopes is not restricted to sources close to the completeness limit. When comparing the luminosity function in arm and inter-arm regions, the difference in slope extends to sources up to $\sim20$ mag (see Fig.~\ref{fig:lumfit_saia}). This is more than 3 magnitudes brighter than the completeness limit and therefore hardly motivated by a difference in completeness. In addition, the completeness test presented in Appendix~\ref{sec:a1} makes use of the scientific frames and, in doing so, takes into account the elevated crowding of the SA region.
A physical interpretation of the difference can be an age difference between arm and inter-arm which would cause the IA luminosity function to be flatter due to the lack of luminous OB stars.
While this is a reasonable possibility, the fact that also the GMC properties are different in the two regions suggest that the difference in the luminosity function is probably due to  environmental effects dominating (see Section~\ref{sec:gmc}).

\subsection{Mass Functions}
\label{sec:massfunc}
We study the mass function, focusing in particular on its high mass end, in sub-regions of M51. We have seen in Paper~I that the mass function of the galaxy, in addition to a $-2$ power-law behavior, presents a drop at high masses, which can be described as an exponential truncation at $M\sim10^5$ \msun\ (i.e. a Schechter functional shape, $dM/dN\propto M^\beta e^{-M/Mc}$, see \citealp{schechter1976}). We investigate now if those properties are the same in all sub-regions of the galaxy. In this analysis we consider a mass-limited sample with $M>5000$ \msun\ and ages $\leqslant200$~Myr. The cut in mass avoids the inclusion of sources with large uncertainties in mass and age derivations. In Appendix~\ref{sec:a1} we show that this constitutes a complete sample. In case of the MR region, instead, we have a mass-limited complete sample for $M>10^4$ \msun\ and ages $\leqslant100$~Myr.
Mass functions are plotted in Fig.~\ref{fig:massfit_mcpop_bins} and Fig.~\ref{fig:massfit_mcpop_saia} both in the binned and cumulative form. In order to verify the agreement on the derived best-fit quantities, we performed the fit of the cluster mass functions using different methods commonly used in the literature. Methods and fit results are described in Sections~\ref{sec:massfunc_maxlk} and \ref{sec:massfunc_bayes}, while in Appendix~\ref{sec:a2} the methods are tested using simulated cluster populations  via Monte Carlo realisations.

\subsubsection{Maximum-likelihood fitting of the mass function}
\label{sec:massfunc_maxlk}
Mass functions were fitted with the maximum-likelihood \texttt{IDL} code \texttt{mspecfit.pro}, implemented by \citet{rosolowsky2005} and also used in the analysis of GMC mass functions in M51 by C14. The code analyses the cumulative mass function considering the possibility that it can be described by a truncated power law, namely:
\begin{equation}
\label{eq:1}
N(M'>M)=N_0\left[\left(\frac{M}{M_0}\right)^{\beta+1}-1\right],
\end{equation} 
where $\beta$ is the power law slope of the differential mass function (i.e., $dN/dM\propto M^\beta$), $M_0$ is the maximum mass of the distribution. At $M=2^{1/(\beta+1)}M_0$ the function deviates from a simple power law, and a truncation is considered statistically significant only if the number of clusters above this limit, $N_0$, is greater than 1, i.e. if also the truncated part of the function is sampled by more than one cluster. More specifically, the code maximizes the likelihood that a set of data (M,N), with the associated uncertainties, is drawn from a distribution of the form of Eq.~\ref{eq:1} with parameters $N_0$,$M_0$ and $\beta$. In order to estimate the uncertainties on the derived parameters, a bootstrapping technique with 100 trials is used to sample the distribution of derived parameters. The uncertainty values that we report in the text are the median absolute deviations of the transformed parameter distribution from the bootstrapping trials. We refer to \citealp{rosolowsky2005} for the formalism of this method.

The recovered results are in Tab.~\ref{tab:massfit_bins}. We consider now all sub-regions except the MR, which, being different in terms of completeness, is discussed later in this section. 
When fitted with a simple power law, the recovered slopes are steeper than $-2$. 
The largest slope difference is observed when comparing SA and IA environments. Similarly to the case of the luminosity function, the SA region has a significantly flatter slope than the one of the IA region.
In all cases the fit results support truncation at high masses (via the high values of the recovered $N_0\gg1$).
The truncation masses do not show substantial variations between the subsamples, as values of $M_0$ are all close to $10^5$ \msun. The largest $M_0$ values are found for SA, Bin1 and Bin4.
Equal area bins show similar results. 

\begin{table*}
\centering
\caption{Results of the fit of the clusters' mass functions divided in subregions with the maximum-likelihood code \texttt{mspecfit.pro}. The functions are fitted both with a truncated power-law and a simple one, using a low-mass cut of: M$>5000$\msun\ (columns 2-5) and M$>10^4$ \msun\ (columns 6-10).
Only clusters younger than 200 Myr (100 Myr for the MR) are considered in the analysis.
The corresponding mass functions are plotted in Fig.~\ref{fig:massfit_mcpop_bins} and Fig.~\ref{fig:massfit_mcpop_saia}.}
\begin{tabular}{lcccccccccccccc}
\hline
\ & \ & \multicolumn{3}{c}{$\rm M>5000$ \msun} & \ & \multicolumn{5}{c}{$\rm M>10^4$ \msun}\\
\hline
\multicolumn{1}{l}{Bin}	& \ & \multicolumn{3}{l}{Truncated PL} 		& \multicolumn{1}{c}{Simple PL}		& \ & \multicolumn{1}{l}{$\rm{N_{YSC}}$}			& \multicolumn{3}{l}{Truncated PL} 			& \multicolumn{1}{c}{Simple PL}		\\
\			& \ & 	$-\beta$			& $M_0$ ($10^5$ \msun)	& $N_0$				& $-\beta$				& \ & \ 	& $-\beta$				& $M_0$ ($10^5$ \msun)	& $N_0$			& $-\beta$				\\
\hline
\hline
nocentr		& \ & $1.91\ _{\pm0.03}$	& $1.33\ _{\pm0.11}$		& $88\ _{\pm18}$	& $2.18\ _{\pm0.04}$ & \ & 868 & $2.17\ _{\pm0.06}$	& $1.74\ _{\pm0.16}$	& $33\ _{\pm5}$ 	& $2.42\ _{\pm0.04}$ \\
SA			& \ & $1.76\ _{\pm0.03}$	& $1.44\ _{\pm0.14}$	& $61\ _{\pm12}$	& $2.07\ _{\pm0.03}$ & \ & 418 & $2.02\ _{\pm0.07}$	& $1.71\ _{\pm0.44}$	& $26\ _{\pm13}$ 	& $2.31\ _{\pm0.02}$ \\
IA			& \ & $1.98\ _{\pm0.03}$	& $1.08\ _{\pm0.18}$	& $45\ _{\pm16}$	& $2.24\ _{\pm0.03}$ & \ & 450 & $2.34\ _{\pm0.03}$	& $1.73\ _{\pm0.46}$	& $10\ _{\pm4}$ 	& $2.51\ _{\pm0.08}$ \\
\hline
MR			& \ & $1.56\ _{\pm0.09}$	& $2.49\ _{\pm0.60}$	& $14\ _{\pm8}$	& $1.85\ _{\pm0.04}$ & \ & 74 & $1.78\ _{\pm0.08}$		& $3.35\ _{\pm0.98}$	& $5\ _{\pm3}$ 		& $2.00\ _{\pm0.11}$ \\
\hline
Bin 1			& \ & $1.77\ _{\pm0.08}$	& $1.67\ _{\pm0.16}$	& $28\ _{\pm8}$ 	& $2.06\ _{\pm0.02}$ & \ & 228 & $2.11\ _{\pm0.08}$	& $2.19\ _{\pm0.53}$	& $9\ _{\pm3}$ 		& $2.33\ _{\pm0.09}$ \\
Bin 2			& \ & $1.86\ _{\pm0.07}$	& $1.08\ _{\pm0.09}$	& $30\ _{\pm9}$ 	& $2.18\ _{\pm0.02}$ & \ & 216 & $2.12\ _{\pm0.10}$	& $1.38\ _{\pm0.30}$	& $12\ _{\pm7}$	& $2.41\ _{\pm0.02}$ \\
Bin 3			& \ & $1.85\ _{\pm0.05}$	& $0.89\ _{\pm0.16}$	& $39\ _{\pm11}$ 	& $2.24\ _{\pm0.06}$ & \ & 220 & $2.22\ _{\pm0.13}$	& $1.24\ _{\pm0.28}$	& $11\ _{\pm4}$	& $2.51\ _{\pm0.07}$ \\
Bin 4			& \ & $1.92\ _{\pm0.08}$	& $1.33\ _{\pm0.39}$	& $20\ _{\pm6}$ 	& $2.14\ _{\pm0.06}$ & \ & 204 & $2.15\ _{\pm0.09}$	& $1.93\ _{\pm0.31}$	& $7\ _{\pm3}$		& $2.35\ _{\pm0.14}$ \\
\hline
EA 1			& \ & $1.81\ _{\pm0.08}$	& $1.46\ _{\pm0.50}$	& $36\ _{\pm18}$ 	& $2.09\ _{\pm0.05}$ & \ & 285 & $2.13\ _{\pm0.06}$	& $1.85\ _{\pm0.24}$	& $12\ _{\pm2}$  	& $2.38\ _{\pm0.11}$ \\
EA 2			& \ & $1.86\ _{\pm0.06}$	& $1.10\ _{\pm0.15}$	& $47\ _{\pm7}$  	& $2.20\ _{\pm0.04}$ & \ & 340 & $2.16\ _{\pm0.08}$	& $1.49\ _{\pm0.28}$	& $16\ _{\pm6}$  	& $2.40\ _{\pm0.08}$ \\
EA 3			& \ & $1.94\ _{\pm0.04}$	& $0.85\ _{\pm0.09}$	& $25\ _{\pm4}$ 	& $2.27\ _{\pm0.06}$ & \ & 161 & $2.19\ _{\pm0.09}$	& $1.08\ _{\pm0.23}$	& $10\ _{\pm4}$ 	& $2.52\ _{\pm0.23}$ \\
EA 4			& \ & $1.82\ _{\pm0.06}$	& $2.00\ _{\pm0.63}$	& $7\ _{\pm3}$ 		& $2.02\ _{\pm0.06}$ & \ & 82 & $2.04\ _{\pm0.09}$	& $4.18\ _{\pm1.72}$		& $2\ _{\pm4}$ 		& $2.15\ _{\pm0.12}$ \\
\hline
\end{tabular}
\label{tab:massfit_bins}
\end{table*}

In order to test the effect of an eventual incompleteness on the mass function, we repeat the analysis considering only clusters with $M>10^4$ \msun. The results are given in Tab.~\ref{tab:massfit_bins}.
The slopes are all steeper than in the previous case, and in particular are all steeper than $-2$. 
The recovered truncation masses does not change significantly in most of the bins, although we notice that the statistical significance of the truncation is reduced because of the smaller number of clusters.

Moving to the analysis of the MR region, we have reported in Tab.~\ref{tab:massfit_bins} the best fits of the mass function only considering clusters with ages $\leqslant100$~Myr, as incompleteness strongly affects older clusters in this region. The slope in the MR is the shallowest of all sub-regions. While, in the case of masses $>5000\ \rm M_{\sun}$, this result can be attributed to partial incompleteness of clusters with low masses, this is not applicable to clusters with masses $\geqslant10^4$ \msun, which we consider to surpass the 90$\%$ completeness limit. Focusing on these latter clusters, the fit suggests that the mass function is different in this region, with a shallow slope $\beta=-1.78 \pm0.08$ ($-2.00 \pm0.11$ in case of a single power-law fit) and possibly no truncation (the statistical significance is low due to the reduced number of clusters in this region).

\begin{figure*}
\centering
\subfigure{\includegraphics[width=0.29\textwidth]{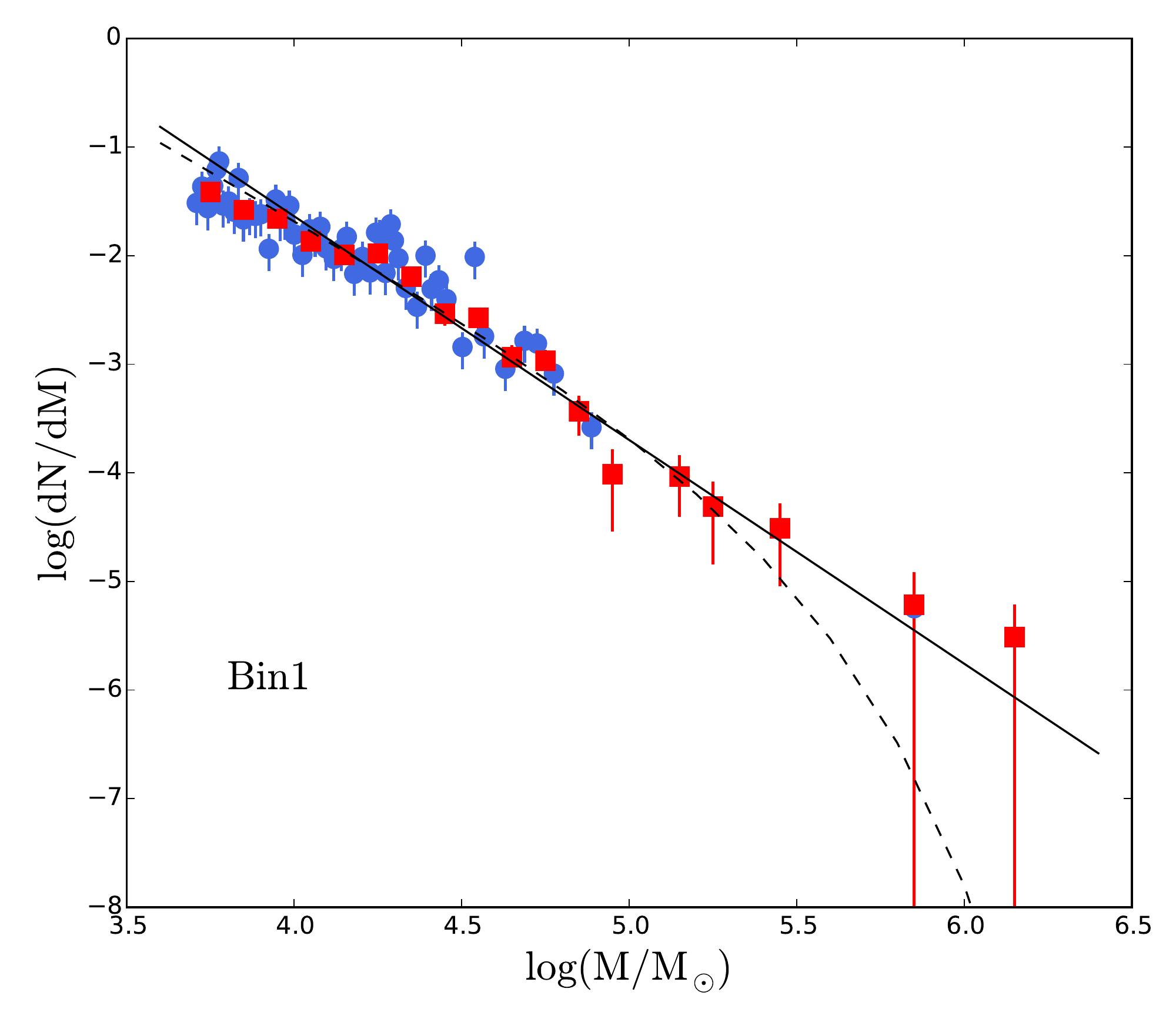}}
\subfigure{\includegraphics[width=0.34\textwidth]{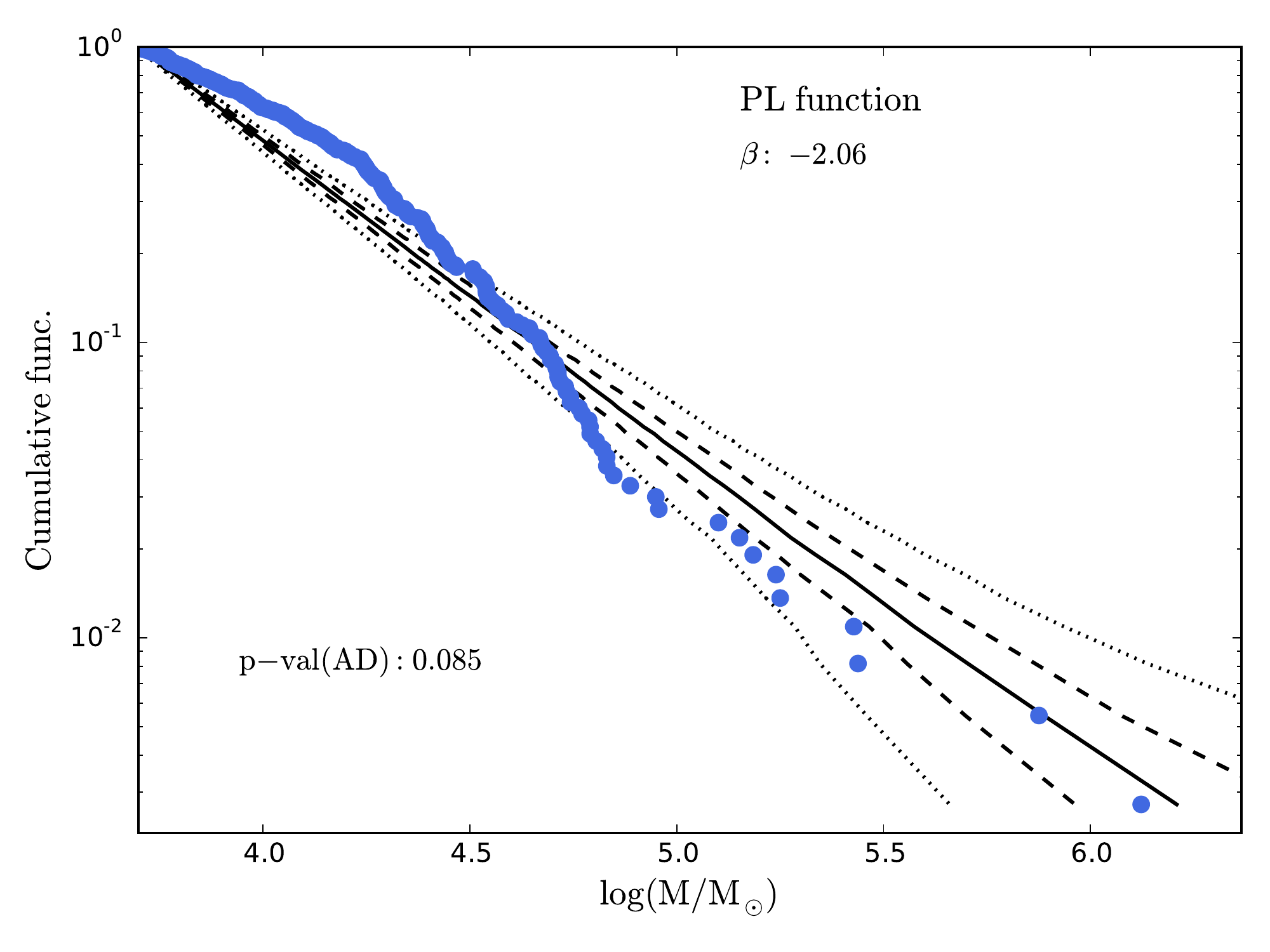}}
\subfigure{\includegraphics[width=0.34\textwidth]{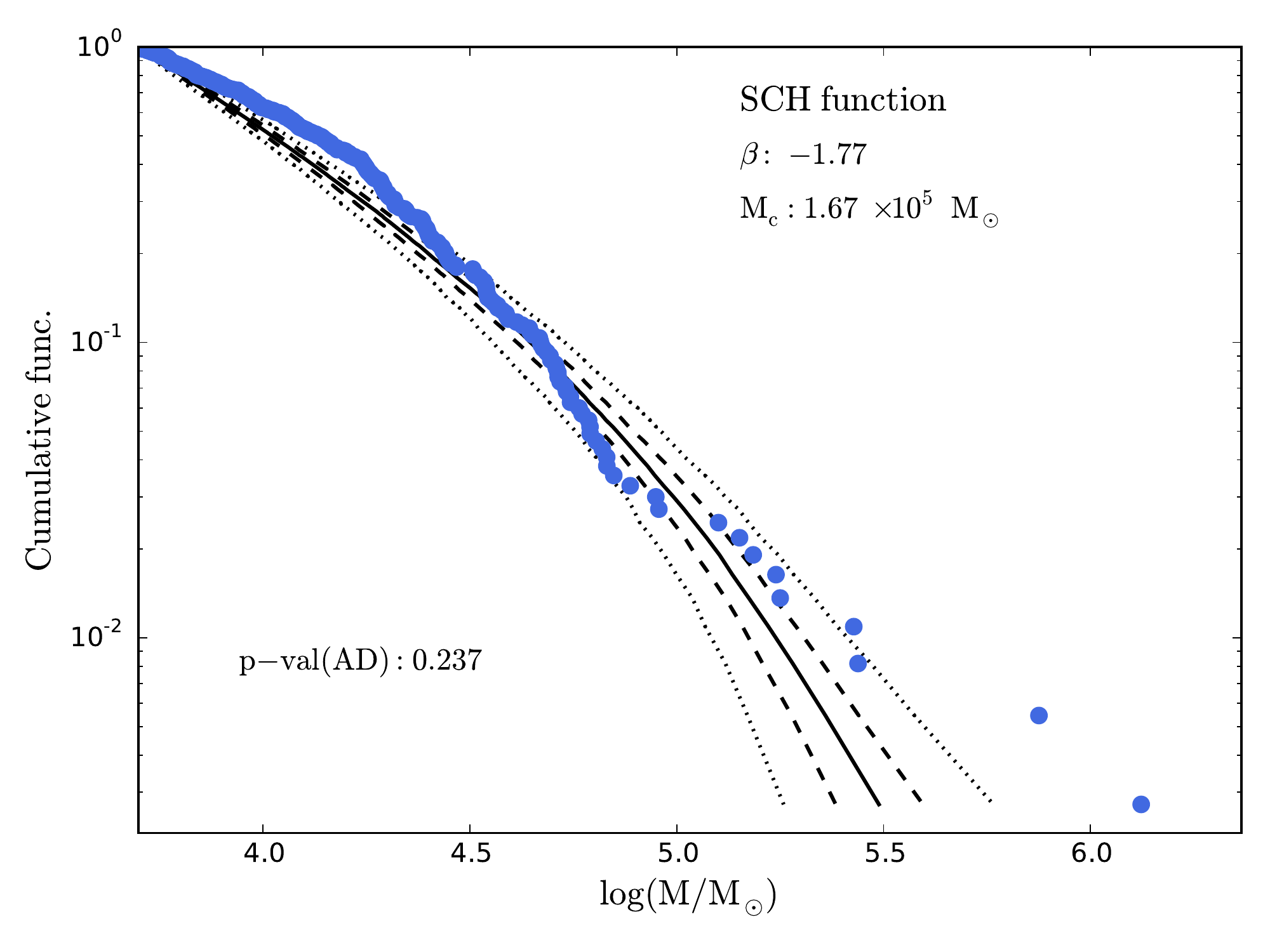}}
\subfigure{\includegraphics[width=0.29\textwidth]{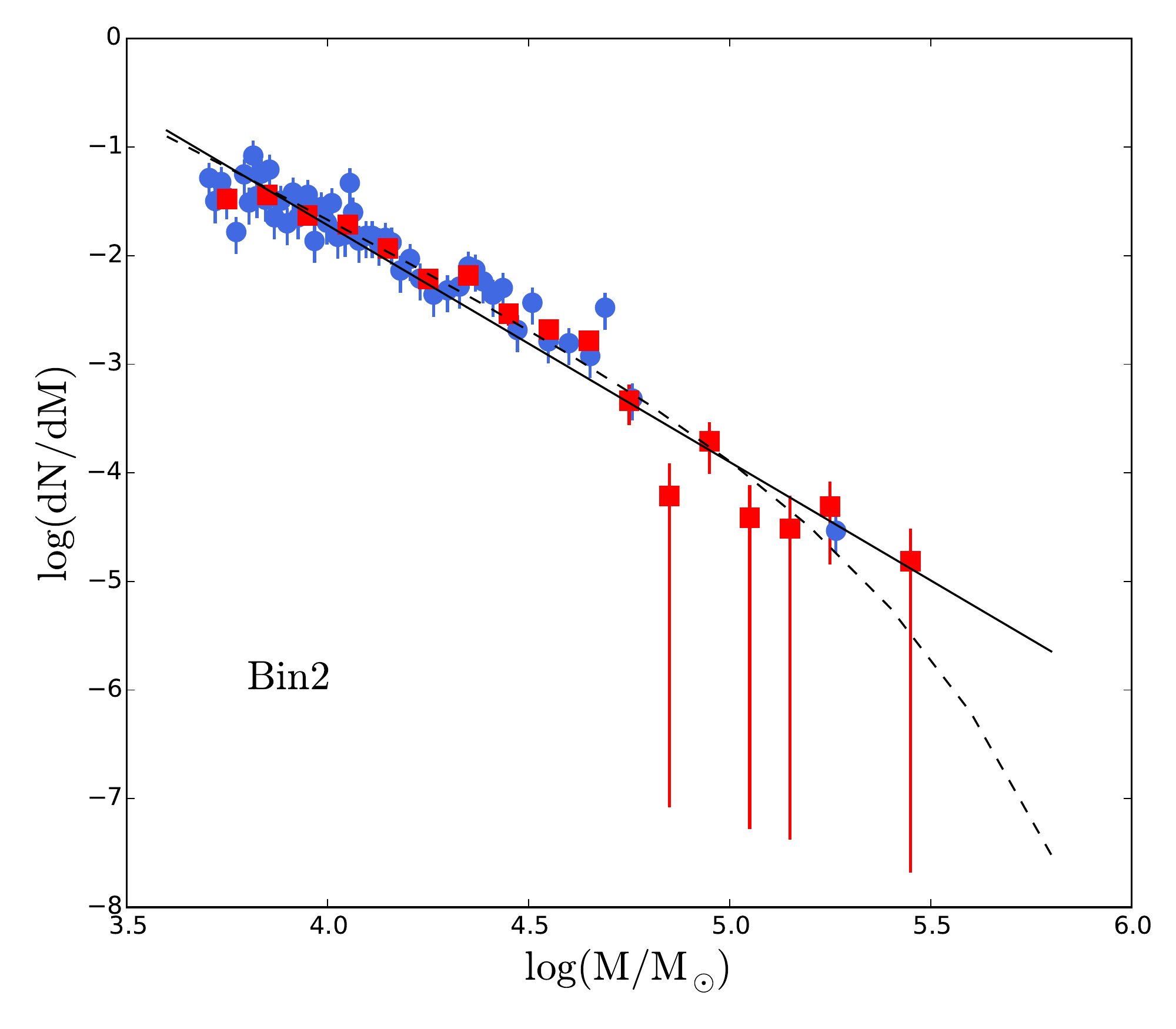}}
\subfigure{\includegraphics[width=0.34\textwidth]{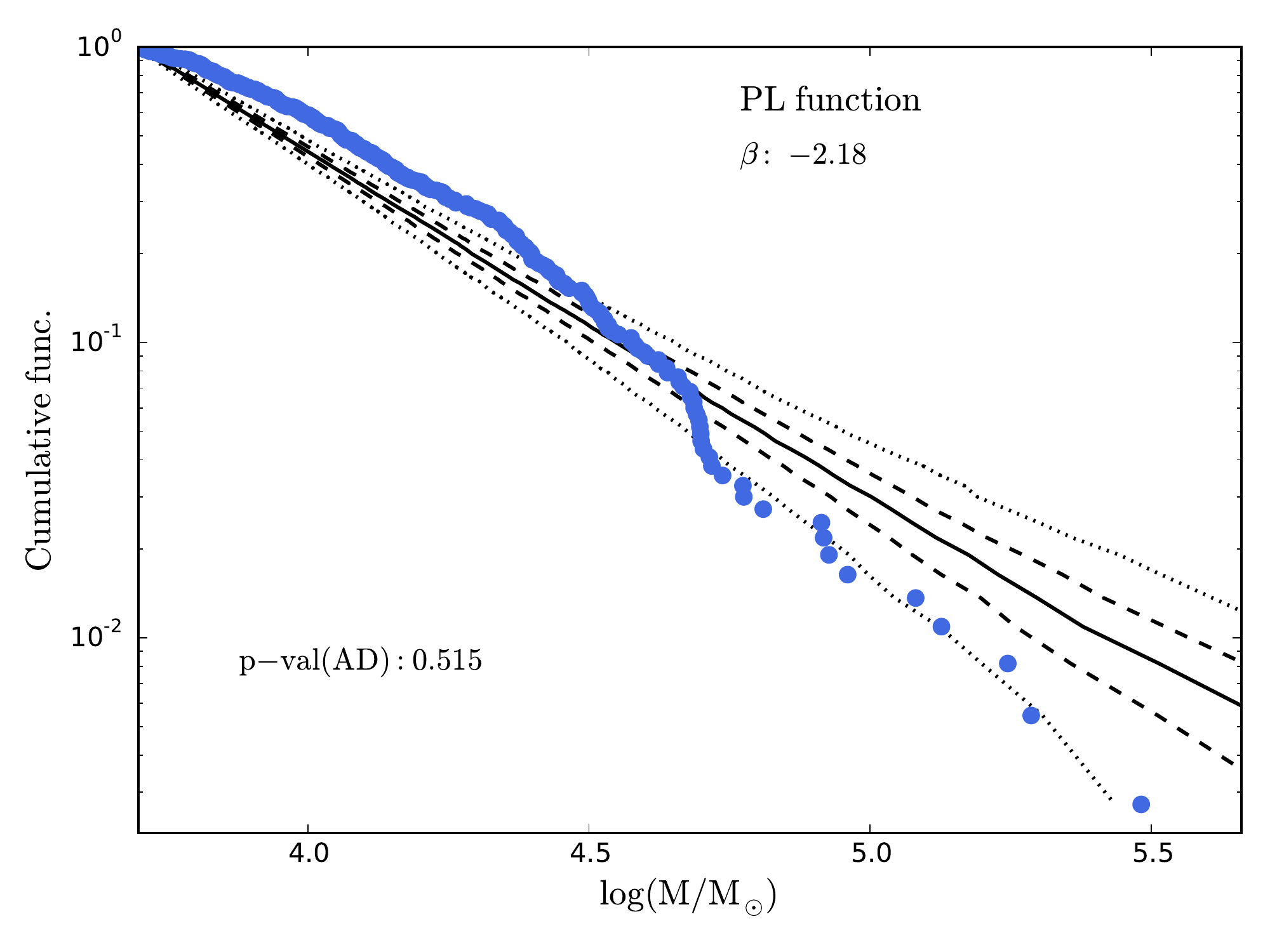}}
\subfigure{\includegraphics[width=0.34\textwidth]{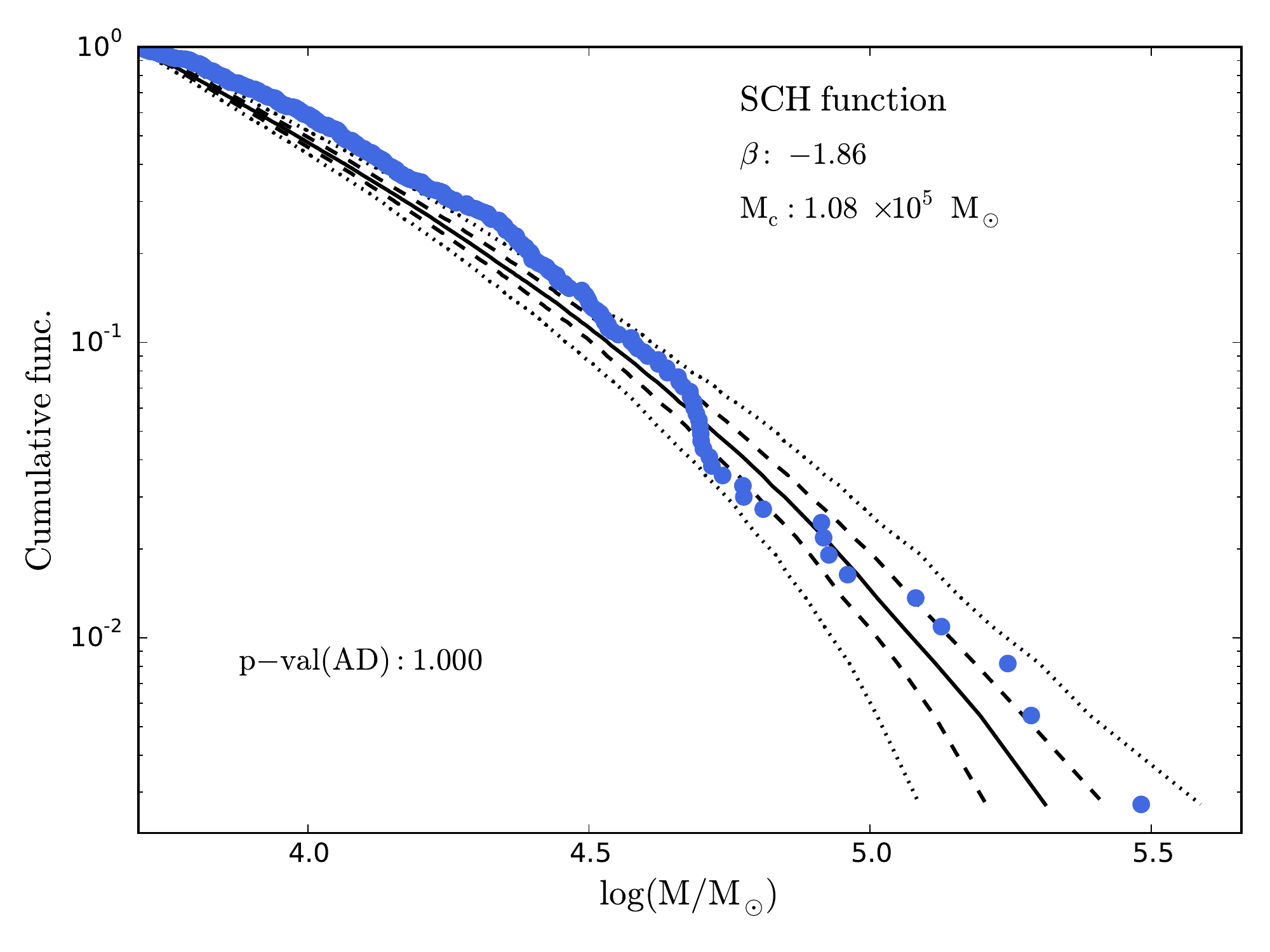}}
\subfigure{\includegraphics[width=0.29\textwidth]{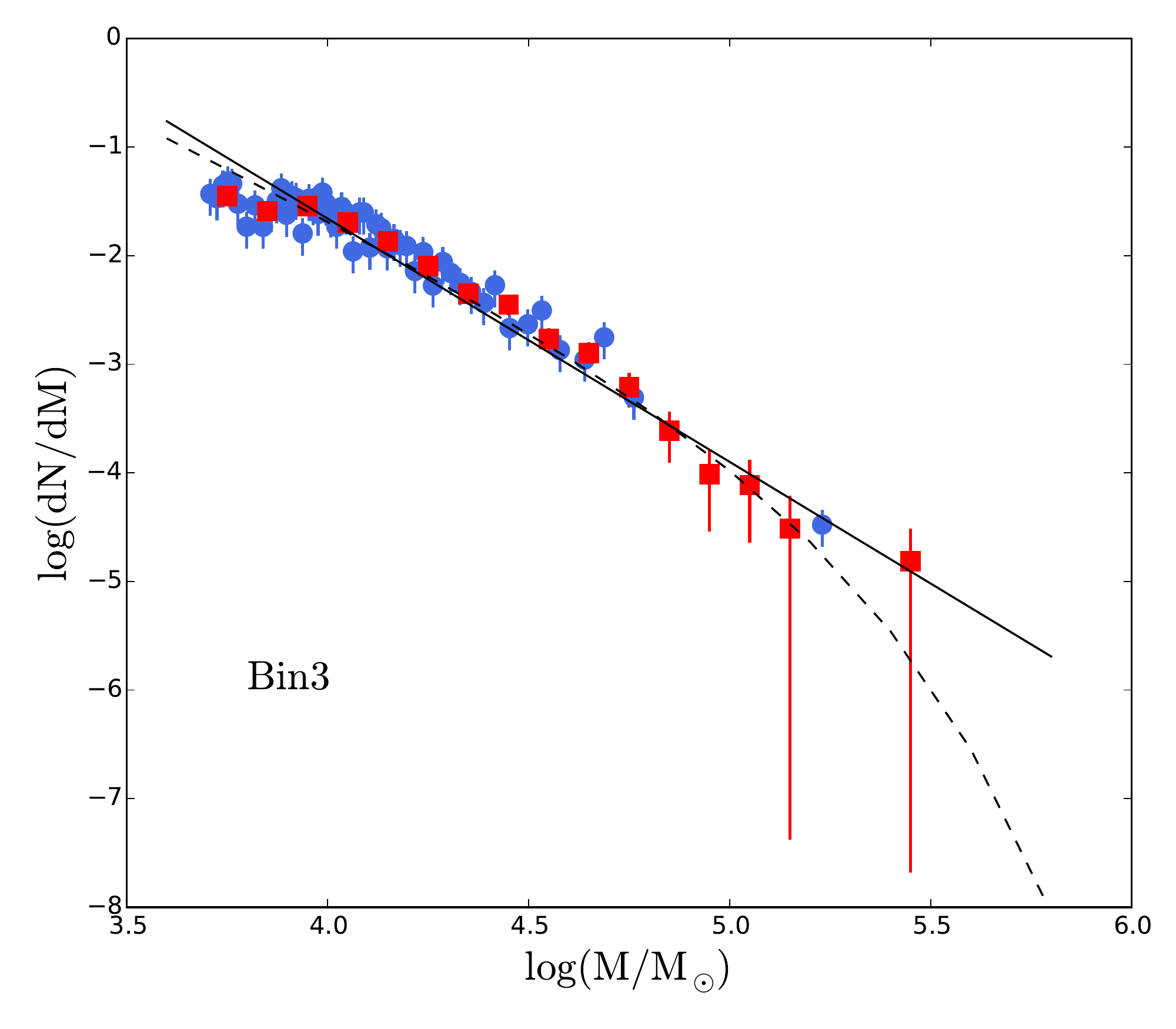}}
\subfigure{\includegraphics[width=0.34\textwidth]{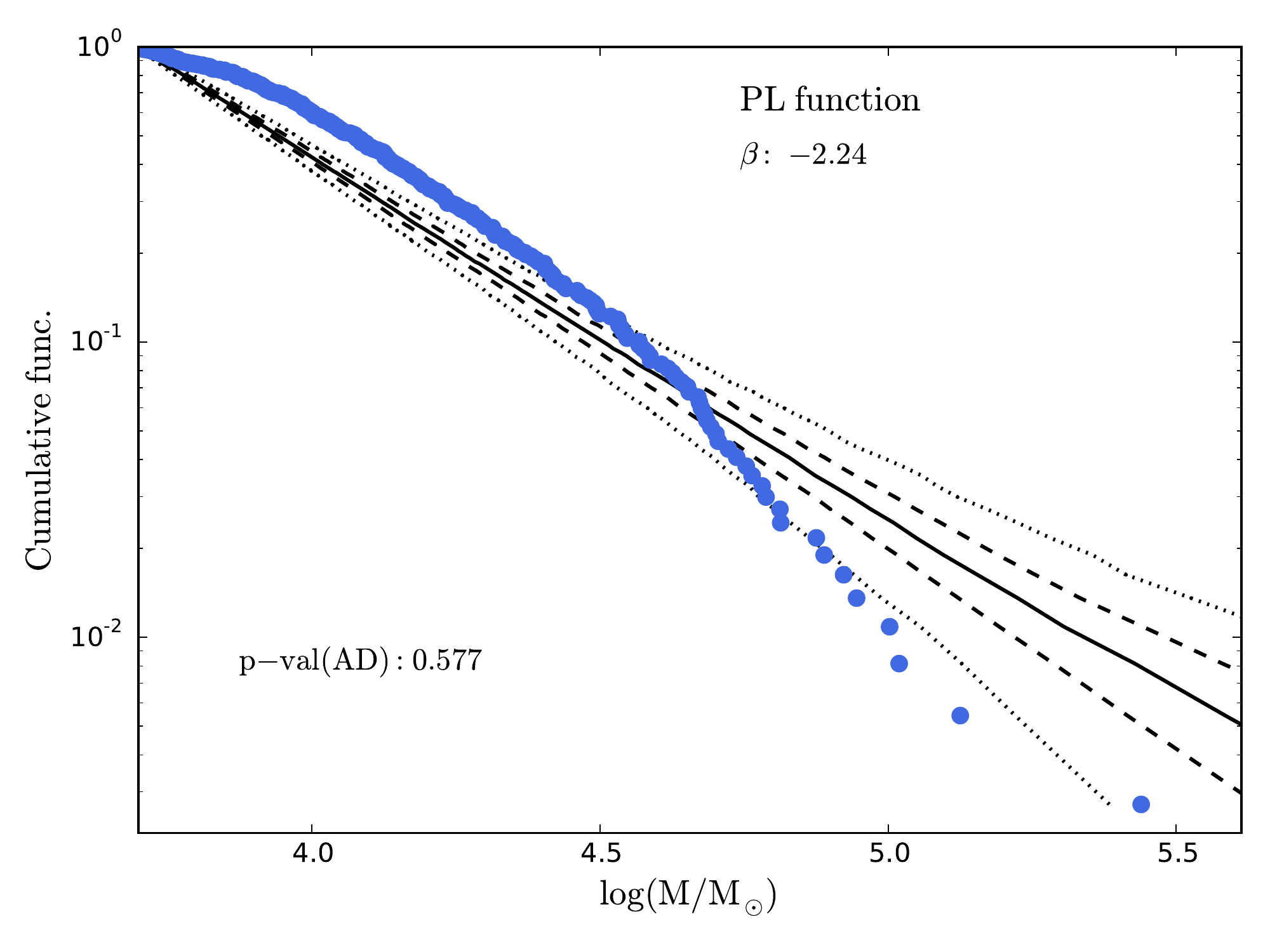}}
\subfigure{\includegraphics[width=0.34\textwidth]{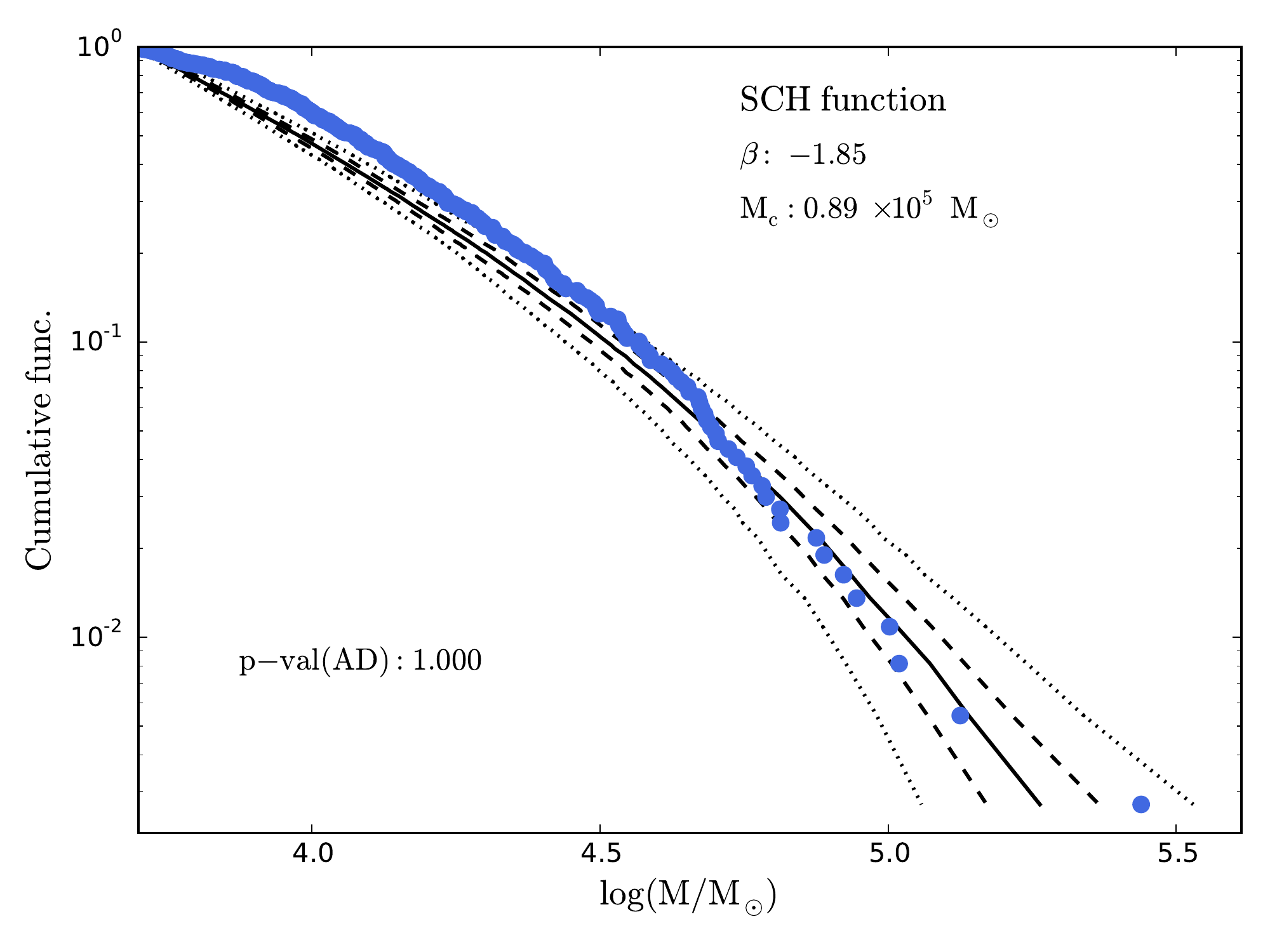}}
\subfigure{\includegraphics[width=0.29\textwidth]{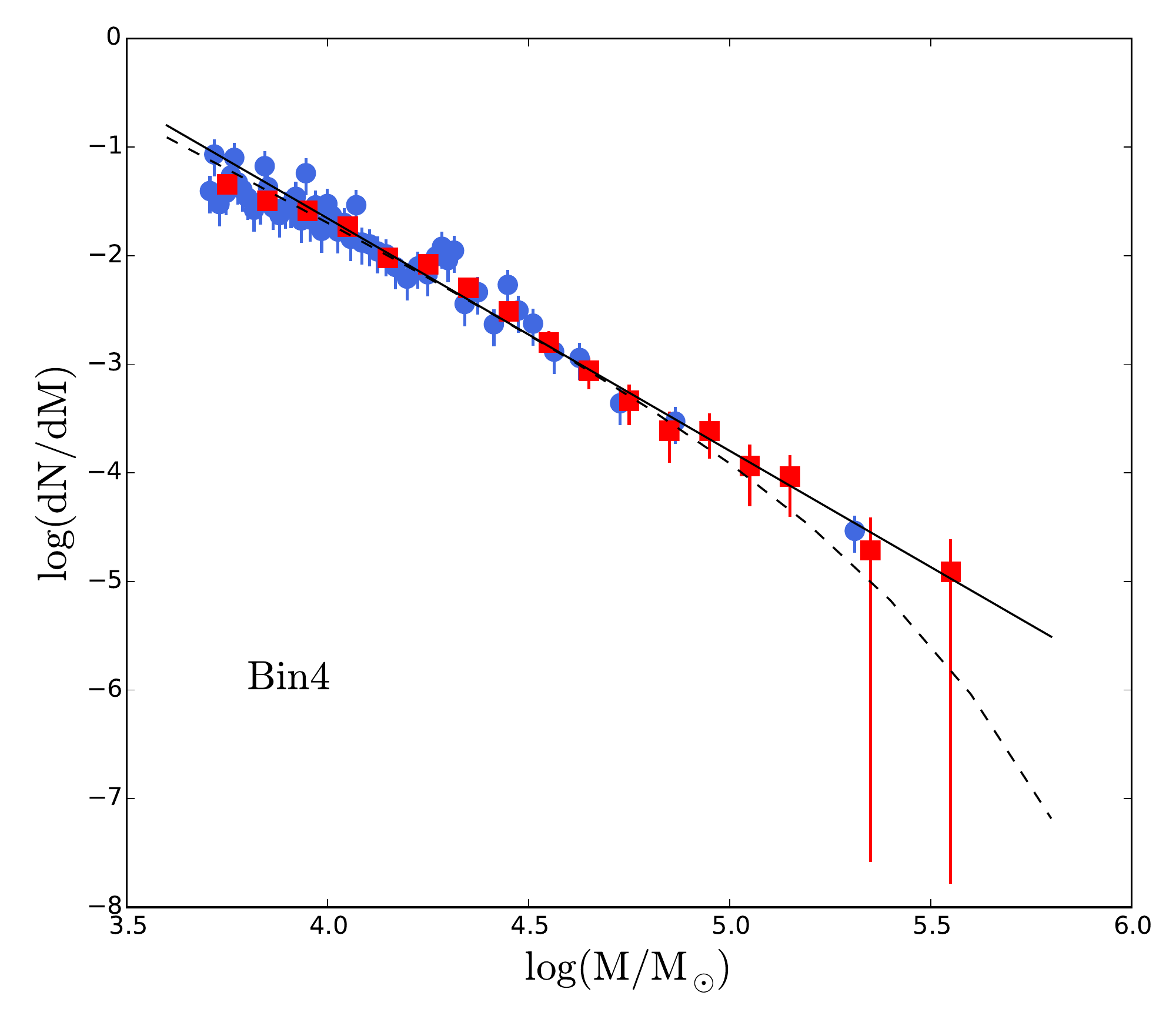}}
\subfigure{\includegraphics[width=0.34\textwidth]{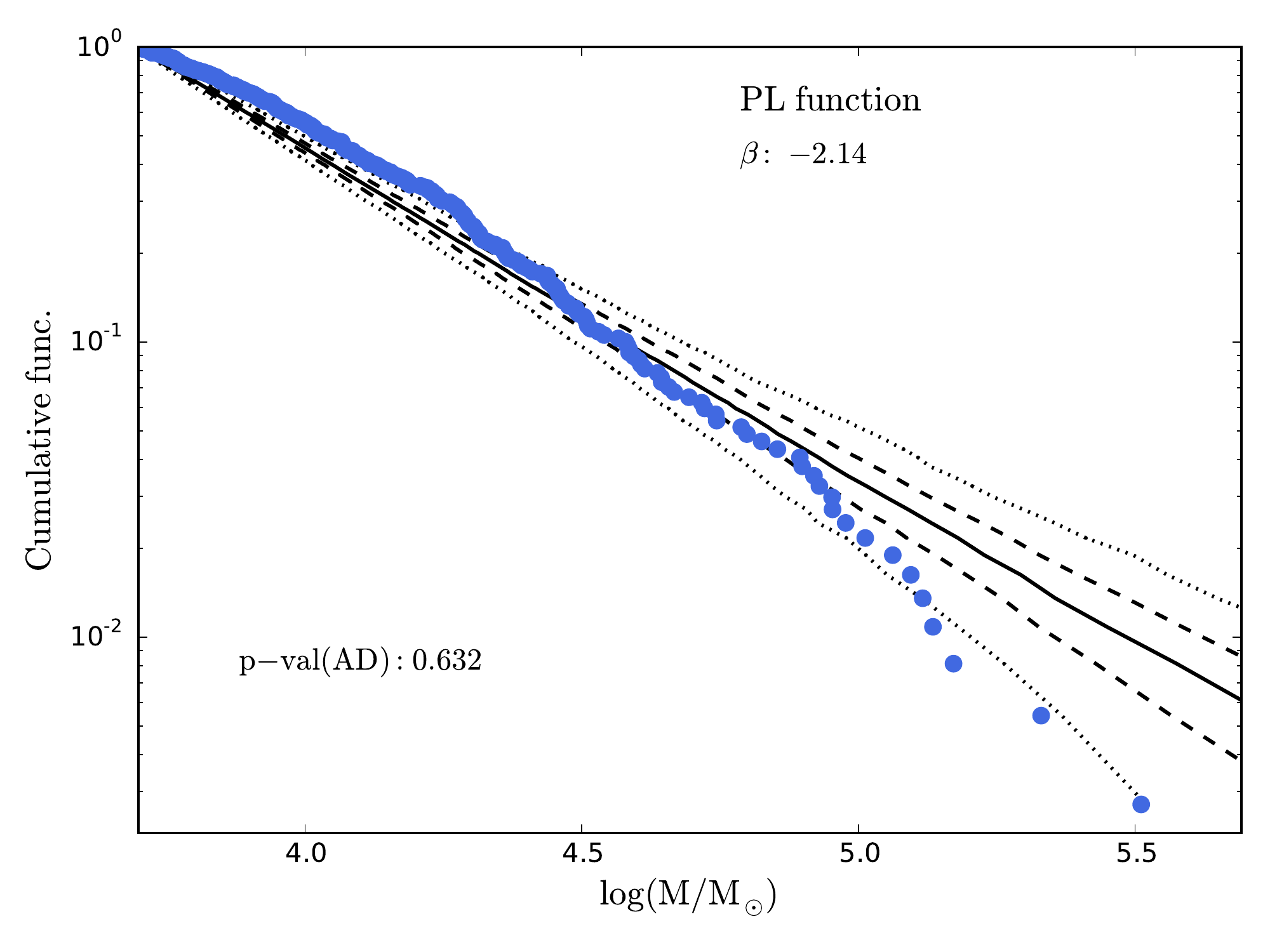}}
\subfigure{\includegraphics[width=0.34\textwidth]{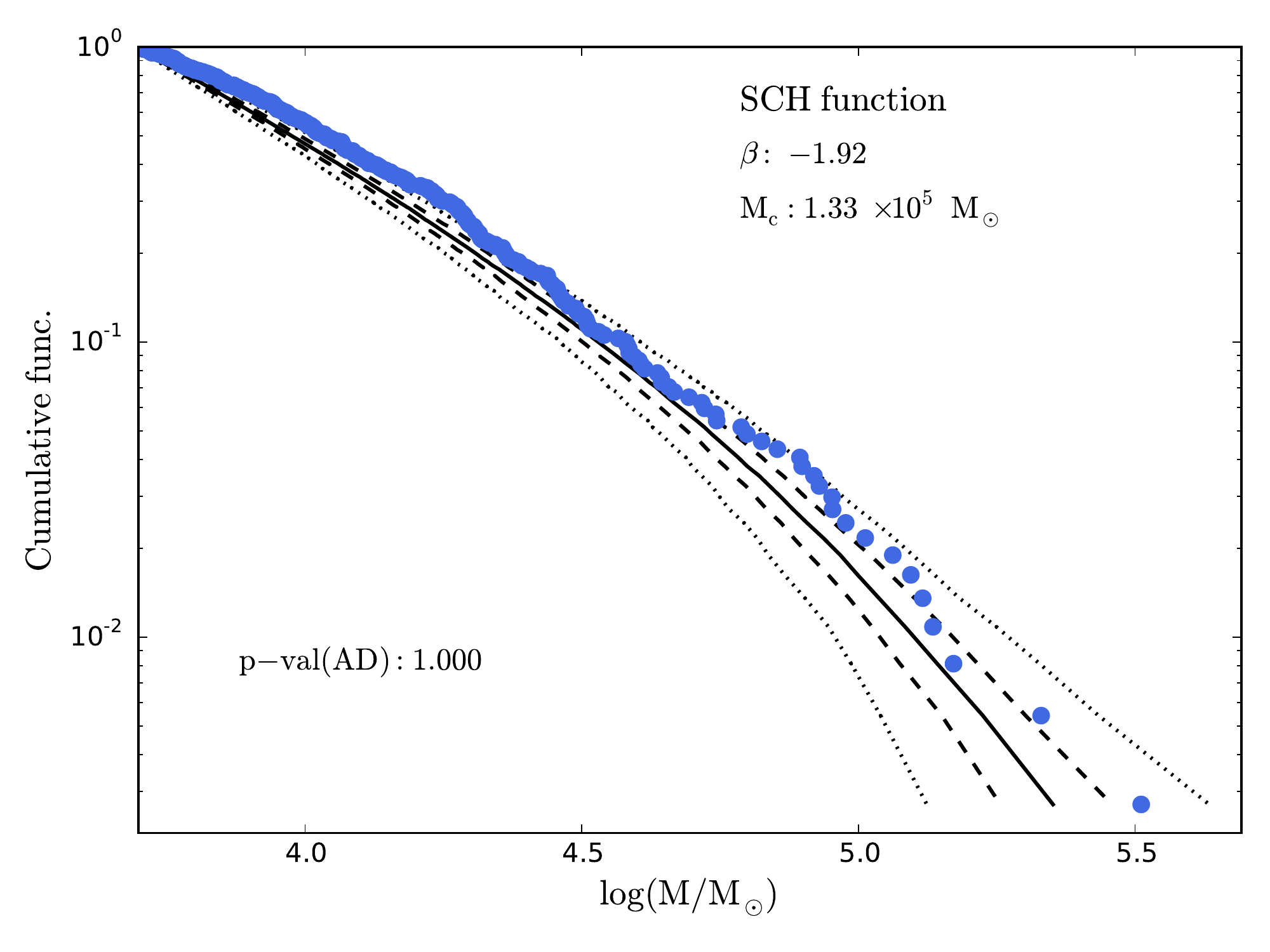}}
\caption{Mass functions in the radial bin regions (Bin 1 to Bin 4 from top to bottom), each row showing a different sub-region. Left column: mass functions in bins containing equal numbers of clusters (blue circles) and in bins of same width (red squares). A pure power law (solid line) and a Schechter function (dashed line), with the best-fit values of Tab.~\ref{tab:massfit_bins} (also reported in the plots of the middle and right columns) are overplotted.
Middle and right columns: the cumulative mass function is compared with a simulated Monte Carlo population drawn from a power law with slope equal to the best fit value (middle) and the best-fit Schechter function (right). 
For each model, 1000 populations are simulated and the median mass function (solid line), along with the lines enclosing 50\% (dashed) and 90\% (dotted) of the simulated mass functions, are plotted over the observed one. P values from the AD test comparing observed and simulated masses above $10^4$ \msun\ are also reported (see text and Tab.~\ref{tab:ad} for more details).}
\label{fig:massfit_mcpop_bins}
\end{figure*}

\begin{figure*}
\centering
\subfigure{\includegraphics[width=0.29\textwidth]{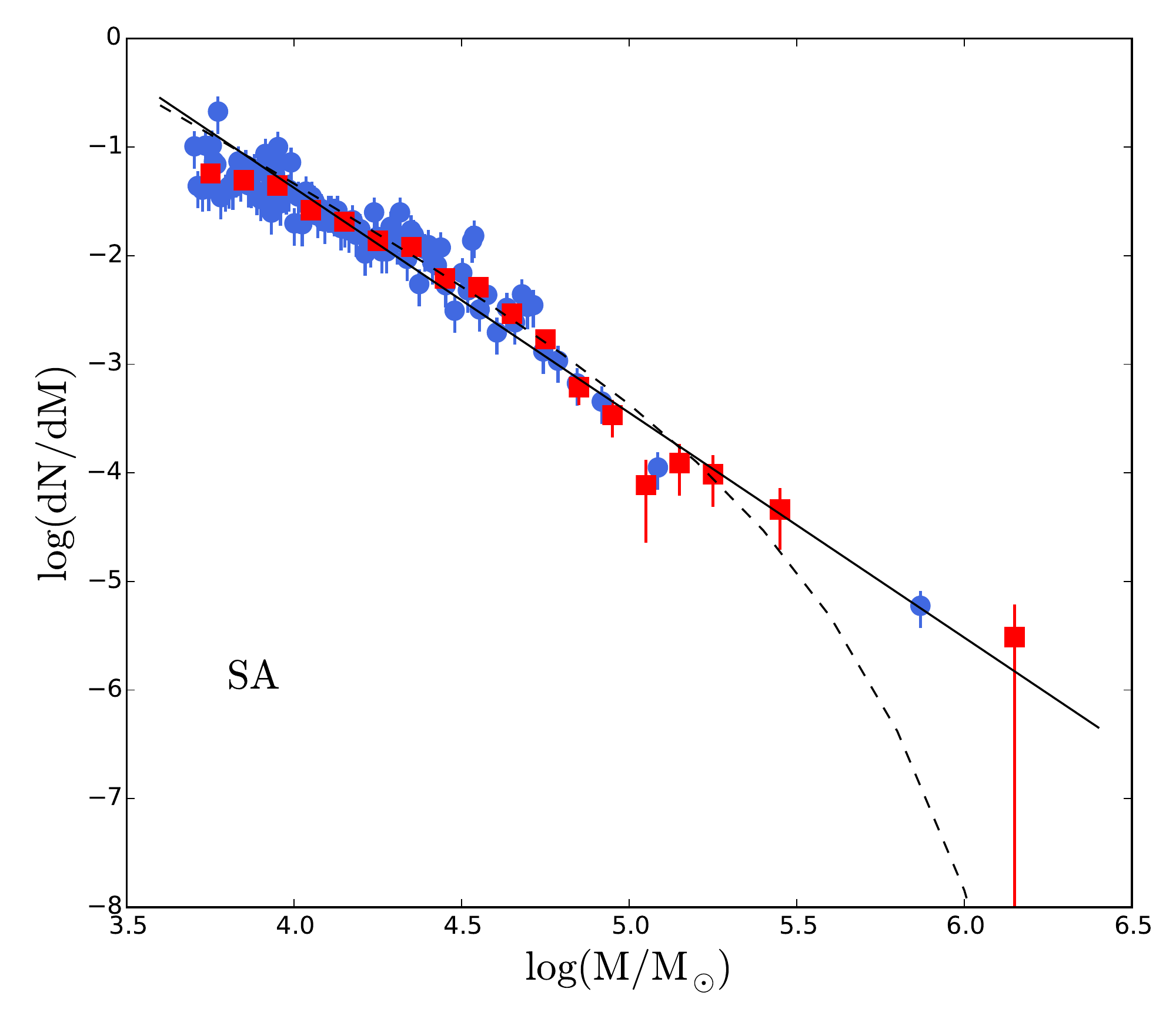}}
\subfigure{\includegraphics[width=0.34\textwidth]{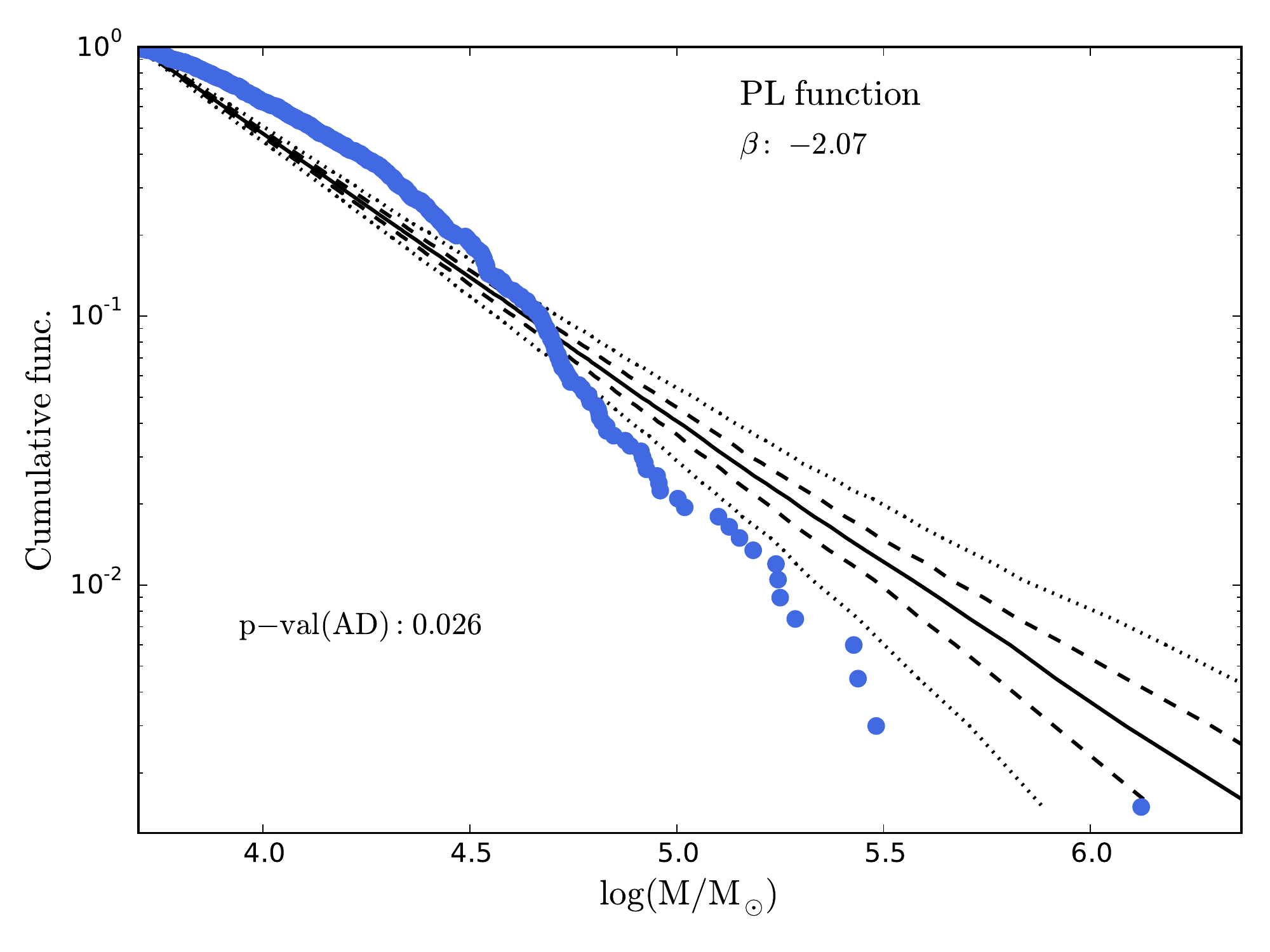}}
\subfigure{\includegraphics[width=0.34\textwidth]{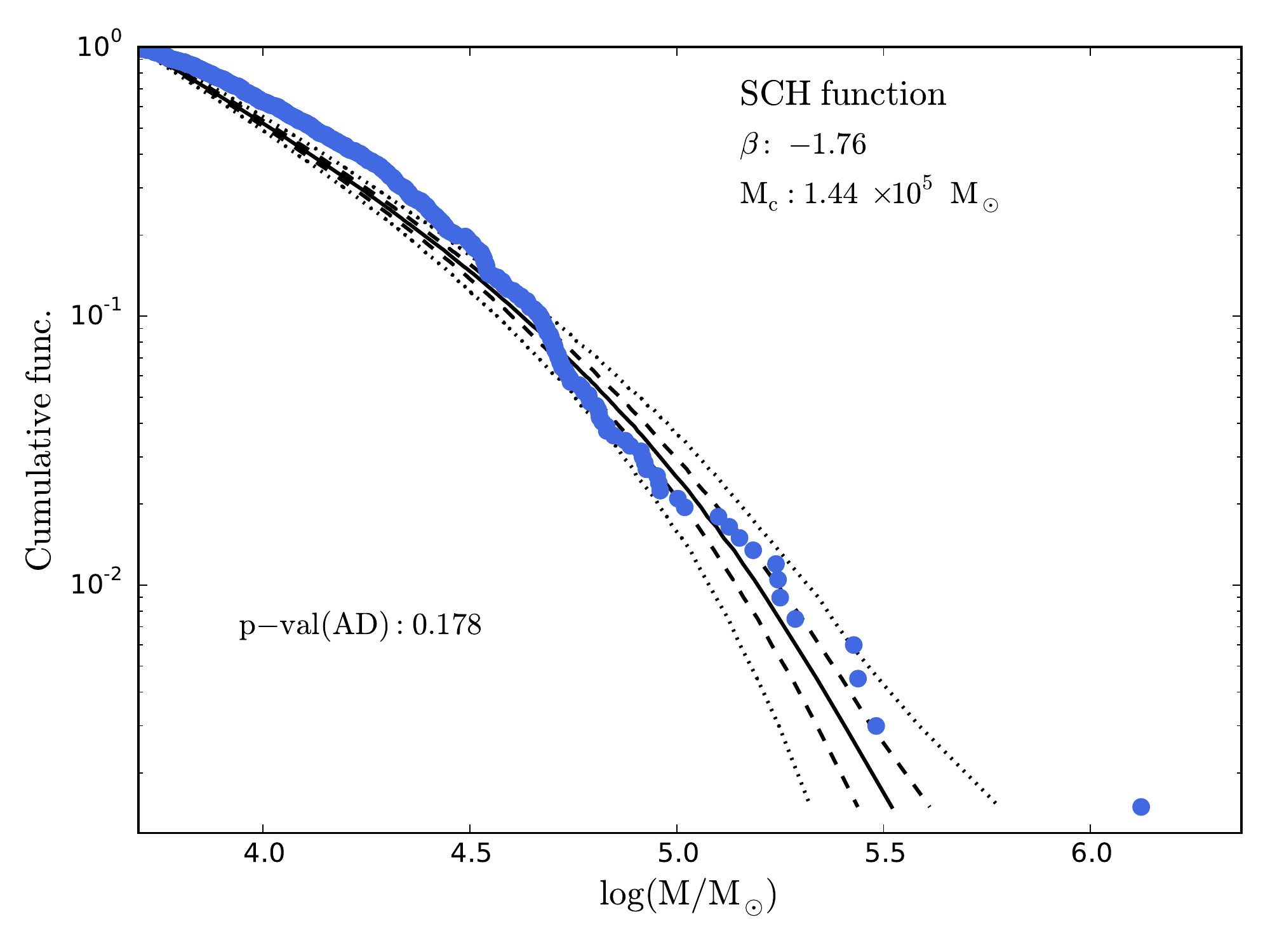}}
\subfigure{\includegraphics[width=0.29\textwidth]{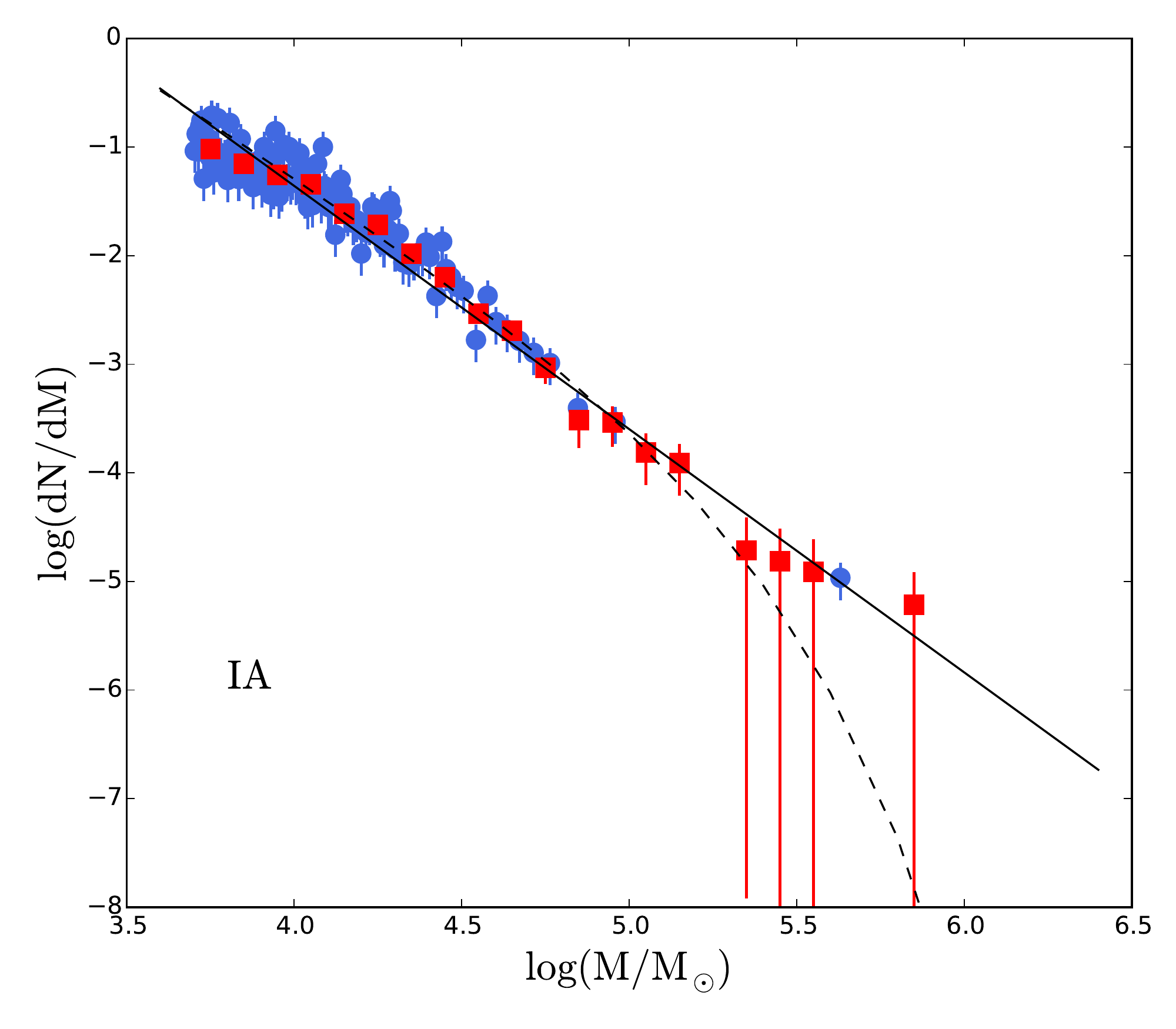}}
\subfigure{\includegraphics[width=0.34\textwidth]{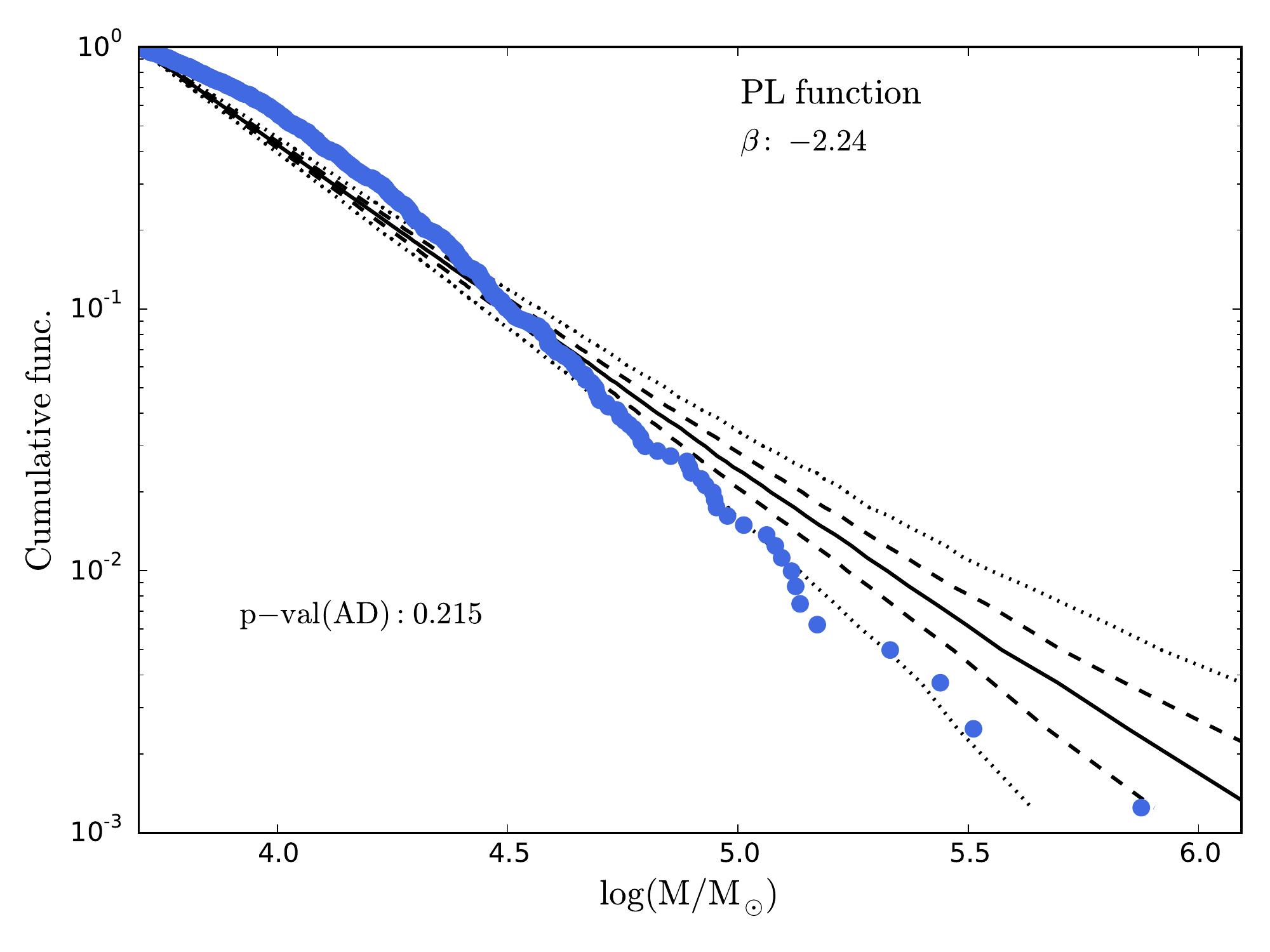}}
\subfigure{\includegraphics[width=0.34\textwidth]{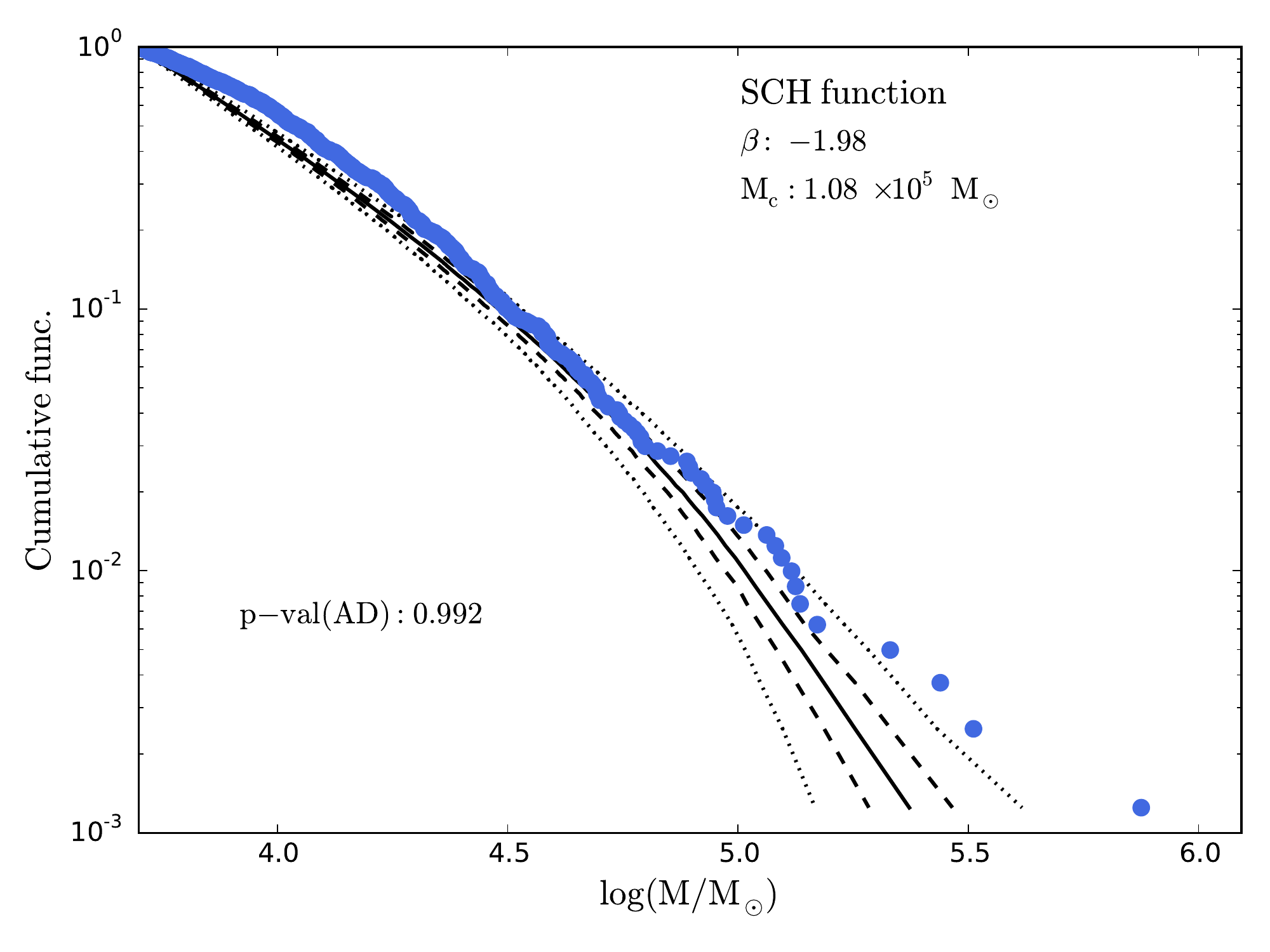}}
\subfigure{\includegraphics[width=0.29\textwidth]{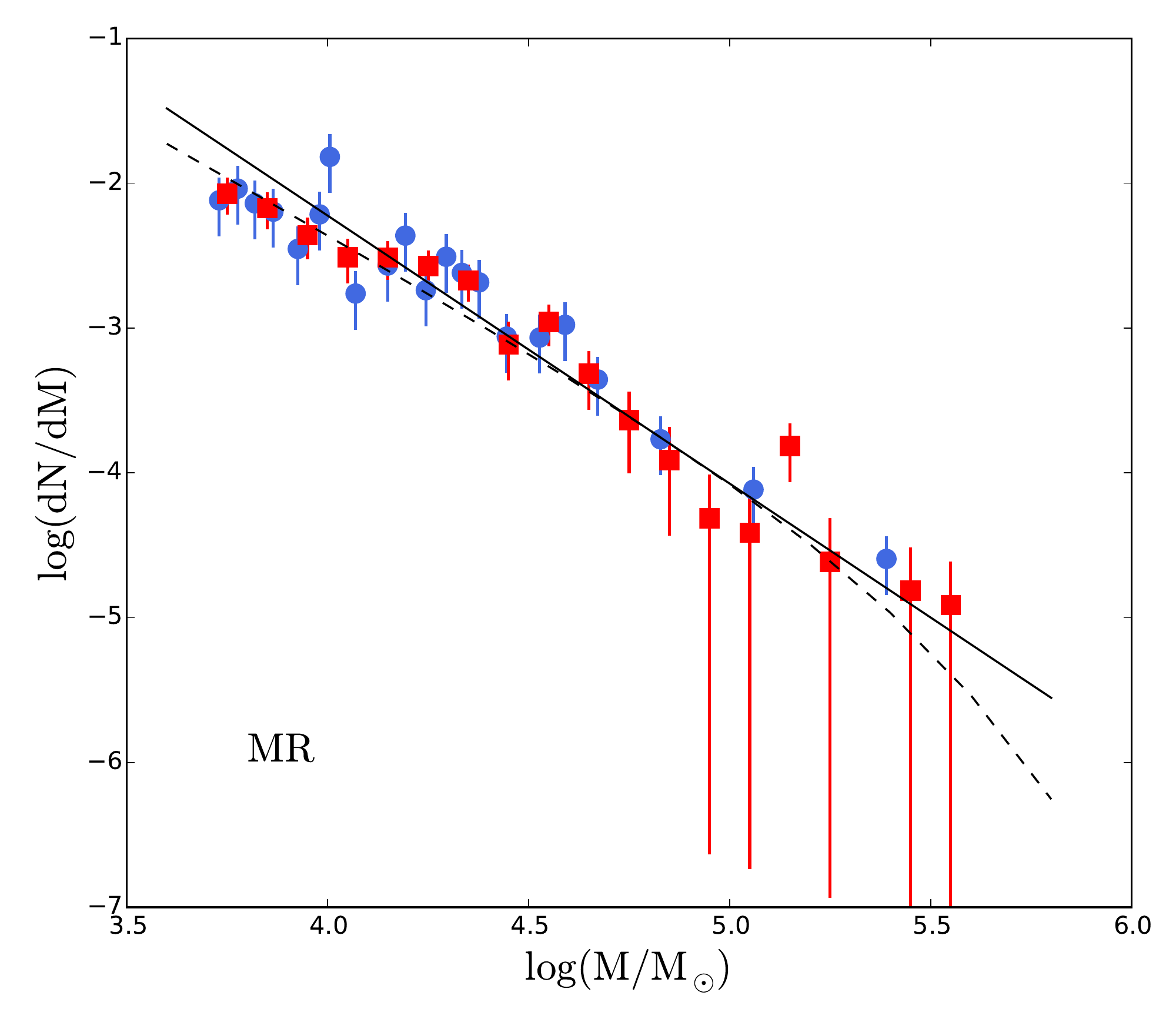}}
\subfigure{\includegraphics[width=0.34\textwidth]{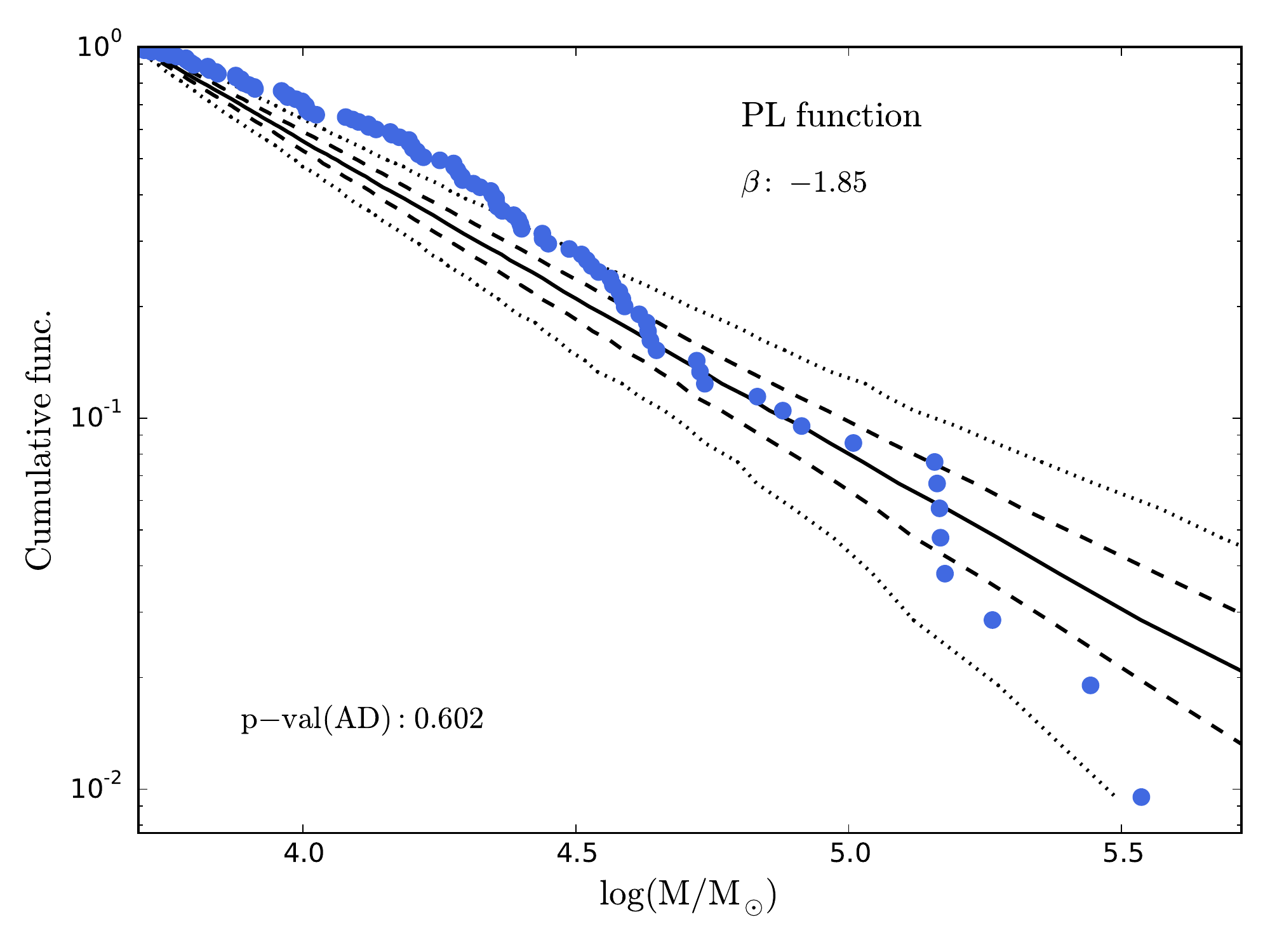}}
\subfigure{\includegraphics[width=0.34\textwidth]{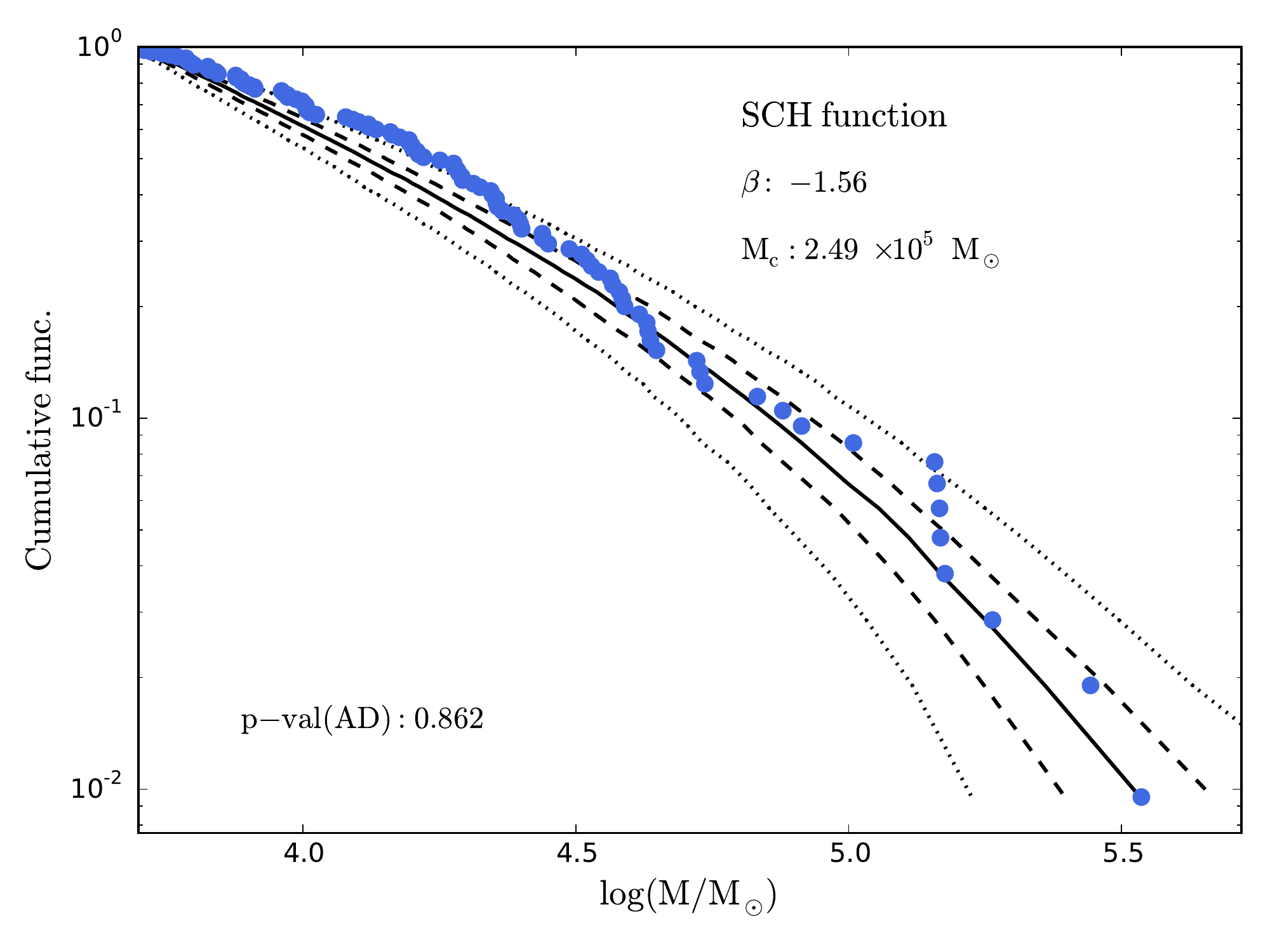}}
\caption{Same as Fig.~\ref{fig:massfit_mcpop_bins} but for the SA, IA  and MR regions. For the mass function of the MR region, only clusters with ages $\leqslant100$~Myr are plotted.}
\label{fig:massfit_mcpop_saia}
\end{figure*}

Following the same methodology as in Paper~I, we compare the observed mass functions with the ones of simulated Monte Carlo populations drawn from 3 different models. The models considered are a pure power-law with $-2$ slope, a simple power-law with slope equal to the best fit value and a Schechter function with slopes and truncation masses given in Tab.~\ref{tab:massfit_bins}. 
For each model 1000 populations were simulated, with the same number of the observed clusters in each bin.
The comparison is shown in Fig.~\ref{fig:massfit_mcpop_bins} and  Fig.~\ref{fig:massfit_mcpop_saia}. The median mass function of the 1000 simulation is plotted over the observed mass function, along with the lines enclosing 50\% and 90\% of the simulated cases.
 In order to test how the high-mass part of the observed mass function is in agreement with the models, we compare the distribution of observed and simulated clusters with $M\geqslant10^4$ \msun\ via the Anderson-Darling (AD) statistics. The AD test returns the probability that the null hypothesis of two samples having been drawn from the same distribution is true \citep{anderson1952,adtest}. Results are displayed in Tab.~\ref{tab:ad}

The Schechter function always shows the best agreement with observations. In many cases also a power law with a slope steeper than $-2$ provides a good description of the data. This result confirms that in all bins the high mass end of the function is steeper than the canonical slope of $-2$, and can be well described by a truncated power law. 
\begin{table}
\centering
\caption{Probability values from the Anderson-Darling (AD) test comparing the observed mass distribution with simulated ones from a simple power law mass function with a slope of $-2$ (PL-2), from a function built with the results of the mass function fit in Tab.~\ref{tab:massfit_bins}, both a simple power law (PL fit) and a Schechter funtion (SCH). The comparison is made only on clusters with $M>10^4$ and age $\leqslant200$ (for MR age $\leqslant100$). The p value reported in the table describes the probability that the null hypothesis (the two samples were drawn from the same distribution) is true.}
\begin{tabular}{llll}
\hline
Bin	& \multicolumn{3}{c}{$p_{AD}$} \\	
\	& PL-2	& PL fit	& SCH \\
\hline
\hline
SA	& 0.022 			& 0.026	& 0.178	\\
IA	& 8$\times10^{-5}$ 	& 0.215	& 0.992	\\	
\hline
MR	& 0.324 			& 0.602	& 0.862	\\
\hline
Bin 1	& 0.069			& 0.085	& 0.237	\\
Bin 2	& 0.031 			& 0.515	& 1.000	\\
Bin 3	& 0.010 			& 0.577	& 1.000	\\
Bin 4	& 0.086 			& 0.632	& 1.000	\\
\hline
\end{tabular}
\label{tab:ad}
\end{table}

\subsubsection{Bayesian fitting of the mass function}
\label{sec:massfunc_bayes}
We also implement a different type of fitting code to the mass function, based on Bayesian inference. This method allows us to find the most probable set of values for the slope and for the truncation mass, and to see the correlation between them.
The Bayesian fitting method is similar to what was done by \citet{johnson2017} in the analysis of M31. We firstly define the likelihood function of an observed cluster with mass $M$ as:
\begin{equation}
p_{cl}(M | \vec{\theta})\equiv\frac{p_{MF}(M | \vec{\theta})}{Z},
\end{equation}
where $p_{MF}(M | \vec{\theta})$ is the cluster mass function, Z the normalization factor, i.e.:
\begin{equation}
Z=\int_{0}^{\infty}p_{MF}(M | \vec{\theta})dM,
\end{equation}
and $\vec{\theta}$ represents the set of parameters which describe a certain shape of the mass function.
We used two possible mass distribution functions, namely a Schechter one:
\begin{equation}
\label{eq:bay_trun}
p_{MF,sch}(M | \vec{\theta})\propto M^{\beta}e^{-M/Mc}\ \Theta(M_{lim}) 
\end{equation}
and a power-law one:
\begin{equation}
\label{eq:bay_untrun}
p_{MF,pl}(M | \vec{\theta})\propto M^{\beta}\ \Theta(M_{lim}) 
\end{equation}
In both cases we limited the study of the mass function to masses above $M_{lim}$. This is indicated by the introduction of the Heaviside step function $\Theta(M_{lim})$.
We use Bayes' theorem to derive the posterior probability distribution function of the parameters $\vec{\theta}$, defined as:
\begin{equation}
p(\vec{\theta} | \{M_i\})\propto p_{cl}(\{M_i\} | \vec{\theta})p(\vec{\theta}),
\end{equation}
where $\{M_i\}$ is the observed mass distribution and $p(\vec{\theta})$ is the prior probability of the parameters $\vec{\theta}$. We choose a flat uninformative top-hat prior probability distribution to cover the range of possible values $-3<\beta<-1$ and $\rm{log}(M_{lim}/M_\odot)<\rm{log}(Mc/M_\odot)<8$.
The same prior distribution has been used for the truncated and un-truncated mass functions (Eq.~\ref{eq:bay_trun} and Eq.~\ref{eq:bay_untrun}) since in the analysis of the previous section outlined that in both cases the recovered slopes are close to $\beta\approx-2$. The limiting values chosen for the prior distributions can therefore be considered safe limits.

For the sampling of the posterior probability distributions we use the \texttt{Python} package \texttt{emcee} \citep{emcee}, which implements a Markov Chain Monte Carlo (MCMC) sampler from \citet{goodman2010}. We use 100 walkers, each producing 600 step chains, and we discard the first 100 burn-in steps of each walker. This results in 50000 independent sampling values for each fit.

The results of the fit are listed in Tab.~\ref{tab:massfit_bayes}. The fit with the Schechter function returns shallower values for the slopes (in range $-\beta=1.20-1.66$). This result points out firstly that this method is very sensitive to the low-mass part of the distribution, and secondly that we may be incomplete around $5000$ \msun. Truncation masses are smaller than what found with the previous method, spanning a range between 0.36 and 0.91 ($\times10^5$ \msun), but in most of the cases are consistent with the previous results within the uncertainties. The trends found with the previous fitting methods are confirmed.
The slope of MR is again shallower than the ones in the other bins but we know that in this region we are strongly limited by incompleteness at those low masses. 
Again, the biggest difference in slopes is between the SA and IA environments. 
The similar trends recovered, compared to the previous fitting method, is expected because, having used flat priors, the posterior probability has reduced to be proportional to the likelihood. 

\begin{figure*}
\centering
\subfigure[Bin 4, $M_{lim}=5000$ \msun]{\includegraphics[width=0.45\textwidth]{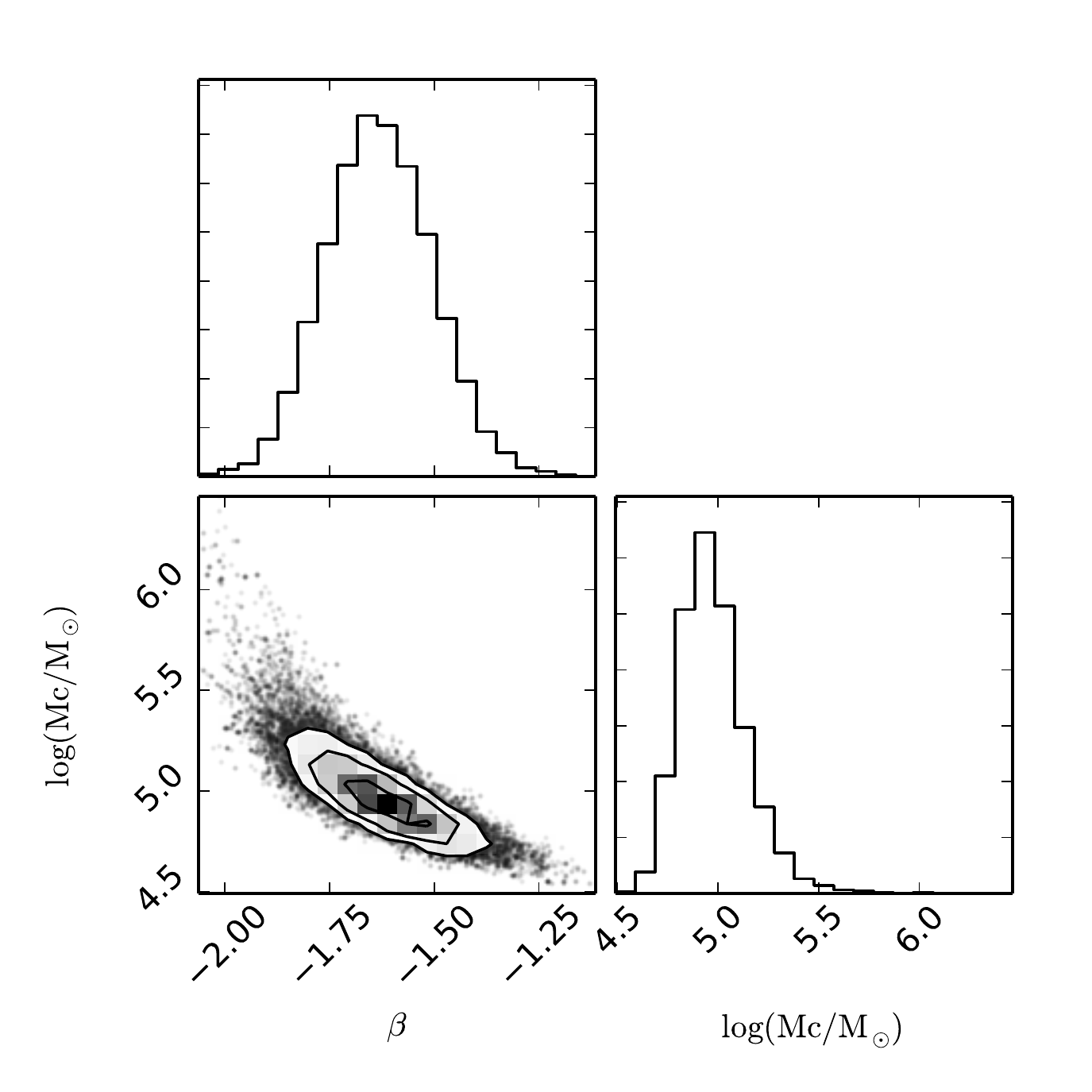}}
\subfigure[Bin 4, $M_{lim}=10^4$ \msun]{\includegraphics[width=0.45\textwidth]{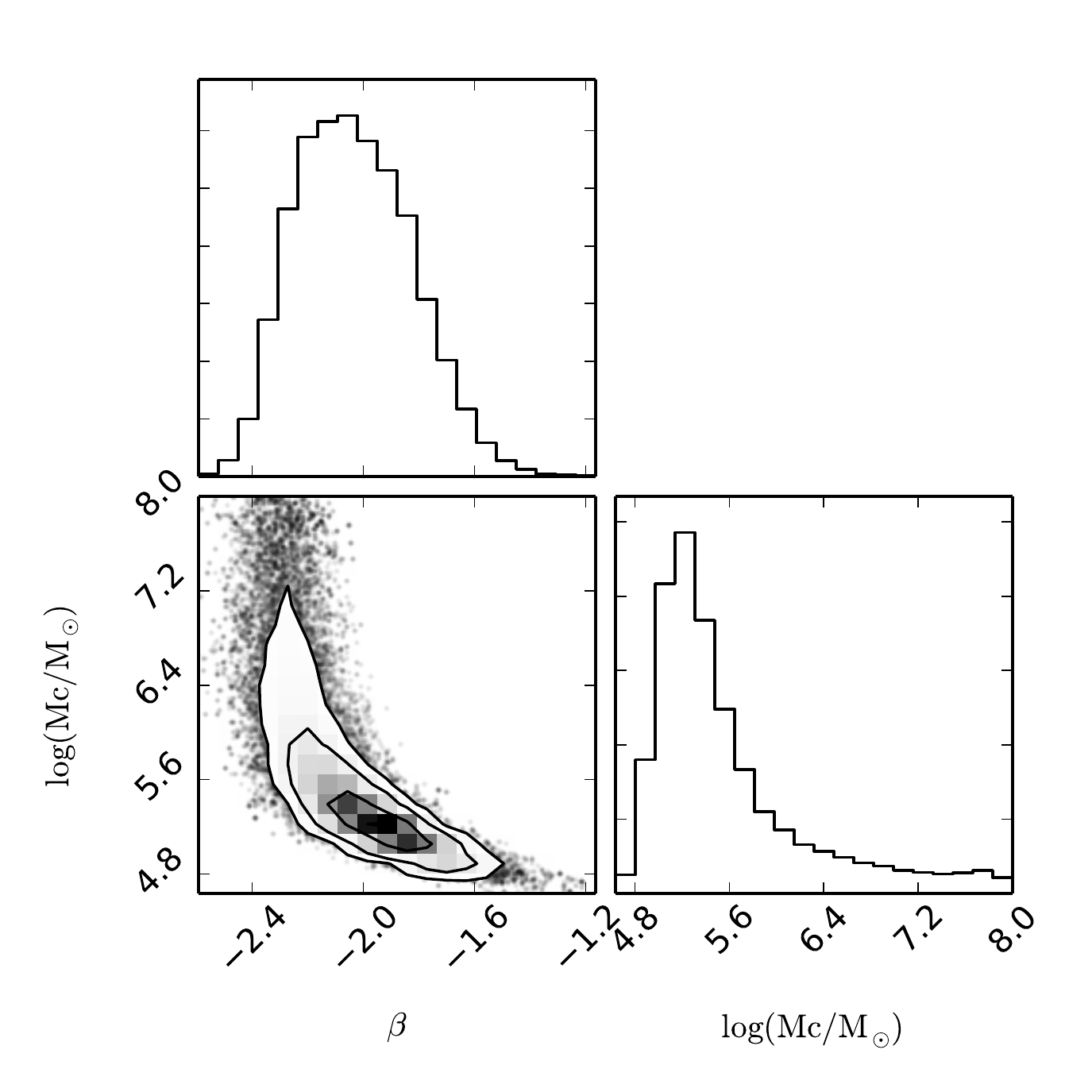}}
\caption{Posterior probability distribution, along with the marginalized distributions, for the mass function in Bin 4. Fit down to two different limiting masses are compared. In the left panel is clear the degeneracy at $\beta\approx-2.3$, suggesting that also the fit with a pure power-law is a good description of the data.}
\label{fig:massfit_bayes_bin4}
\end{figure*}

In order to focus only on the high mass part of the distributions, we have repeated the analyses considering only $M>10^4$ \msun. 
We notice, however, that in this case statistics are worse due to the low number of clusters. In particular, in some bins we do not get enough sampling of the function to be able to derive a meaningful value for the truncation mass, as the large uncertainties reveal.
As already noticed, when a pure power-law is fitted at those high masses the slopes are steeper than $-2$. 
As an example the posterior distribution for Bin 1 is shown in Fig.~\ref{fig:massfit_bayes_bin4}, comparing the fit down to masses $M=5000$ \msun\ (left panel) and $M=10^4$ \msun\ (right panel). The posterior distributions in the other bins show similar shapes. 

In conclusion, the Bayesian fit confirms many of the findings pointed out in the analysis of the cumulative mass function: similar ranges of truncation masses and slopes in the radial bins (within uncertainties), a difference between arm and inter-arm cluster mass functions and a mass function steepening at high masses. It also highlights a shallow mass function slope around $\sim5000$ \msun, possibly caused by partial incompleteness. The MCMC posterior sampling, plotted in Fig.~\ref{fig:massfit_bayes_bin4}, highlights also the correlation between truncation masses and slopes.

\begin{table*}
\centering
\caption{Results of the Bayesian fitting, considering clusters with: $\rm M>5000$ \msun\ (columns 2-4) and $\rm M>10^4$ \msun\ (columns 5-7).
An age cut at 200 Myr (100 Myr for the MR region) is applied. Those are the same age and mass cuts applied in Tab.~\ref{tab:massfit_bins}. Examples of the posterior probability distributions obtained are shown in Fig.~\ref{fig:massfit_bayes_bin4}.
}
\begin{tabular}{lccccclc}
\hline
\ & \multicolumn{3}{c}{$\rm M>5000$ \msun} & \ & \multicolumn{3}{c}{$\rm M>10^4$ \msun} \\
\hline
\multicolumn{1}{l}{Bin}				& \multicolumn{2}{c}{Schechter} 			& \multicolumn{1}{c}{simple PL}	 & \multicolumn{1}{c}{}		& \multicolumn{2}{c}{Schechter} 			& \multicolumn{1}{c}{simple PL}	\\
\			& $-\beta$		& Mc ($10^5$\msun)		& $-\beta$		& \ & $-\beta$		& Mc ($10^5$\msun)			& $-\beta$				\\
\hline
\hline
SA			& $1.42\ ^{+0.09}_{-0.08}$	& $0.93\ ^{+0.24}_{-0.16}$	& $1.90\ ^{+0.03}_{-0.03}$ 	& \ & $1.79\ ^{+0.13}_{-0.13}$	& $1.86\ ^{+1.09}_{-0.57}$	& $-2.15\ ^{+0.05}_{-0.06}$ \\
IA			& $1.66\ ^{+0.09}_{-0.09}$	& $0.83\ ^{+0.24}_{-0.17}$	& $2.08\ ^{+0.04}_{-0.04}$ 	& \ & $2.35\ ^{+0.13}_{-0.1}$	& $7.76\ ^{+99.39}_{-5.01}$	& $-2.44\ ^{+0.07}_{-0.07}$\\
\hline
MR	& $1.29\ ^{+0.16}_{-0.17}$	& $1.51\ ^{+1.30}_{-0.49}$	& $1.74\ ^{+0.07}_{-0.07}$	& \ & $1.58\ ^{+0.26}_{-0.25}$	& $2.63\ ^{+9.12}_{-1.31}$	& $-1.93\ ^{+0.10}_{-0.11}$ \\
\hline
Bin 1			& $1.66\ ^{+0.09}_{-0.09}$	& $2.51\ ^{+1.66}_{-0.81}$	& $1.90\ ^{+0.05}_{-0.05}$ 	& \ & $2.05\ ^{+0.12}_{-0.10}$	& $13.8\ ^{+115.02}_{-9.23}$	& $-2.13\ ^{+0.07}_{-0.08}$ \\
Bin 2			& $1.38\ ^{+0.14}_{-0.14}$	& $0.54\ ^{+0.19}_{-0.12}$	& $2.01\ ^{+0.05}_{-0.05}$ 	& \ & $1.93\ ^{+0.25}_{-0.24}$	& $1.17\ ^{+1.71}_{-0.47}$	& $-2.34\ ^{+0.09}_{-0.09}$ \\
Bin 3			& $1.20\ ^{+0.12}_{-0.14}$	& $0.36\ ^{+0.09}_{-0.06}$	& $2.02\ ^{+0.05}_{-0.05}$ 	& \ & $1.90\ ^{+0.28}_{-0.27}$	& $0.83\ ^{+0.99}_{-0.32}$	& $-2.42\ ^{+0.10}_{-0.10}$ \\
Bin 4			& $1.63\ ^{+0.13}_{-0.12}$	& $0.91\ ^{+0.44}_{-0.25}$	& $2.04\ ^{+0.05}_{-0.05}$ 	& \ & $2.05\ ^{+0.23}_{-0.20}$	& $2.29\ ^{+12.84}_{-1.14}$	& $-2.29\ ^{+0.09}_{-0.09}$ \\
\hline
\end{tabular}
\label{tab:massfit_bayes}
\end{table*}

\subsubsection{Comparison with GMC Masses}
\label{sec:gmc}
The work of C14 on the GMC properties in M51 showed that the GMC mass function is not universal inside the galaxy. \citet{huges2013} showed that the GMC mass function distribution in M51 are shallower in regions of brighter CO emission, suggesting a tight relation between the distribution of molecular gas inside the galaxy  and the properties of single GMCs. In the same work, the authors compared properties of young (age$<10^7$ Myr) clusters from the catalogue by \citet{chandar2011} to GMC properties, finding that mass functions slopes of YSCs and GMCs are in good agreement in many subregions.
The fits of the mass function in the previous paragraphs seem to suggest that, also in the case of YSCs, the mass function varies at sub-galactic scales.
Following up the work of \citet{huges2013}, we investigate here the possibility of a direct relation between GMC and cluster mass functions using our YSC catalogue and the the GMC results reported by C14. 

In the work by C14, the mass function is fitted with the code \texttt{mspecfit.pro} in the same way as we did for the star clusters, and their results are listed in our Tab.~\ref{tab:massfit_gmc_clusters}. 
In this comparison, we are limited by the small area covered by PAWS. The GMC population extends only to a partial fraction of Bin 2. We anyway divide the GMC population in MR, Bin 1 and Bin 2 sub-samples.

\begin{table*}
\centering
\caption{Fit results for the mass function of GMCs and YSC in Bin 1, Bin 2 and in the dynamical regions defined by \citet{colombo2014a} (see the text and Fig.~\ref{fig:dynregions} for the description of the division of SA and IA into dynamical subregions). 
Columns (2) $-$ (4) display the results of the fit of the GMCs: for the MR, SA, and IA regions the results are directly taken from \citet{colombo2014a}, while the fits in Bin 1 and 2 were performed by us using the same code. Columns (5) $-$ (8) display the results of the fit of the YSCs. Only clusters up to 100~Myr were considered in this analysis. $^{(*)}$The fit results for Bin 1, Bin 2 and MR are taken directly from Tab.~\ref{tab:massfit_bins}. In the case of Bin 1 and Bin 2 clusters up to 200~Myr old are considered.}
\begin{tabular}{lllllllll}
\hline
\		& \multicolumn{3}{l}{GMCs} 				& \multicolumn{1}{l}{\ }				& \multicolumn{4}{l}{YSCs}											\\
\multicolumn{1}{c}{Region} 	& \multicolumn{1}{c}{$-\beta$}	& \multicolumn{1}{c}{$M_0$  ($10^6$ \msun)}	& \multicolumn{1}{c}{$N_0$}	& \multicolumn{1}{l}{\ } & \multicolumn{1}{c}{$N_{YSC}$}	& \multicolumn{1}{c}{$-\beta$}	& \multicolumn{1}{c}{$M_0$  ($10^5$ \msun)}		& \multicolumn{1}{c}{$N_0$}		\\
\multicolumn{1}{c}{(1)}  	& \multicolumn{1}{c}{(2)}	& \multicolumn{1}{c}{(3)}	& \multicolumn{1}{c}{(4)}	& \multicolumn{1}{l}{\ } & \multicolumn{1}{c}{(5)} 	& \multicolumn{1}{c}{(6)}	& \multicolumn{1}{c}{(7)}	& \multicolumn{1}{c}{(8)}		\\
\hline
\hline
Bin 1		& $1.97\ _{\pm0.08}$	& $11.5\ _{\pm0.8}$		& $34\ _{\pm7}$ 	& \	& 367$^{(*)}$	& $1.77\ _{\pm0.08}$	& $1.67\ _{\pm0.16}$	& $28\ _{\pm8}$ \\
Bin 2		& $2.69\ _{\pm0.16}$	& $21.7\ _{\pm12.4}$	& $1\ _{\pm1}$ 		& \	& 367$^{(*)}$	& $1.86\ _{\pm0.07}$	& $1.08\ _{\pm0.09}$	& $30\ _{\pm9}$	\\
\hline
MR		& $1.63\ _{\pm0.17}$	& $15.0\ _{\pm3.2}$		& $26\ _{\pm20}$ 	& \	& 105$^{(*)}$	& $1.56\ _{\pm0.09}$	& $2.49\ _{\pm0.60}$	& $14\ _{\pm8}$	\\
SA-DWI	& $1.75\ _{\pm0.20}$	& $12.2\ _{\pm1.8}$		& $15\ _{\pm12}$ 	& \	& 83			& $1.57\ _{\pm0.17}$	& $1.22\ _{\pm0.65}$	& $18\ _{\pm11}$ \\
SA-DWO	& $1.79\ _{\pm0.09}$	& $11.8\ _{\pm0.9}$		& $24\ _{\pm9}$ 	& \	& 65			& $1.67\ _{\pm0.12}$	& $1.04\ _{\pm0.48}$	& $10\ _{\pm2}$ \\
SA-MAT	& $2.52\ _{\pm0.20}$	& $158.6\ _{\pm7.4}$	& $0\ _{\pm2}$ 		& \	& 66			& $2.05\ _{\pm0.28}$	& $1.11\ _{\pm0.48}$		& $3\ _{\pm8}$ \\
IA-UPS	& $2.44\ _{\pm0.40}$	& $9.3\ _{\pm4.0}$		& $2\ _{\pm3}$ 		& \	& 58			& $1.96\ _{\pm0.34}$	& $0.47\ _{\pm0.11}$		& $8\ _{\pm8}$ \\
IA-DNS	& $2.55\ _{\pm0.23}$	& $8.3\ _{\pm1.9}$		& $5\ _{\pm4}$ 		& \	& 197		& $1.84\ _{\pm0.07}$	& $1.33\ _{\pm0.40}$	& $14\ _{\pm6}$ \\
\hline
\end{tabular}
\label{tab:massfit_gmc_clusters}
\end{table*}
The GMC mass function properties appear to change significantly across different environments of the galaxy.
However, as observed for the clusters, the MR has the shallowest slope and Bin 2 the steepest, with Bin 1 having a value in between the two. $M_0$ for Bin 2 in the GMC sample is almost a factor of 3 larger than for Bin 1, but the relative error on that value is more than 50\%. This is likely caused by small number statistics due to the restricted number of GMCs falling in Bin 2. The truncation mass recovered in the MR and in Bin 1 are similar.

\begin{figure}
\centering
\includegraphics[width=\columnwidth]{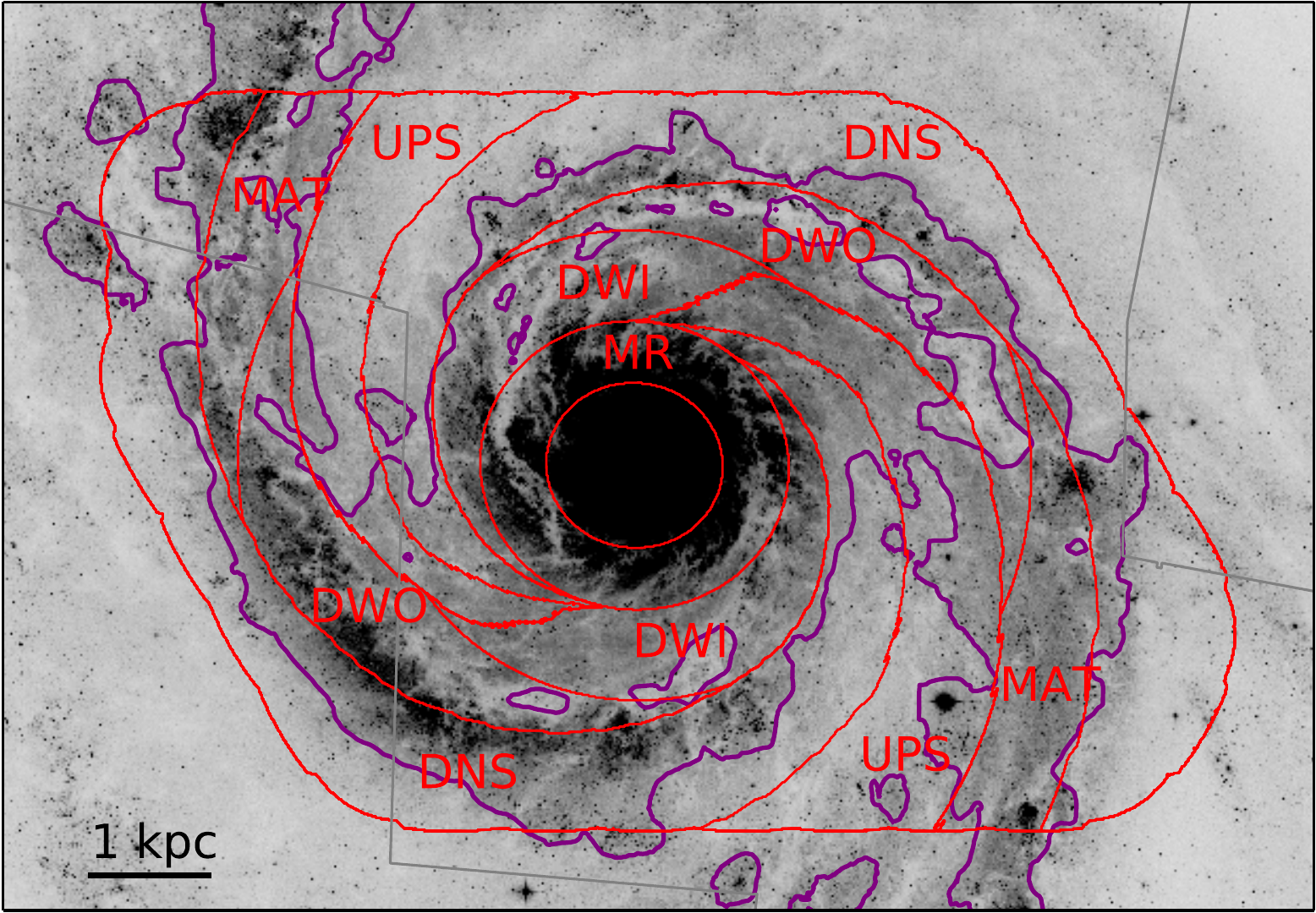}
\caption{Dynamical regions as defined by \citet{colombo2014a} for the PAWS project (red contours), overplotted on the $V$ band frame and compared to the SA-IA division presented in Section~\ref{sec:environment} (purple contours).}
\label{fig:dynregions}
\end{figure}
This comparison shows that the mass distributions of GMCs and clusters have similar radial trends in this central part of the galaxy.
However, since the biggest differences between GMCs mass function in C14 are found comparing different dynamical regions, we use their same division to analyse the clusters found in each of their sub-regions. 
We divide the spiral arms into inner density-wave spiral arms (DWI), outer density-wave spiral arms (DWO) and material arms (MAT), while the inter-arm zone is divided into downstream (DNS) and upstream (UPS) regions relative to the spiral arms. We point out that the arm/inter-arm division used in C14 is not the same as our SA/IA division, due to the shift between the peaks of optical and radio emissions \citep[see, e.g.,][]{schinnerer13,schinnerer2017}. This is illustrated in Fig.~\ref{fig:dynregions}. 
Considering the limited size of these sub-regions and the fact that clusters survive for much longer timescales than GMCs, possibly moving from their natal place, we consider in this analysis only clusters with ages $\leqslant100$~Myr. This timescale also appears to be, in a hierarchical structure, the typical scale for young stellar complexes to dissolve \citep{gieles2008,bastian2009,gouliermis2015}. We expect clusters younger than this to be located close to their original birthplace.

Results of the mass function fit are given in Tab.~\ref{tab:massfit_gmc_clusters}.
Similarly to the GMC results, we find power-law slopes flatter than $-2$ in the DWI and DWO regions and a slope steeper than $-2$ in the MAT region. The slope in the inter-arm region is steeper than in the DWI and DWO regions of the arm. The truncation mass remains unconstrained in some of the sub-regions ($N_0$ values consistent with 1 within uncertainties) because of the low number of YSCs.
The YSC mass function follows the same general trends of GMC mass function in the same regions. 
A difference in the mass function between the internal spiral-arm ($R_{gal}<85$'', DWI and DWO) and the part outside a radius of 85'' (MAT) is observed in both GMCs and clusters. A reason for this difference, as suggested by C14, may be that the MAT region is defined to be beyond the radius where the torque associated with the density wave spiral goes to zero (see \citealp{meidt2013,querejeta2016}). This means that the gas in the MAT region behaves like in flocculent galaxies, where arms are formed by gas over-densities in rotation with the rest of the disk. The implication on the mass function is that its shape is similar to what is found in the inter-arm environment. 

Differences in the GMC mass function as a function of arm and inter-arm environments have been also studied in simulations. As described by C14, the mass function of GMCs  from the simulations of \citet{dobbs2013} is shallow in the arm environment and steep in the inter-arms, when considering a two-armed spiral galaxy (Figure 7 of C14). 
The physical process causing this difference should be able not only to move the gas to the arms (where most of the star formation activity happens), but also to prevent the fragmentation of massive clouds there. Streaming motions associated with the spiral potential have been proposed as a possibility to lower the gas pressure outside the GMCs, leading to higher stable GMCs masses in the arms \citep{meidt2013,jog2013}.
From what we derived in the analyses of this section, we suggest that the processes that regulate the gas motion inside the galaxy, via the regulation of GMC masses are also consequently able to influence the cluster mass distribution. Indeed high pressure will lead GMC forming compact clusters, while low pressure will allow the GMC to form more dispersed stars (e.g., \citealp{elmegreen2008}). The main features seen in the arm and inter-arm clusters are also seen in GMCs and therefore a possible explanation of why the clusters are on average more massive in the spiral arms is the fact that they originate from more massive clouds.

\subsection{Age Functions}
\label{sec:agefunc}
The age distribution of clusters is regulated by the combination of the star and cluster formation history and of cluster disruption. Disentangling the two effects is possible only by knowing the star formation history by other means. We instead study YSC disruption in M51 assuming a constant star formation history and analysing the drop in the number of clusters going to older ages, via the fit of their age functions, $dN/dt$. 
A constant star formation history is usually a good assumption for spiral galaxies, which keep the same SFR over long periods. We know however that M51 is an interacting systems and that galaxy interactions have been proved to enhance the star formation \citep{pettitt2017}. The interaction in M51 started around roughly 300-500~Myr ago \citep{salo2000,dobbs2010_m51} and we assume that the star formation rate has not changed drastically over the period of the interaction. As explained in the next paragraphs, we are looking only at a very short period of the galaxy's life (namely the latest 200 Myr) and therefore we expect the star formation history in this age range not to be affected by the interaction.
We also assume that the constancy of the star formation history is not spatially-dependent. Both these assumptions are validated by recent photometric studies of stars in M51 \citep{cooper2012,eufrasio2017}. 

Age functions are built by dividing each subsample into age bins of 0.5 log(age/Myr) width and taking the number of sources in each bin normalised by the age range spanned by the bin. For a constant star formation history, they are expected to show a flat profile in case of no disruption (i.e. same number of cluster per age interval). On the other hand, in case of cluster disruption they are expected to display a declining profile, with a shape depending on the strength and type of disruption process (see \citealp{lamers2009} for a description of the expected age function shapes for different disruption models).

\begin{figure}
\centering
\includegraphics[width=\columnwidth]{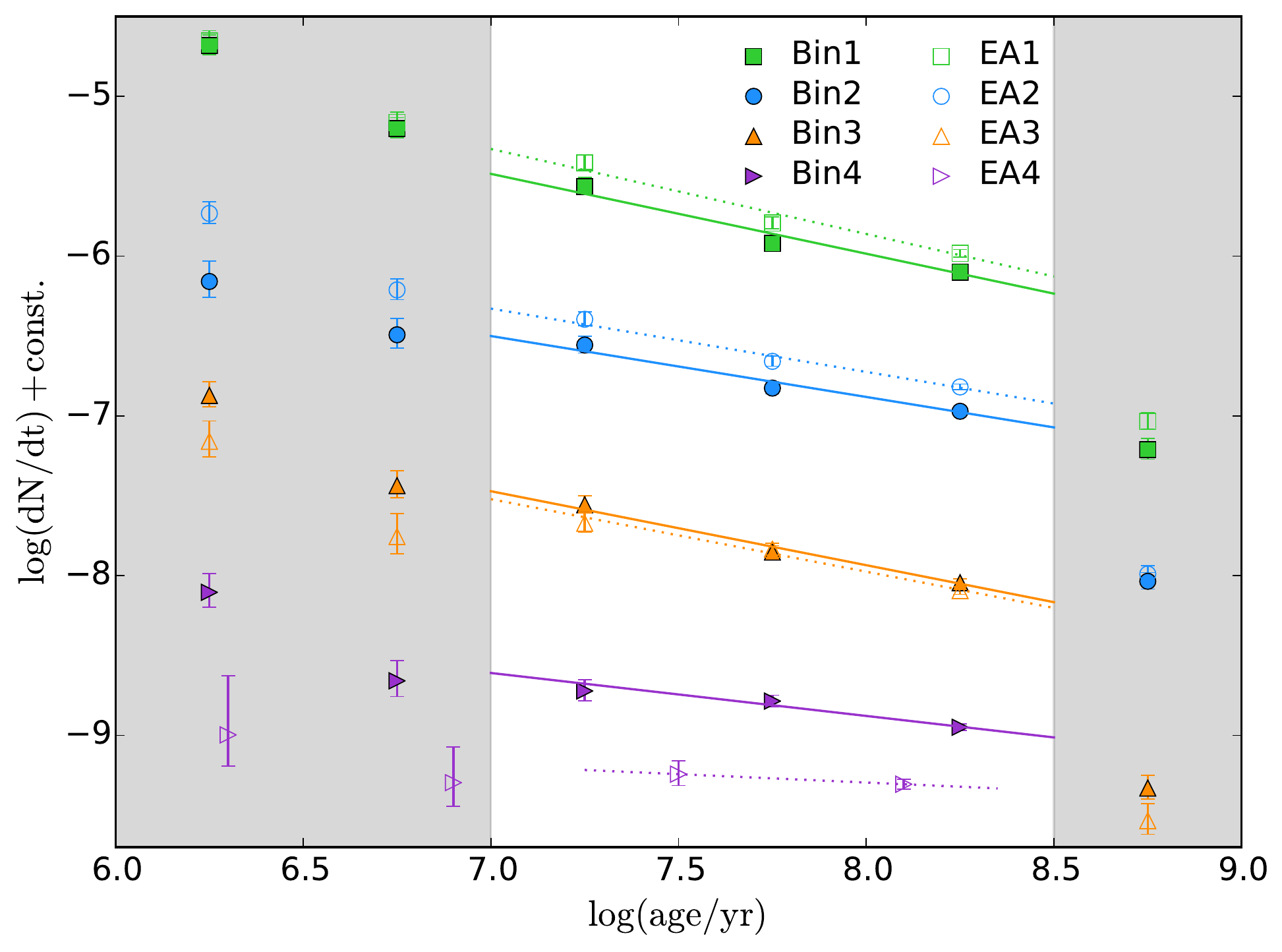}
\caption{Age functions of the radial subsamples. Age bins are 0.5 dex wide, and the fit was performed only for points in the range log(age)$=7-8.5$. In the case of EA4 the clusters in that age range have been divided in only 2 bins due to low number statistics. Gray shaded regions mark the age ranges excluded from the fit. Fit results are given in Tab.~\ref{tab:agefit_bins}}
\label{fig:agefit_bins}
\end{figure}

\begin{figure}
\centering
\includegraphics[width=\columnwidth]{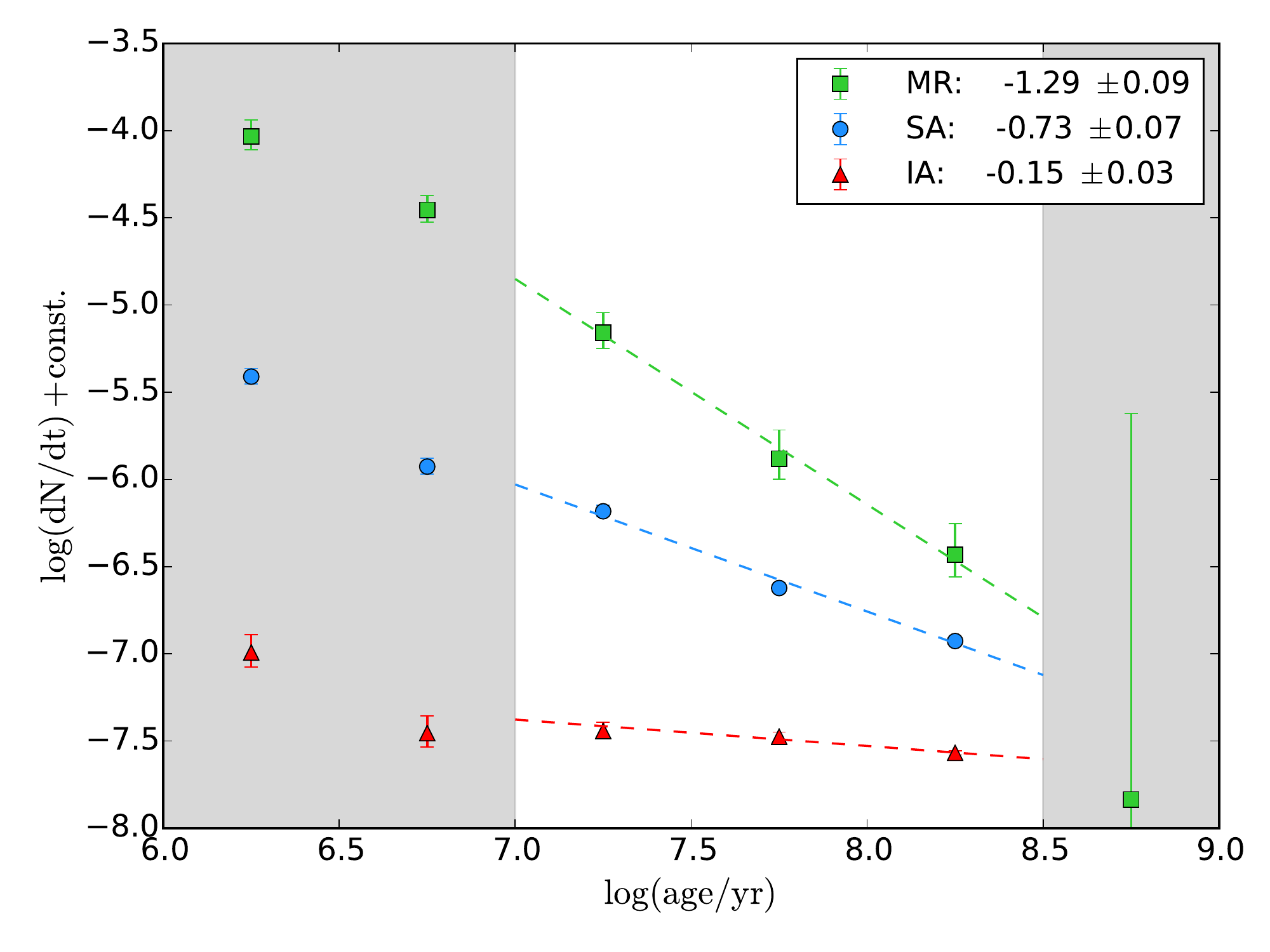}
\caption{Same as Fig.~\ref{fig:agefit_bins} for the molecular ring (MR), arm (SA) and inter-arm (IA) regions.} 
\label{fig:agefit_saia}
\end{figure}

Age functions are shown in Fig.~\ref{fig:agefit_bins} for the radial binning and in Fig.~\ref{fig:agefit_saia} for the SA and IA division. Incompleteness affects the sources older than 200~Myr, causing a drop in the number of sources detected at those ages, and consequently also a steepening in the age function. On the other hand, the sample at young ages could be contaminated due to the difficulty to assess the dynamical status of the sources we are studying. Assuming a typical cluster radius of a few parsecs \citep{ryon2015,ryon2017} we can also infer that sources older than $\sim10$~Myr have ages older than their crossing time. This is not true for younger sources, which may be unbound systems quickly dispersing during the first Myr of their life. We are interested in how the gravitationally-bound systems evolve and therefore those young sources are considered contaminants. 

Neglecting sources older than 200~Myr because of incompleteness and younger than 10~Myr because of contamination, we are left with age functions in the age range log(age/Myr)$=7-8.5$, which we fit with power-laws. The power-law fit of the age function, $dN/dt\propto t^{\gamma}$, is commonly used in the study of cluster populations and the recovered $\gamma$ slopes are used to describe the strength of the cluster disruption process (see e.g. Section~3 of the review by \citealp{adamobastian2015}).
The fit results are listed in Tab.~\ref{tab:agefit_bins}.
The innermost bin has the steepest age function, with a slope $\gamma=-0.50 \pm0.09$, while the outermost bin has the shallowest one, $\gamma=-0.27 \pm0.06$. Bin 2 and 3 have values in between, $-0.38 \pm0.07$ ($1\sigma$ consistent with Bin 4) and $-0.46 \pm0.06$ ($1\sigma$ consistent with Bin 1). The differences between the bins are within $2\sigma$. We note that the age functions at log(age/Myr)$=6.75$ lie on the best fit lines, therefore the slopes we recover are representative for the age functions down to $\sim3$~Myr.
In all bins the slopes vary around the value $\gamma_{tot}=-0.30 \pm0.06$ found for the entire sample, without significant differences. 
Very similar results are retrieved if bins of equal area are considered. For the outermost bin, EA4, the number of sources was too small, therefore only 2 age bins of width 0.6 dex were considered in the age range log(age/Myr)$=7-8.5$.

The MR again behaves dramatically, as the recovered slope there is even steeper than $-1$ ($\gamma=-1.29 \pm0.09$). We expect the sample here to be partially incomplete at ages $\sim200$~Myr (log(age)$\sim8.3$) but the age function in Fig.~\ref{fig:agefit_saia} seems to keep the same slope also up to the last fitted bin (log(age)$\sim8.5$). The steepness of the slope suggests that, particularly in this region, it can be the case that the hypothesis of a constant star formation history is not valid, and that the SFR increased during the most recent Myrs. 

\begin{table}
\centering
\caption{Results of the fit of the age function with a power law function. Only clusters with M$>5000$ \msun\ were considered (except in the MR, where the limit used was M$>10^4$ \msun). Functions are plotted in Fig.~\ref{fig:agefit_bins} and Fig.~\ref{fig:agefit_saia}.}
\begin{tabular}{lclc}
\hline
Bin	& $\gamma$	& Bin	& $\gamma$ \\
\hline
\hline
MR	& $-1.29\ _{\pm0.09}$	& \		& \				\\
Bin 1	& $-0.50\ _{\pm0.09}$	& EA 1	& $-0.53\ _{\pm0.10}$ \\
Bin 2	& $-0.38\ _{\pm0.07}$	& EA 2	& $-0.40\ _{\pm0.06}$ \\
Bin 3	& $-0.46\ _{\pm0.06}$	& EA 3	& $-0.45\ _{\pm0.06}$ \\
Bin 4	& $-0.27\ _{\pm0.06}$	& EA 4	& $-0.15$ \\
\hline
SA	& $-0.73\ _{\pm0.07}$	& IA		& $-0.15\ _{\pm0.03}$ \\
\hline
\end{tabular}
\label{tab:agefit_bins}
\end{table}

\begin{figure}
\centering
\includegraphics[width=\columnwidth]{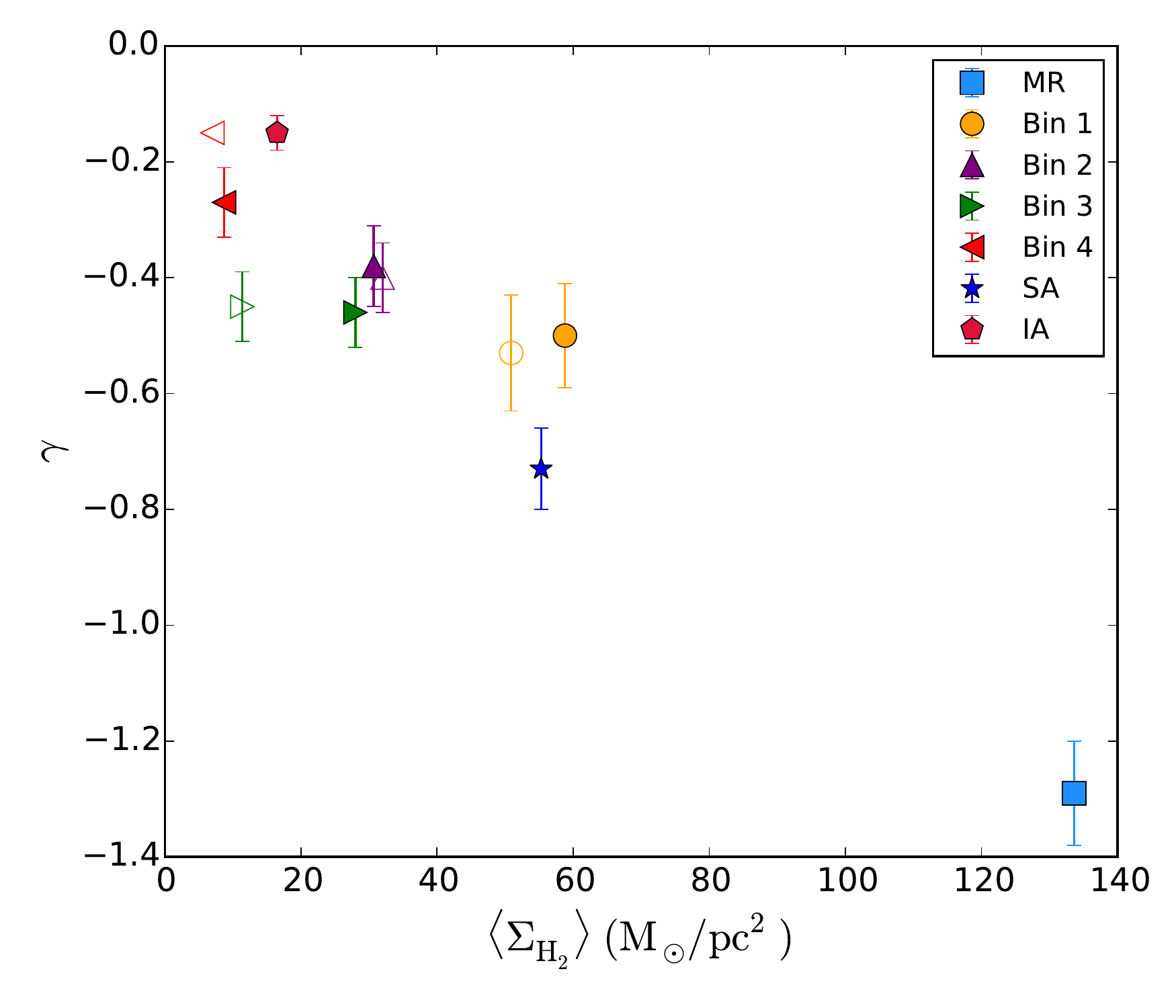}
\caption{Best-fit values for the slope of the age function, $\gamma$, as listed in Tab.~\ref{tab:agefit_bins} plotted in function of the $H_2$ surface density of each sub-region. The open symbols represent the bins of equal area.} 
\label{fig:gammaVSsigmagas}
\end{figure}

The division in SA and IA (Fig.~\ref{fig:agefit_saia}) confirms that those regions have very different disruption strengths.
The disruption seems therefore to depend on the environment and, considering the average gas densities in the regions, to be stronger in denser environments, as modeled by \citet{elmegreen2010}, \citet{kruijssen2011} and \citet{miholics2017}. 
The relation between the $\gamma$ slope and the average \sigmah\ in M51 sub-regions is illustrated in Fig.~\ref{fig:gammaVSsigmagas}.
The large difference between the age functions of the SA and IA regions could also be caused by the migration of clusters. If the majority of clusters are formed in the arm, such clusters may be old as they reach the inter-arm, contributing to make the age function appearing flat. It should however be considered that, in M51, clusters seem to stay distributed along the spiral arms until an age of $\sim200$ Myr, as was pointed out in Section~4.3 of Paper I and explored more deeply in Shabani et al., submitted to MNRAS.

\begin{figure*}
\centering
\includegraphics[width=0.7\textwidth]{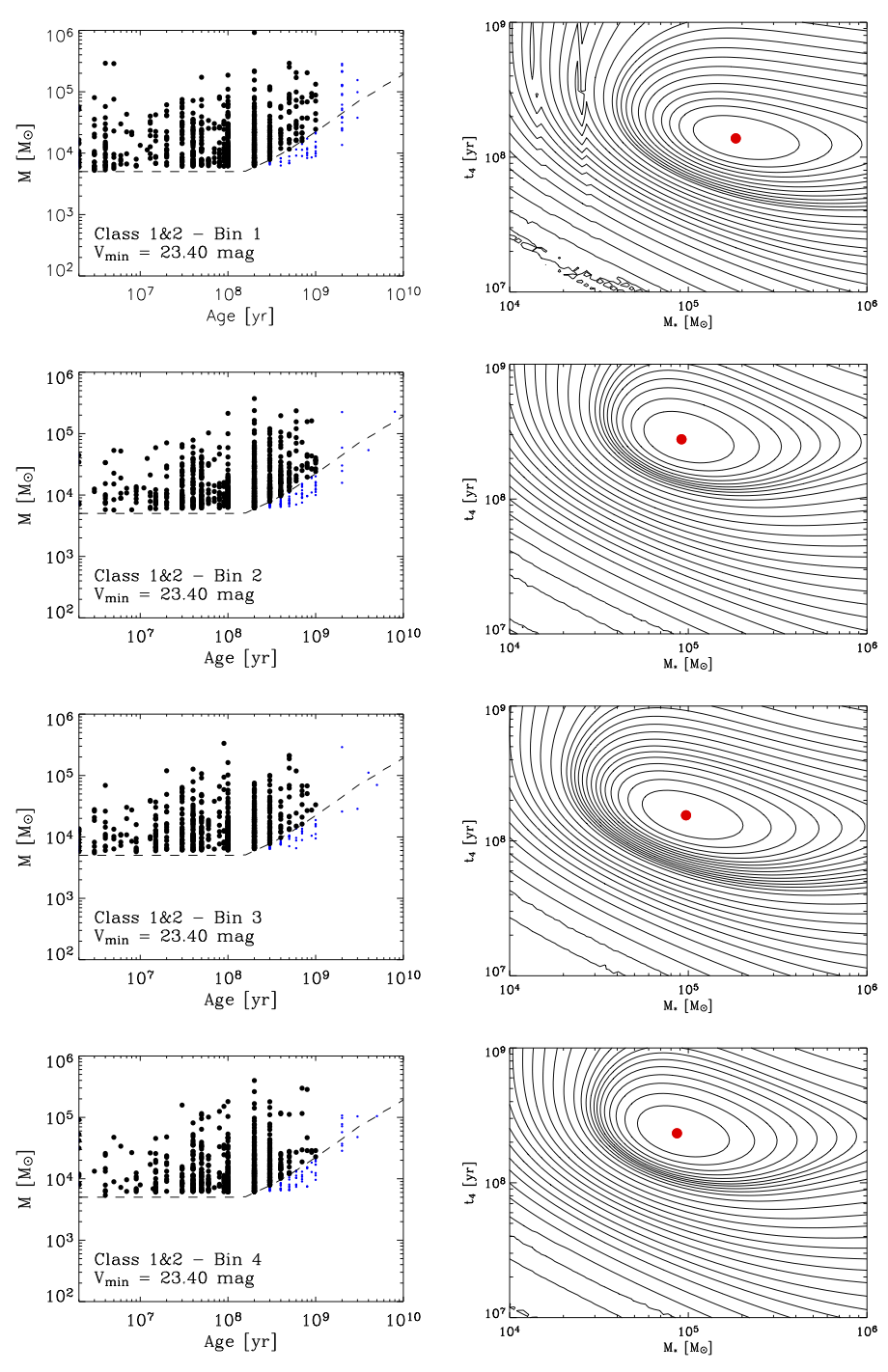}
\caption{Maximum-likelihood fit for the mass cut $M^*$ and the typical timescale $t_4$, assuming a mass-dependent disruption time. The bins 1 to 4 are shown from top to bottom. The left panels show the clusters used in the analysis for each bin (black points). The dashed lines indicate the completeness cut used, combining both a mass cut at $M_{lim}=5000$ \msun\ and a $V$ band magnitude cut at $V_{min}=23.4 $ mag. Blue points are clusters left out of the analysis by this completeness cut and by an age cut at $10^3$~Myr.
The right panels show the maximum-likelihood value as a red dot in the $M^*-t_4$ space, as well as likelihood contours. Likelihood values are calculated on a grid covering the plotted $M^*$ and $t^4$ intervals and the contours shown are chosen so that $3\%$ of the resulting likelihood values are enclosed between each two consecutive contours. Parameters associated with the maximum-likelihood are listed in Tab.~\ref{tab:maxlike_bins}.} 
\label{fig:maxlike_bins}
\end{figure*}

\begin{figure*}
\centering
\includegraphics[width=0.7\textwidth]{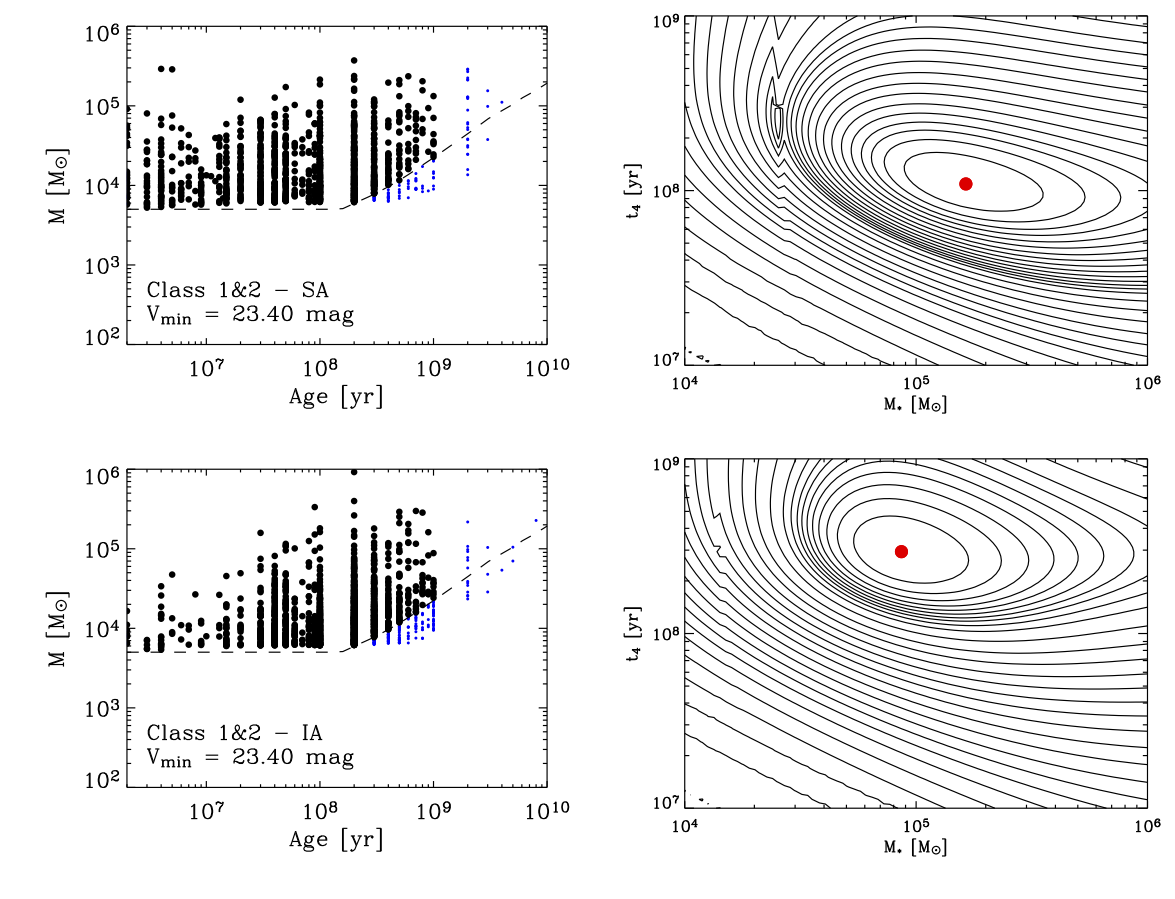}
\caption{Same as Fig.~\ref{fig:maxlike_bins} but for the arm, inter-arm and molecular ring regions.} 
\label{fig:maxlike_saia}
\end{figure*}

Similarly to what was done in Paper I, we estimate the typical disruption time for $10^4$ \msun\ clusters, $t_4$, in different regions inside the galaxy based on the hypothesis where the disruption time depends on the cluster mass as $t_{dis}\propto M^{0.65}$ \citep{lamers2005}, therefore assuming that clusters with smaller masses have shorter disruption timescales. We use a maximum-likelihood code introduced by \citet{gieles2009}, assuming an initial cluster mass function described by a power law with a slope of $-2$ and with an exponential truncation at $M_0$ that evolves as a function of the strength of the disruption, given by the timescale $t_4$. Both $M_0$ and $t_4$ are free parameters in the analysis.
Our results are presented in Tab.~\ref{tab:maxlike_bins} and Fig.~\ref{fig:maxlike_bins} and \ref{fig:maxlike_saia}. The most-likely values for $M_0$ are generally smaller than $10^5$ \msun\ but consistent with it within 2$\sigma$. They are very similar in Bins 2 to 4 and IA regions, while in Bin 1 and SA the value is slightly larger.  
We find shorter disruption times in the SA environment and in Bins 1 and 3 compared to the other regions. The differences are within a factor $2-3$ but in all cases the values of $t_4$ are larger than 100~Myr. These long timescales can explain why in some regions we see very little disruption in the age range $10-200$~Myr of the age functions. The analysis of the MR was strongly limited by the low number of clusters in the region, and was therefore neglected. 

\begin{table}
\centering
\caption{Results of the maximum-likelihood fit of the mass function, with the mass cut $M_0$ and the disruption time of a $10^4$ \msun\ cluster $t_4$ as free parameters. In all cases sources in the age range $1-10^3$~Myr have been considered, limited by the mass cut $M>5000$ \msun. Incompleteness at high masses due to the magnitude cut $V_{mag}<23.4$ mag has also been considered in the fitting analysis. The MR is neglected in this analysis because, due to the low number of clusters, the code is not converging to a result.}
\begin{tabular}{lcc}
\hline
Bin	& $M_0\ (10^5$ \msun)		& $t_4\ (10^8$ yr)	\\
\hline
\hline
SA	& $1.64\ ^{+0.18}_{-0.10}$	& $1.09\ ^{+0.06}_{-0.07}$ 	\\
IA	& $0.86\ ^{+0.05}_{-0.11}$		& $2.94\ ^{+0.17}_{-0.18}$ 	\\
\hline
Bin 1	& $1.84\ ^{+0.20}_{-0.35}$	& $1.38\ ^{+0.08}_{-0.08}$ \\
Bin 2	& $0.92\ ^{+0.10}_{-0.18}$	& $2.77\ ^{+0.31}_{-0.34}$ \\
Bin 3	& $0.97\ ^{+0.11}_{-0.12}$		& $1.55\ ^{+0.17}_{-0.09}$\\
Bin 4	& $0.86\ ^{+0.10}_{-0.11}$		& $2.33\ ^{+0.26}_{-0.29}$\\
\hline
\end{tabular}
\label{tab:maxlike_bins}
\end{table}


\subsection{Cluster Formation Efficiency}
\label{sec:gamma}
Another cluster property that has been predicted to depend on the galactic environment is the fraction of star formation happening in bound clusters. This is known as cluster formation efficiency, CFE ($\Gamma$). In the literature it has been proposed that $\Gamma$ should change as a function of the gas pressure, traced by \sigmah (or \sigmasfr), with denser environments hosting a higher fraction of bound clusters (see the model of \citealp{kruijssen2012}). We can test these predictions in the environment of M51, using the observed variations in \sigmah\  from the centre to the outskirts (see Fig.~\ref{fig:h2sfr}).	

We derive the CFE with the same approach used in Paper~I. In each bin, cluster masses are summed to provide a total mass in bound clusters with $M>5000$ \msun. This value is then corrected to find the total expected mass in clusters down to 100 \msun. In order to make this correction, an assumption of the shape of the mass function is necessary. We assume that mass functions, from 100 \msun\ to the most massive clusters observed, can be described by a power law with an exponent of  $-2$, exponentially truncated corresponding to the $M_0$ value derived by the best fit in Tab.~\ref{tab:massfit_bins}. In the calculation of $\Gamma$, only clusters with ages in the range $10-100$~Myr were considered. As pointed out in Section~\ref{sec:agefunc}, younger sources can be already unbound at birth. Their inclusion would artificially increase the derived value of $\Gamma$, whereas we are interested in the bound clusters only.
In order to derive a cluster formation rate, the total stellar mass in clusters is divided by the age range considered.
Finally, the cluster formation rate is divided by the SFR to obtain a cluster formation efficiency. We have used the SFR from the $FUV$+$24$ \mum\ measurement.
The derived values of $\Gamma$ are listed in Tab.~\ref{tab:gamma_bins}. In order to test how the age range selected affects $\Gamma$, we derived the CFE also for sources in age ranges $1-10$ and $1-100$~Myr. We use the SFR derived from \ha$+24$ \mum\ in the calculation of the CFE between $1-10$~Myr and the SFR obtained from $FUV$+$24$ \mum\ for the CFE in the age range $1-100$~Myr.
This method of deriving the CFE is only weakly affected by incompleteness, as it focuses only on the high-mass clusters (which are above the completeness limit) and corrects for the missing mass of low-mass clusters by the assumption of a power-law mass function with slope $-2$.

\begin{table*}
\centering
\caption{Star formation rate (SFR), cluster formation rate (CFR) and cluster formation efficiency ($\Gamma$) in each of the radial regions. The value derived in the age range $10-100$~Myr is considered the reference one, as it is the least affected by systematics. For comparison also values of $\Gamma$ derived over age ranges $1-100$~Myr and $1-10$~Myr are reported. For each age range the table also gives the star and cluster formation rates. Uncertainties of 10\% in the SFR are considered.}
\begin{tabular}{lcclcclccl}
\hline
Bin	& \multicolumn{3}{c}{Ages $10-100$} & \multicolumn{3}{c}{Ages $1-100$} & \multicolumn{3}{c}{Ages $1-10$} \\
\	& SFR		& CFR 		& $\Gamma$ 	& SFR		& CFR 		& $\Gamma$ 	& SFR		& CFR 		& $\Gamma$	\\
\	& [\msun/yr]	& [\msun/yr] 	& [\%] 		& [\msun/yr]	& [\msun/yr] 	& [\%] 		& [\msun/yr]	& [\msun/yr] 	& [\%] 			\\
\hline
\hline
MR	& 0.220 	& $0.047\ _{\pm0.017}$ & $\textbf{21.3}\ _{\pm7.9}$ & 0.220	& $0.086\ _{\pm0.030}$ & $39.2\ _{\pm14.3}$ &0.176 & $0.437\ _{\pm0.167}$ & $248.1\ _{\pm98.3}$ 	\\
\hline
Bin 1	& 0.524 	& $0.100\ _{\pm0.028}$ & $\textbf{19.1}\ _{\pm5.7}$	& 0.524 	& $0.135\ _{\pm0.037}$ & $25.8\ _{\pm7.4}$ & 0.404 & $0.442\ _{\pm0.137}$ & $109.5\ _{\pm35.6}$	\\
Bin 2	& 0.396 	& $0.089\ _{\pm0.025}$ & $\textbf{22.4}\ _{\pm6.8}$	& 0.396	& $0.095\ _{\pm0.028}$ & $23.9\ _{\pm7.3}$ & 0.314	 & $0.144\ _{\pm0.044}$ & $45.8\ _{\pm14.6}$\\
Bin 3	& 0.385 	& $0.124\ _{\pm0.038}$ & $\textbf{32.1}\ _{\pm10.3}$& 0.385 	& $0.130\ _{\pm0.039}$ & $33.8\ _{\pm10.7}$ & 0.373 & $0.177\ _{\pm0.055}$ & $47.5\ _{\pm15.6}$	\\
Bin 4	& 0.331 	& $0.097\ _{\pm0.027}$ & $\textbf{29.3}\ _{\pm8.7}$	& 0.331	& $0.103\ _{\pm0.027}$ & $31.1\ _{\pm8.9}$ & 	0.346 & $0.148\ _{\pm0.055}$ & $42.7\ _{\pm16.6}$	\\
\hline
EA 1	& 0.595 	& $0.097\ _{\pm0.026}$ & $\textbf{16.3}\ _{\pm4.7}$	&  0.595 	& $0.117\ _{\pm0.033}$ & $19.7\ _{\pm5.9}$ & 0.448 & $0.292\ _{\pm0.108}$ & $65.1\ _{\pm25.0}$	\\
EA 2	& 0.641 	& $0.117\ _{\pm0.030}$ & $\textbf{18.3}\ _{\pm5.0}$	&  0.641	& $0.123\ _{\pm0.030}$ & $19.1\ _{\pm5.1}$ & 0.578	 & $0.161\ _{\pm0.073}$ & $27.8\ _{\pm13.0}$\\
EA 3	& 0.286 	& $0.059\ _{\pm0.021}$ & $\textbf{20.5}\ _{\pm7.6}$	&  0.286 	& $0.062\ _{\pm0.021}$ & $21.6\ _{\pm7.8}$ & 0.293 & $0.084\ _{\pm0.120}$ & $28.8\ _{\pm41.0}$	\\
EA 4	& 0.114 	& $0.024\ _{\pm0.010}$ & $\textbf{21.2}\ _{\pm8.8}$	&  0.114	& $0.023\ _{\pm0.010}$ & $20.5\ _{\pm8.7}$ & 0.117 & $0.014\ _{\pm0.055}$ & $11.9\ _{\pm47.1}$	\\
\hline
\end{tabular}
\label{tab:gamma_bins}
\end{table*}

Different sources of uncertainty are considered in the calculation. 
Both the uncertainties in the derived ages and masses and in the fits of the mass function will affect the value of $\Gamma$. In both cases we used simulated populations to assess the propagation of those uncertainties. We considered errors on ages and masses of 0.1 dex. For the mass function parameters, instead, we considered a $\sigma$ of 0.1 for the slope of $-2$ and a $\sigma$ equal to the uncertainty found in Tab.~\ref{tab:massfit_bins} for the truncation mass. 
A Poisson error due to the finite number of sources, used to calculate the total cluster mass, is also considered as a source of uncertainty.
Finally, the uncertainty associated with the SFR is 10\%.

\begin{figure}
\centering
\includegraphics[width=\columnwidth]{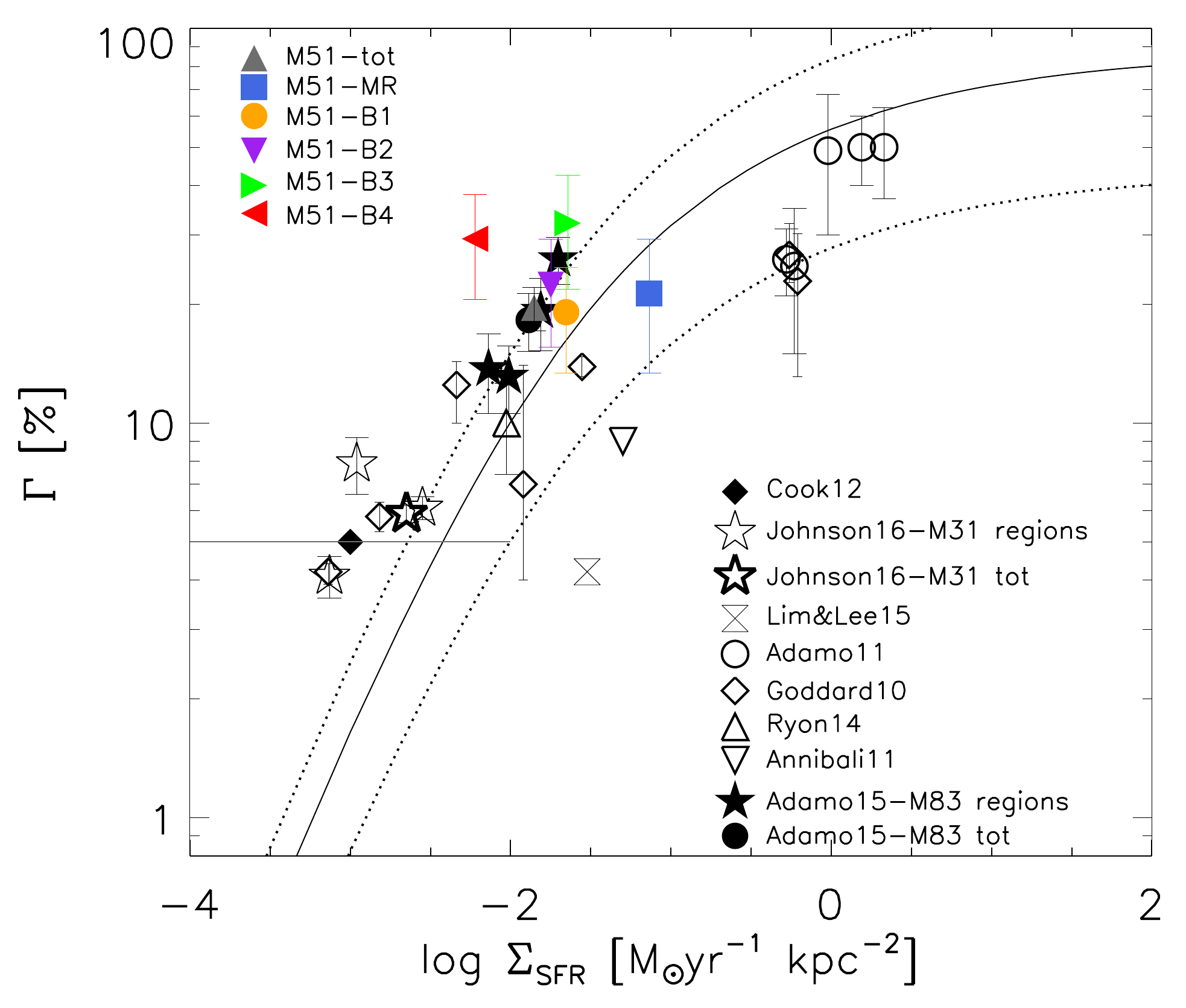}
\caption{CFE for the whole M51 compared to the CFE values retrieved in the radial bins (listed in Tab.~\ref{tab:gamma_bins}). The CFE values are plotted as a function of the average \sigmasfr\ and the fiducial model of \citet{kruijssen2012} is overplotted (solid line) with its 3$\sigma$ uncertainty (dotted lines). Literature CFE values of other galaxies are also shown.}
\label{fig:gamma_bins}
\end{figure}

The recovered CFEs in the subregions are all within $\sim$20\% and $\sim$30\%. Bin 1 and the MR region have lower CFEs even though they are the densest regions. 
As can be seen in Fig.~\ref{fig:gamma_bins}, the CFE does not show any trend with the average \sigmasfr\ of each bin. 
If compared to the values derived with the model by \citet{kruijssen2012}, we observe variations generally within a factor of 2 (a factor 3 in the case of Bin 4). However, we note that Fig.~\ref{fig:gamma_bins} concerns a comparison to the fiducial (i.e.~simplified) \citet{kruijssen2012} model, which assumes a relation between the galactic rotation curve and the gas surface density profile. When using the complete model, which treats the gas surface density and the angular velocity as independent variables, the scatter around the prediction is smaller (see Section~\ref{sec:discussion}).

Considering the entire age range down to 1~Myr ($1-100$~Myr) does not noticeably change the values of $\Gamma$. On the other hand, if only clusters younger than 10~Myr are considered $\Gamma$ reaches larger values. This could be expected because of the contamination of young unbound sources. 
In this age range, the mass calculation relies on a small number of sources and the final uncertainties are therefore much bigger than in the other two cases. 
The $\Gamma$ values in bins of same area (also in Tab.~\ref{tab:gamma_bins}) do not show significant differences. 
We will discuss these results in Section~\ref{sec:discussion}, comparing our observations with model predictions.

\section{A Self-consistent Model for Cluster Formation}
\label{sec:discussion}
The analyses of the cluster mass function, age distribution, and formation efficiency suggest that the largest differences in the environments of M51 can be found 
when comparing the cluster population in the spiral arms of the galaxy to the one in the inter-arm regions. However, some of these differences seem to be washed out when the sample is averaged over annular bins at different galactocentric distances.

The model of \citet{reinacampos2017} (hereafter \mrc) studies the dependence on the gas surface density and angular velocity to predict the maximum GMC and cluster mass scales that can form in a galaxy. 
We apply the model to our data, in order to provide predictions of the maximum GMC and cluster masses from the gas properties. These predictions are then compared to the truncation masses observed. We refer the reader to \mrc\ for a detailed description of the model, but we summarize here briefly the motivation for this model and its main points.
It has been recently suggested that GMC and cluster maximum masses could have a common origin related to the Toomre mass \citep{kruijssen2014}, i.e., the maximum mass of gas that can gravitationally collapse against centrifugal forces in the disk of a galaxy \citep{toomre1964}.
The idea that the maximum collapsing gas mass is set by shearing motions has been used to explain the maximum masses of GMC and clusters in local galaxies (e.g., \citealp{adamo2015} and \citealp{freeman2017}) as well as determining the maximum size for the coherence of star formation \citep{grasha2017b}.
\mrc\ argue that, under some conditions, feedback activity from young stars can become effective before the gas cloud has entirely collapsed, interrupting the mass growth of the forming GMCs and any clusters forming within them. 
The method quantifies the competition between both mechanisms, 
to establish whether the maximum mass that can collapse into a cloud (considered to be the maximum GMC mass achievable, $M_{GMC,max}$), corresponds to the mass enclosed in the unstable region (Toomre mass) or to a fraction of it. 

The model is self-consistent and depends only on 3 parameters: the gas surface density $\Sigma_g$, the epicyclic frequency $\kappa$ and the velocity dispersion of the gas $\sigma$. 
We can therefore apply the model to the M51 radial bins. We have already calculated the average H$_2$ surface densities (values reported in Tab.~\ref{tab:properties}), to which we add the surface density of atomic gas (HI) from \citet{schuster2007} to derive the total gas surface density, $\Sigma_g$. The surface density of HI is almost negligible, compared to \sigmah, in the centre of the galaxy but starts having a noticeable effect from R$_{gal}\sim4$ kpc outwards (see Fig.~\ref{fig:h2sfr}).
Using the 2nd moment maps of the $^{12}$CO(1-0) gas from IRAM single-dish observations\footnote{Retrievable on the PAWS website: http://www2.mpia-hd.mpg.de/PAWS/PAWS/Data.html}, we also calculate the average velocity dispersion of the molecular gas inside each radial bin. The epicyclic frequency is derived from the rotation curve of the galaxy in \citet{garciaburillo1993}.

\begin{figure}
\centering
\includegraphics[width=\columnwidth]{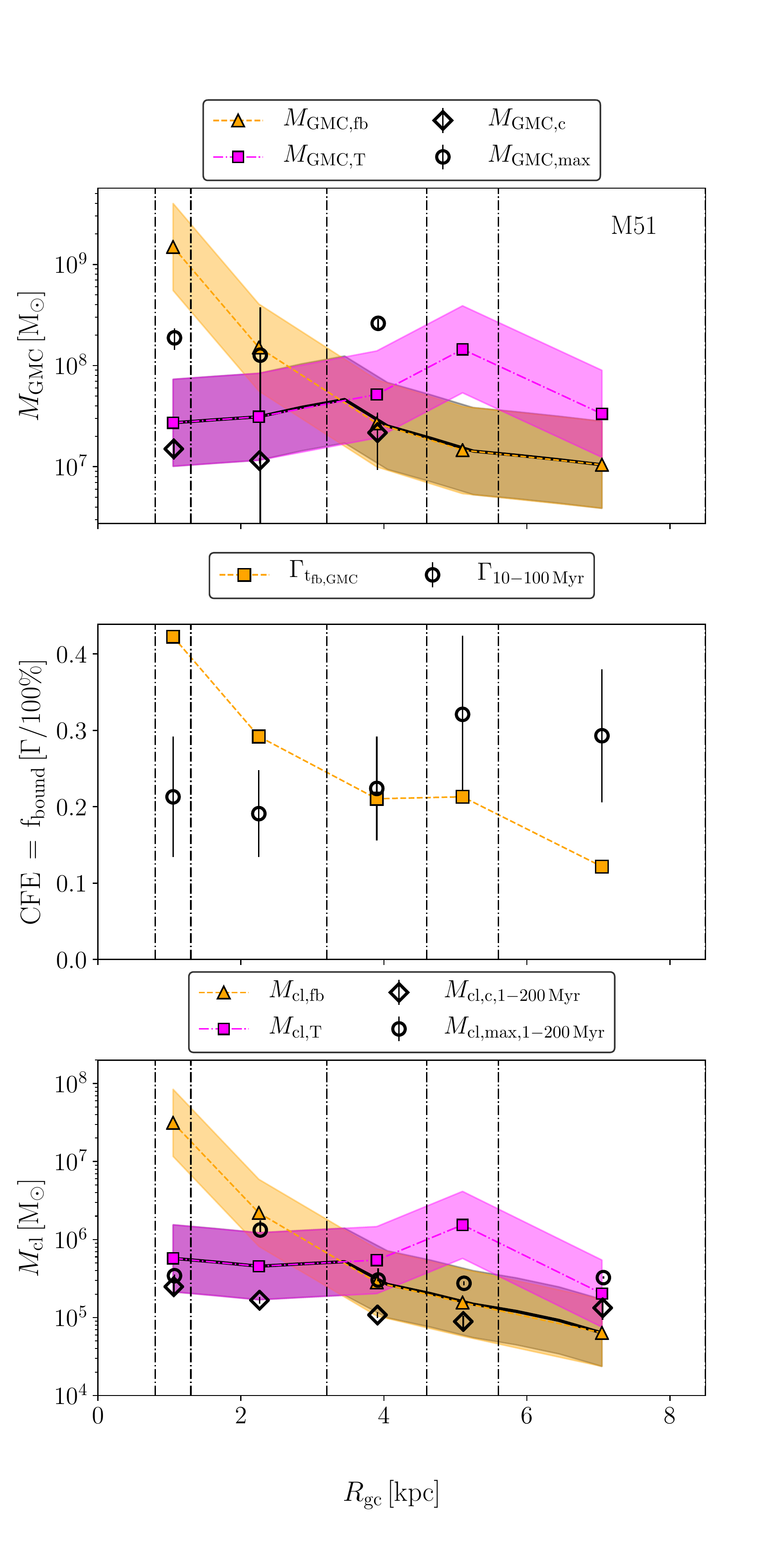}
\caption{
Comparison between the observed maximum and truncation masses of GMCs and YSCs derived in this work and the predictions made with the \citet{reinacampos2017} model, as a function of galactocentric distance. \textit{Top panel:} Maximum mass-scales of GMCs. Predictions for the feedback(shear)-limited regimes are given by the orange (pink) symbols. The black solid line marks the predicted maximum mass-scale at each radial bin; observed values should lie on the dark-shaded area in order for the predictions to be 1$\sigma$ consistent with observations. \textit{Middle panel:} CFE derived in this work (black symbols) and predicted using \citet{kruijssen2012} model at $t=t_{fb}$ (orange squares). \textit{Bottom panel:} Maximum mass-scales of stellar clusters. We use the same colour coding as the top panel. Along the x-axes we show the galactocentric distance $R_{\rm gc}$, with vertical lines showing the binning used. Predictions are plotted with filled symbols within shadowed areas (which represent the uncertainties), whereas values derived from observed data are plotted as empty symbols.
}
\label{fig:marta}
\end{figure}

Results for the maximum GMC masses are shown in the top panel of Fig.~\ref{fig:marta}. Shear and centrifugal forces determine the maximum GMC mass in the internal $\sim3.5$ kpc of the galaxy, where $M_{GMC,max}$ is therefore the Toomre mass (pink band in the top panel of the Figure). At larger galactocentric distances, however, the feedback time becomes shorter than the collapsing timescale of the Toomre mass, and feedback is therefore able to stop the collapse, reducing the amount of mass that can collapse (orange band in the figure). As a result, the model does not predict a large variation in the maximum expected GMC mass (black solid line) at different galactocentric distances, in line with what we observe. The model underestimates the maximum GMC mass observed at all considered radii, but is consistent within the errors with the truncation mass of the mass functions, $M_0$ (from Tab.~\ref{tab:massfit_gmc_clusters}). 

Following \citet{kruijssen2014}, the maximum cluster mass can be derived from $M_{GMC,max}$ taking into account the fraction of gas converted into stars, i.e., the star formation efficiency $\epsilon$, and the fraction of star formation happening in clustered form, i.e., the cluster formation efficiency $\Gamma$, such that:
\begin{equation}
M_{cl,max}=\epsilon\Gamma M_{GMC,max}
\end{equation}
The cluster formation efficiency can, in turn, be derived from the gas properties, under the assumption that star formation is halted by the onset of feedback activity. The second panel of Fig.~\ref{fig:marta} shows the predicted $\Gamma$ using the model by \citet{kruijssen2012} at $t=t_{fb}$ (see \mrc\ for details). The predicted $\Gamma$ deviate significantly from the estimated one (from Tab.~\ref{tab:gamma_bins}) in some of the radial bins. 
In the inner bins (MR and Bin1) this discrepancy could be caused by cluster disruption, which strongly affects clusters in the denser environment, lowering the number of observed clusters in these regions and therefore also the value of the estimated CFE.

We evaluate the expected maximum cluster mass assuming a star formation efficiency of $\epsilon=0.05$\footnote{This value is lower than the fiducial one of the RC\&K17 model ($\epsilon=0.1$), but was chosen to match the typical star formation efficiency found in nearby star-forming regions \citep{lada2003}, whereas RC\&K17 adopted an elevated value to accommodate higher star formation efficiencies in high-redshift clumps.}, and using the predicted $\Gamma$.
The resulting $M_{cl,max}$ predicted by the model are compared to observations in the bottom panel of Fig.~\ref{fig:marta}.
For the cluster maximum masses and mass truncations $M_0$ we use the values listed in Tab.~\ref{tab:massfit_bins}.
The model predicts an almost flat radial profile for $M_{cl,max}$, consistent with the absence of radial variation in the recovered truncation masses (see Section~\ref{sec:massfunc}). This result suggests that the radial profile for $M_{cl,max}$ can be set by the average gas properties at the galactic sub-scales considered.

We can compare these results with the analysis of another local spiral galaxy, M83. Like M51, M83 has been studied radially, but, unlike our case, the maximum GMC and cluster masses in M83 appear to be determined only by shear and centrifugal forces, resulting in a monotonic decrease with increasing increasing $R_{gal}$ in line with the predictions (Figure 9 in \mrc).

\section{Conclusions} 
\label{sec:conclusions}
We divided the galaxy M51 in subregions and studied the cluster sample in each region, looking for a possible dependence of the cluster properties on the galactic environment. The cluster catalogue production was described in a previous paper \citep{paper1} in which the cluster population as a whole was analysed. In this follow-up work, the galaxy has been divided in radial annuli containing equal number of clusters (Bin 1 to Bin 4, from the centre of the galaxy to the outskirts). Another division, in radial annuli of equal area was used to check the dependence of the results on the binning choice. In order to study the difference between the dense spiral arm environment (SA) and the inter-arm region (IA), we divided the galaxy in two environments of different background luminosity. 
The environment of each of the regions considered was characterized by its value of H$_2$ and SFR surface densities. Those quantities allowed a comparison between the observed cluster properties and predictions from models.
The analysis of cluster properties led to the following results:
\begin{enumerate}
\item The luminosity function shows a dearth of bright clusters in all 4 bins if a single power law fit is assumed. In the 2 outermost bins, however, a single power-law is a good fit of the function. The slopes recovered from the fit present variations among the radial bins. The biggest difference is found when comparing the arm and inter-arm environments, because the luminosity function has different slopes up to a magnitude of $\sim20$. This difference suggests that also the underlying mass function may differ noticeably in arm and inter-arm environments.
\item The mass function is similar in all radial annuli. Power-law slopes are all compatible within $2\sigma$. In all bins the high-mass part of the function is steeper and can be described by an exponential truncation. Truncation masses span the range $M_0=0.89-1.67$ ($\times10^5$ \msun). Both these results suggest that the mass distribution is on average similar at all radii inside M51. The molecular ring (MR), a dense region around the centre of the galaxy, is the only one showing a different mass distribution, flat and un-truncated. This difference may in part be caused by partial incompleteness. MFs in SA and IA regions are both well-fitted by a truncated function ($M_0\sim10^5$ \msun\ for IA region and $M_0\sim1.5\times10^5$ \msun\ for SA region). Truncation is more statistically significant in the SA region. In addition, they have different slopes, with IA having a significantly steeper slope. An analysis focusing only on the high-mass part of the function (M$>10^4$ \msun) confirms these findings.
\item A comparison with the giant molecular cloud (GMC) catalogue published by \citet{colombo2014a} shows that the mass functions of the two objects seem to behave similarly. In particular, dividing the samples using the M51 dynamical regions defined by \citet{meidt2013}, we recover for both clusters and GMCs mass functions that are shallow in the spiral arms and steep in the inter arm region.
This comparison suggests that the shape of the cluster mass function is not universal at sub-galactic scale and can be influenced by the mass shape of the GMCs, which in turn depends on the galaxy dynamics. This can be the cause of the difference in the mass function in the arm and inter-arm regions.
\item The study of the age distribution reveals regions with elevated cluster disruption (Bin 1 and the SA region), but also regions consistent with little disruption (Bin 4 and the IA region). The age function in the very gas-dense molecular ring drops quickly towards older ages, sign of an elevated disruption rate. The age function seems to strongly depend on the galactic environment, and in particular to have a steeper slope (more effective cluster disruption) in denser environments, as expected from models \citep[e.g.][]{elmegreen2010,kruijssen2011}. 
\item The fraction of stars forming in bound clusters, or cluster formation efficiency (CFE) is found to be in the range $\sim20-30\%$. Deeper analyses accounting for \sigmah\ reveal discrepancies with predicted CFE values. Cluster disruption is a possible cause for the observed discrepancies in the inner bins, but further analyses are needed to reconcile predictions from models and observations.
\item A self-consistent model \citep[by][]{reinacampos2017}, based on the gas density, velocity dispersion and shear, is used to predict the maximum cluster mass in each bin. The model suggests that shear is stopping star formation in GMCs up to 4 kpc and the stellar feedback regulates star formation in the outer part of the galaxy. As a result the model predicts a lack of radial trend in the maximum cluster mass, consistently to what is observed. 
\end{enumerate}
In conclusion, in this work we showed that properties of young star clusters can vary on sub-galactic scales. These variations depend on the environment in a non-trivial way, i.e. while for example the strength of cluster disruption (studied via the age function) shows a direct correlation with local \sigmah, and varies radially, the same is not true for the mass function, which instead shows a dependence on the dynamical properties of the gas and a deep correlation with the GMC properties. These results suggest that studies of clusters at sub-galactic scales, in the comparison with local environments and with GMC properties, are necessary in order to constraint models of cluster formation and evolution.

\section*{Acknowledgements}
We are thankful to the anonymous referee for comments and suggestions that helped improving the manuscript.
A.A. and G.\"{O}. acknowledge the support of the Swedish Research Council (Vetenskapsrådet) and the Swedish National Space Board (SNSB).
B.G.E acknowledges the HST grant HST-GO-13364.14-A. 
D.A.G. kindly acknowledges financial support by the German Research Foundation (DFG) through program GO 1659/3-2. 




\bibliographystyle{mnras.bst}
\bibliography{biblio_p2}



\appendix
\section{Completeness limits of the sub-regions}
\label{sec:a1}
The criteria used to define the final cluster sample of M51 imply a completeness limit in either luminosity or mass which is not trivial to define. The interplay of cuts in different filters was already discussed in Paper~I \citep{paper1}, where it was pointed out that the completeness of the sample is mainly set by the exclusion of cluster candidates with an absolute $V$ band magnitude fainter than $-6$ mag. We already showed that the completeness limit can change inside the galaxy, and that is brighter in the central part of the galaxy (see Section 3.3 of Paper~I). 
In this appendix we analyse how the completeness limit varies in the galaxy sub-regions defined in Section~\ref{sec:environment}, in order to understand how completeness can affect the study of the mass function of Section~\ref{sec:massfunc}.\\

\begin{figure}
\centering
\includegraphics[width=\columnwidth]{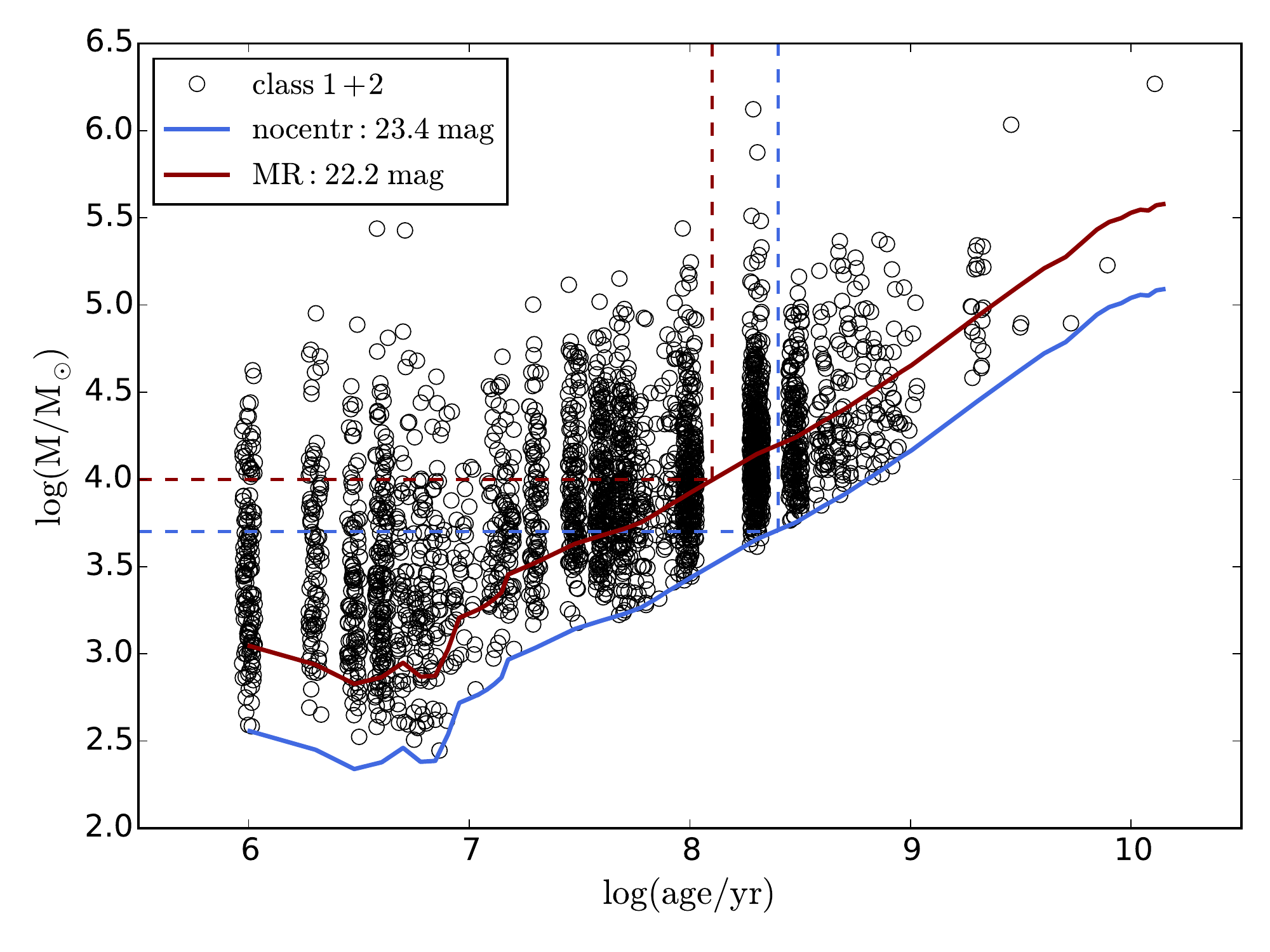}
\caption{Age-mass diagram for the clusters in our sample (black circles). Solid lines represent the magnitude limit in the V band at 23.4 mag (for the regions outside the centre) and at 22.2 mag (for the MR). the dashed lines enclose the mass-limited samples above 5000 and $10^4$ \msun.}
\label{fig:agemass}
\end{figure}

The $V$ band was used as reference frame for cluster production (Section~\ref{sec:data}) and we therefore use it as reference also for the completeness analysis. Synthetic clusters of effective radii in the range $1-5$ pc and magnitudes in range $20-26$ mag are added to the scientific $V$ band frame. The resulting image is processed following the same steps as for the real cluster catalogue, i.e. sources are extracted using \texttt{SExtractor} \citep{sextractor} and then photometrically analysed. A Galactic reddening correction is also applied to all sources. From a comparison between the number of simulated and of recovered clusters, we can estimate a completeness fraction at each magnitude. The completeness is a decreasing function with magnitude, and we decide to take the magnitude at which completeness goes below 90\% as the reference completeness limit. The $V$ band 90\% completeness limits in M51 subregions are displayed in Tab~\ref{tab:completeness}.
Excluding for the moment the MR region (it will be discussed separately later), in all regions the 90\% limit is fainter than the 23.4 mag (equal to $-6$ in absolute mag) cut applied on the data. It is therefore the applied cut that sets the completeness limit of the $V$ band in all the sub-regions. 
We convert this magnitude limit into an age-mass limit and we plot it in Fig.~\ref{fig:agemass}. From the plot we see that we have a complete sample of clusters more massive than 5000 \msun\ at ages $\leqslant200$~Myr.
We consider a cluster with a mass of 5000 \msun\ and an age of 200~Myr as the faintest element of our mass-limited sample, and using the same models as for the cluster SED fitting, we calculate its expected apparent magnitude in each of the bands. These values are reported as $M_{200}^{5e3}$ in Tab.~\ref{tab:completeness}.
\begin{table}
\centering
\caption{90$\%$ completeness limit in the five bands. As explained in the text, the $V$ band is the reference one used for the completeness test, i.e. is the only band where also cluster extraction is executed. $M_{200}^{5e3}$ and $M_{100}^{1e4}$ represent the magnitudes of a 200~Myr old cluster of 5000 \msun\ and of a 100~Myr old  cluster of $10^4$ \msun\ respectively.}
\label{tab:completeness}
\begin{tabular}{lccccc} 
\hline
Region 	& $V$ 	& $NUV$ 	& $U$ 	& $B$ 	& $I$ 		\\
\hline
\hline
Bin 1		& 23.72 	& 23.61 	& 23.81 	& 24.75 	& 23.90 		\\
Bin 2		& 23.67 	& 23.62 	& 23.81 	& 24.50 	& 23.96 		\\
Bin 3		& 23.47 	& 23.50 	& 23.63 	& 24.64 	& 23.94 		\\
Bin 4		& 23.74 	& 23.65 	& 23.82	& 24.87 	& 24.37 		\\
SA		& 23.65 	& 23.40 	& 23.70 	& 24.52 	& 23.78 		\\
IA		& 23.72 	& 23.66 	& 23.83 	& 24.80 	& 24.25 		\\
$M_{200}^{5e3}$& 23.30 	& 23.25 	& 23.23 	& 23.43 	& 22.83 		\\
\hline
MR		& 22.20	& 21.72	& 23.10	& 23.57	& 22.89		\\
$M_{100}^{1e4}$& 21.99 	& 21.45 	& 21.59 	& 22.09 	& 21.57 		\\
\hline
\end{tabular}
\end{table}

Since, in the catalogue production process, cluster candidates that are not detected with a magnitude error smaller than 0.3 mag in at least 4 bands are discarded from the sample, we check if clusters of 5000 \msun\ and  200~Myr are detectable in bands other than $V$. 
We run photometry on the synthetic clusters in all the bands, keeping only the ones detected with an error smaller than 0.3 and we derive the magnitude at which we reach the 90\% completeness (Tab.~\ref{tab:completeness}). In all filters and in all sub-regions the completeness limit is fainter than $M_{200}^{5e3}$. We therefore conclude that the mass-limited sample considered in the analyses of this work is complete in all sub-regions. 
We also notice that in some sub-regions the completeness is on average worse (for example Bin 3 and the SA environment) and that the recovered completeness limits there can be close to $M_{200}^{5e3}$ (especially in the $V$ and $U$ bands). Since we are considering the 90\% completeness limit, and the completeness function is not a step function, there is the possibility that those regions are partially affected by incompleteness. We have considered this possibility when discussing the results of the analyses in the text.

A similar process was used for the MR region. The main difference there is that the $V$ band completeness is worse than the 23.4 mag cut applied to the data. More specifically we find a 90$\%$ completeness limit at 22.20 mag in the $V$ band. From Fig.~\ref{fig:agemass} we can see that this limit implies that we can consider a mass-completed sample at masses above $10^4$ \msun\ at ages $\leqslant 100$~Myr. We derive the magnitude of a cluster with $10^4$ \msun\ and $100$~Myr, $M_{100}^{1e4}$, and verify again that is detectable in all bands other than $V$ in the MR region. Since the $90\%$ completeness limits in the MR are all fainter than that we are complete in the MR region choosing these limiting values.

\section{Comparing different fitting methods}
\label{sec:a2}
In this appendix we compare different ways of fitting the cluster mass function. The benefit of this exercise is to help understanding the differences in the recovered properties of the same (known) mass function, when different methods of fitting it are considered. Many are the ways in which mass functions are fitted in literature, sometimes leading to different conclusions, especially concerning the presence or absence of a truncation (e.g. the mass function analysis of M51, see \citealp{chandar16} and \citealp{paper1}). We consider here 3 approaches:
\begin{enumerate}
\item fit on the binned function (as in \citealp{chandar16});
\item fit on the cumulative function, using the maximum-likelihood code \texttt{mspecfit.pro} by \citet{rosolowsky2005};
\item Bayesian fit (as in the analysis of the cluster mass function by \citealp{johnson2017}).
\end{enumerate}
The difference between a binned and a cumulative approach, were already pointed out in Section 5 of Paper~I. Here we apply the different fitting approaches on two simulated mass distributions, namely: 
\begin{enumerate}
\item a mass distribution drawn from a Schechter mass function with slope $\beta=-2$ and exponential truncation mass $Mc=10^5$ \msun;
\item a mass distribution drawn from a power-law with slope $\beta=-2$.
\end{enumerate}
In both cases the simulated distributions are built in order to have a number of sources close to 1200 (i.e. close to the number of observed clusters in M51 in Paper~I) and mass values up to $10^7$ \msun. Each of the functions is fitted with all 3 approaches listed above and for each approach both a truncated function and a pure power-law are fitted.

\begin{figure*}
\centering
\subfigure{\includegraphics[width=\textwidth]{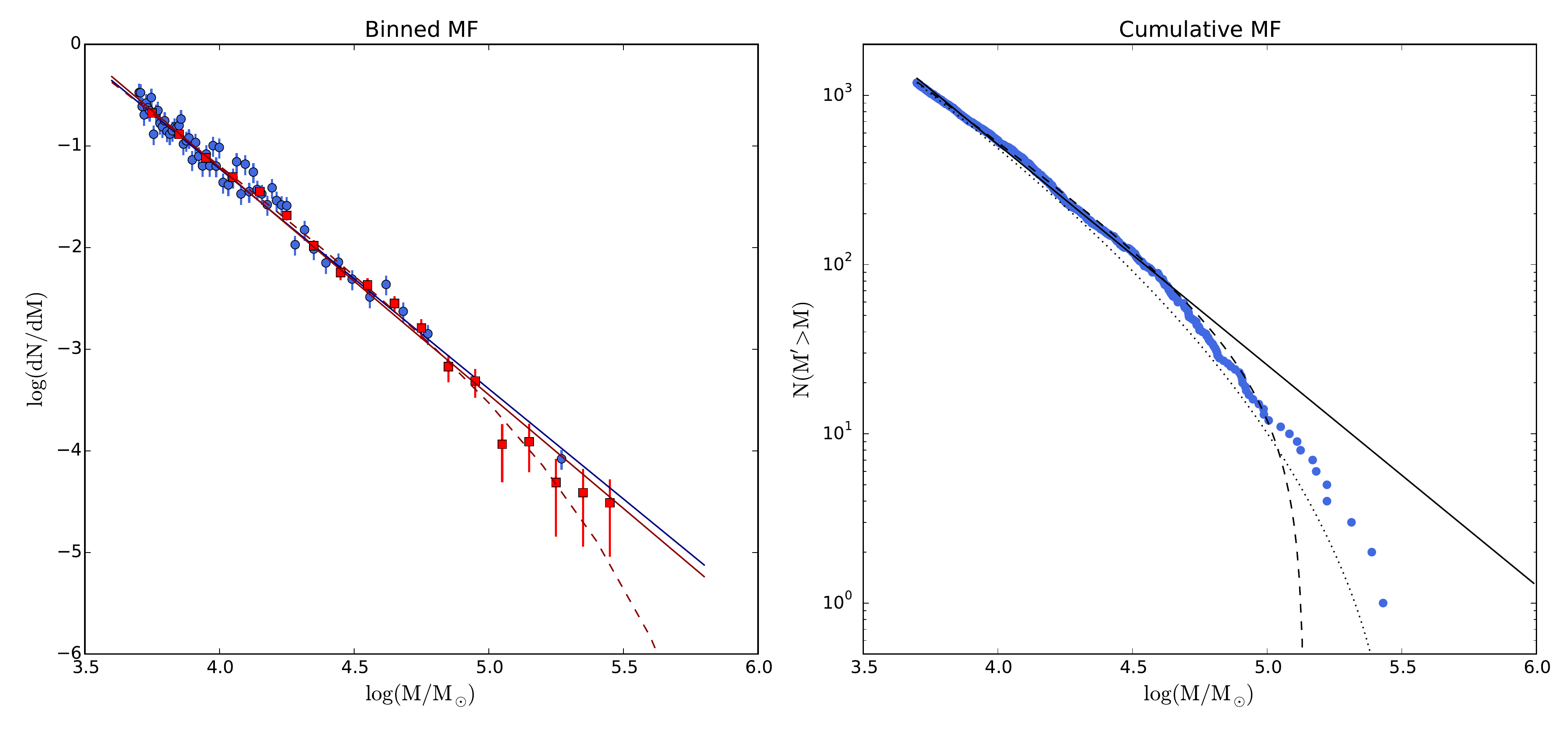}}
\subfigure{\includegraphics[width=0.5\textwidth]{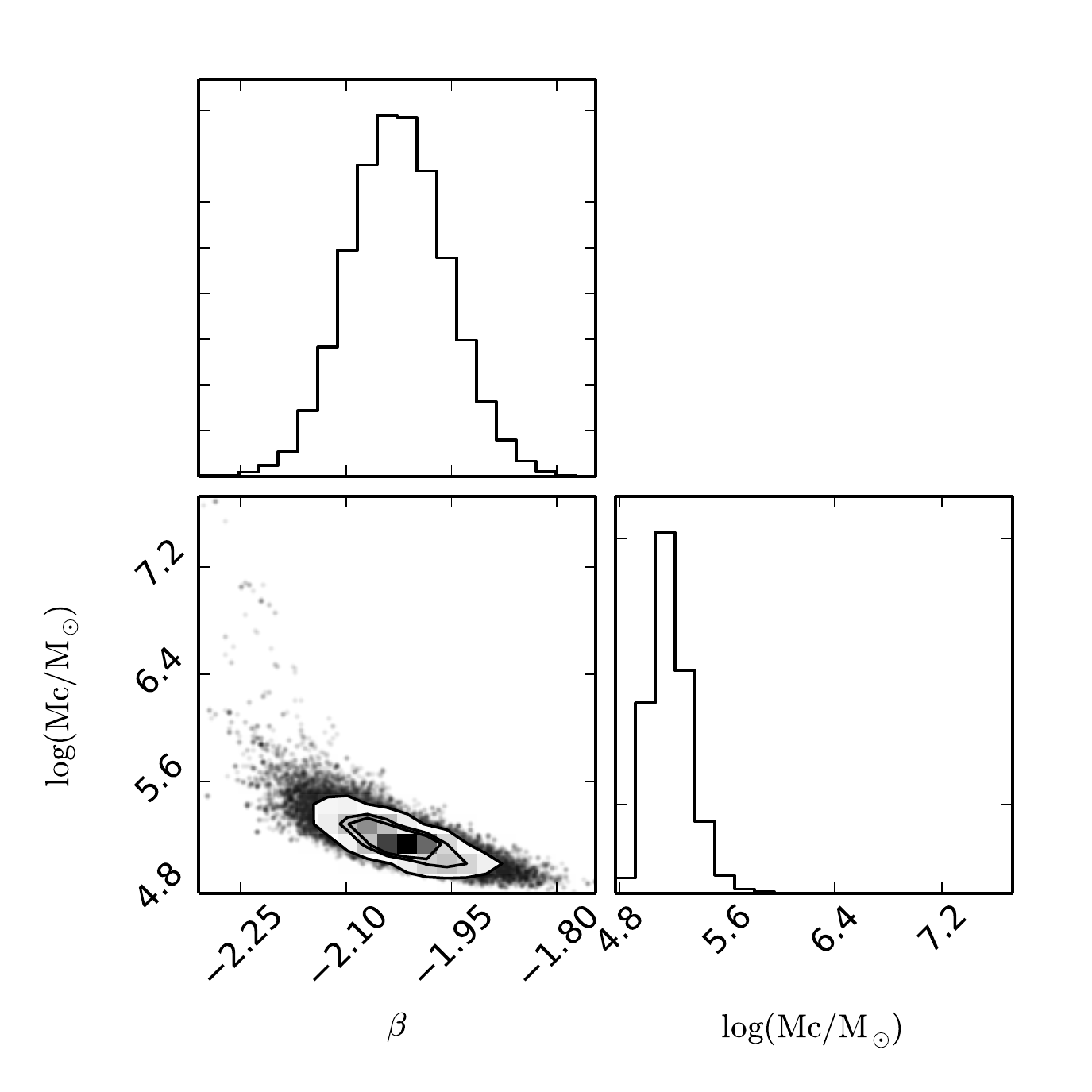}}
\subfigure{\includegraphics[width=0.4\textwidth]{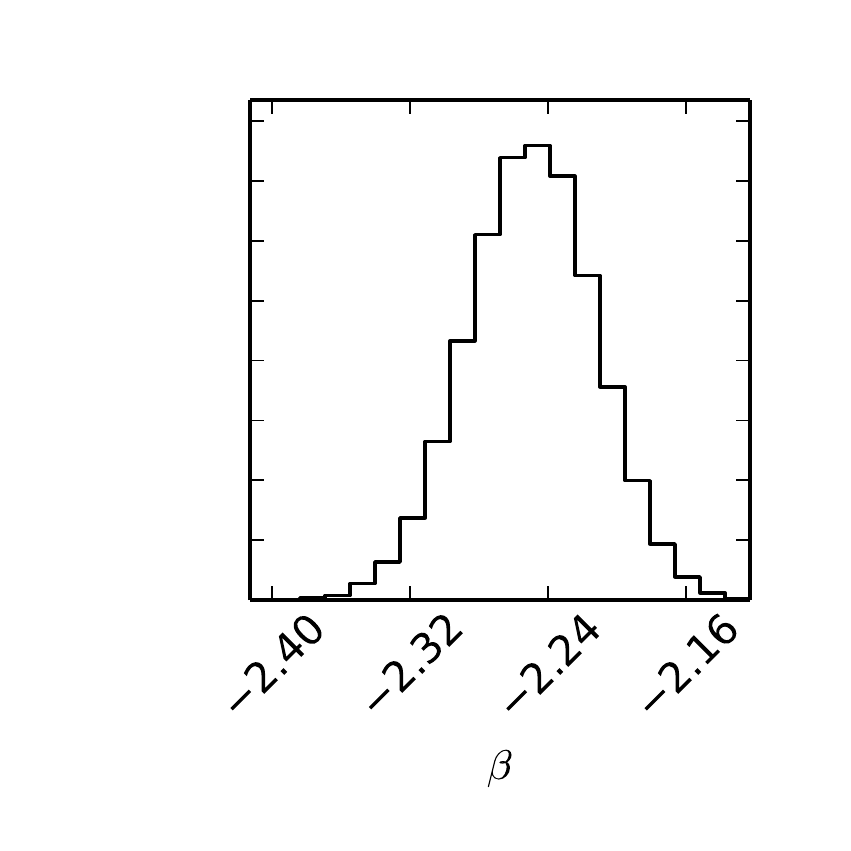}}
\caption{Simulated mass distribution drawn from a Schechter function with slope $\beta=-2$ and truncation mass $Mc=10^5$ \msun. Different fits and representations are plotted. Top left: Binned function. Both bins containing equal number of clusters (blue circles) and bins of equal width (red squares) are shown. The single power-law fit and Schechter fit are shown (as solid and dashed lines, respectively). Top right: Cumulative function with the results from the fit with a single power-law (dashed line) and with a truncated function (solid line). The dotted line shows the cumulative function if, instead of using the fit results in the formalism of \citet{rosolowsky2005} (Eq.~\ref{eq:1}), the cumulative mass function is drawn from a Schechter function with truncation mass $Mc=M_0$, as done in the comparison between the observed function and Monte Carlo populated ones in Section~\ref{sec:massfunc}. Bottom left: posterior probability distribution resulting from the Bayesian fitting, along with the marginalized distributions of $\beta$ and $\rm{log}(Mc/M_\odot)$. Bottom right: posterior probability distribution of $\beta$ in the Bayesian fit with a pure power-law.}
\label{fig:massfit_simsch}
\end{figure*}

We start analyzing the mass distribution drawn from a Schechter function. The distribution is binned in bins of equal area and in bins containing equal number of clusters. In both cases the function is fitted with a least-$\chi^2$ method and the best-fit results, along with the uncertainties, are displayed in Tab.~\ref{tab:massfit_appendix_binned} and plotted in Fig.~\ref{fig:massfit_simsch}. Despite the known truncation, a single power-law provides a good fit to the data ($\chi^2_{red}=1.09$). However, the recovered slopes are steeper than the simulated $-2$ slope. A Schechter function also provides a good fit to the data but does not add a statistical improvement. The recovered truncation mass values have big uncertainties, but the slopes are closer to the simulated value. \\
The fit with the \texttt{mspecfit.pro} code returns a value of $N_0=27\pm6$, suggesting that the mass distribution has a statistically significant truncation at $M_0=(1.37 \pm0.22)\times10^5$ \msun\ (see Tab.~\ref{tab:massfit_appendix_cumulative}). The recovered slope is steeper than $-2$, and it steepens even further if a pure power-law fit is considered ($\beta=-2.30 \pm0.04$). The plot shows that the truncated function deviates from the single power-law around $10^{4.5}$ \msun\ and is a better description of the data up to $\sim10^5$ \msun. At masses $M>10^5$ \msun, however, this parametrization deviates from the observations. If $M_0$ is used as the exponential truncation of a Schechter function, as done in the comparison between observed mass functions and Monte Carlo simulated populations in Section~\ref{sec:massfunc} of this paper, the shape of the function better resembles the distribution of the data points at masses $M>10^5$ \msun\ (dotted line in the top-right panel of Fig.~\ref{fig:massfit_simsch}).\\
The Bayesian fit returns, as median values, $\beta=-2.03 \pm0.07$ and $\rm{Mc}=(1.45^{+0.64}_{-0.37})\times10^5$ \msun, both compatible with the simulated values within $2\sigma$ (Tab.~\ref{tab:massfit_appendix_bayesian}). If the fit is made forcing the probability distribution to be a single power-law, the median slope of the posterior distribution is consistently steeper than $-2$ ($\beta=-2.25 \pm0.04$).
The posterior distribution of the fitted parameters (Fig.~\ref{fig:massfit_simsch}) clearly shows that the mass function is truncated. In addition it shows the correlation between the recovered values of $\beta$ and $Mc$.
In conclusion, the analysis of a simulated Schechter function suggests that a fit performed binning the function may hide a truncation at high masses, while both a fit on the cumulative function with \texttt{mspecfit.pro} and a Bayesian fitting are able to recognize the truncation.

In order to check that the fitting techniques are able to fit correctly a pure power-law distribution, we repeat the analyses on a simulated power-law with a slope of $-2$.
In this case the fit on the binned function correctly recovers the input slope. As expected, fitting with a Schechter function does not improve the results. 
The Schechter fit with \texttt{mspecfit.pro} returns a $M_0$ value close to the maximum simulated mass, but $N_0=7 \pm6$ indicates that this truncation is not statistically significant. The same code returns the correct value for the slope, when a single power-law fit is performed.
The bayesian fit also recovers the correct slope of the function. In case of the Schechter fit it also returns a median truncation mass value of $\rm{Mc}=(74.13^{+257.00}_{-47.22})\times10^5$ \msun, bigger than the maximum simulated mass, which makes the recovered value unreliable. Looking at the posterior distribution, the truncation mass distribution is degenerate, and the maximum-likelihood value not well-constrained. In addition, the maximum likelihood value in this case depends on the starting point of the walkers in the MCMC sampling process, and is also affected by the cut at $10^8$ \msun\ manually imposed (such high values of Mc are not physical in this case where the most massive cluster of the sample has $\rm{M}\approx2\times10^6$\msun). On the other hand the maximum-likelihood value of the posterior distribution of Mc in the previous case of a simulated Schechter function is stable and do not depend on the starting point of the MCMC sampling.
This is another indication of the impossibility to find a truncation mass in the current case. 
In all cases we are therefore able to recover both the right shape of the input function and the correct value of the slope. Fitting the distribution with a Schechter shape does not artificially introduce a `fake' recovered truncation, as all methods are helpful in establishing if the mass truncation is significant or not.

\begin{table*}
\centering
\caption{Results of the fit on the binned mass functions. Both a binning with an equal number of clusters per bin and a binning with bins of equal width were used. The input parameters for the simulated functions are PL: $\beta=-2$; SCH: $\beta=-2$, $\rm{Mc}=10^5$ \msun.}
\begin{tabular}{llccccc}
\hline
\multicolumn{1}{l}{Function} & \multicolumn{1}{l}{N$_s$} & \multicolumn{3}{c}{Schechter}  & \multicolumn{2}{c}{Pure PL} \\ 
\multicolumn{1}{l}{\ } & \multicolumn{1}{l}{\ } & \multicolumn{1}{c}{$-\beta$}  & \multicolumn{1}{c}{$\rm{Mc}$ ($10^5$\msun)}  & \multicolumn{1}{c}{$\chi^2_{red}$} & \multicolumn{1}{c}{$-\beta$}  & \multicolumn{1}{c}{$\chi^2_{red}$}  \\ 
\hline
\hline
\multicolumn{1}{l}{EQUAL NUMBER} & \multicolumn{6}{l}{\ } \\ 
Simulated SCH		& 1189	& $2.07\ _{\pm0.07}$	& $3.51\ _{\pm2.56}$ & $1.07$ 		& $2.17\ _{\pm0.04}$	& $1.09$\\
Simulated PL		& 1232	& $-$				& $-$ 			& $-$ 		& $1.97\ _{\pm0.03}$	& $0.98$\\
\hline
\multicolumn{1}{l}{EQUAL WIDTH} & \multicolumn{6}{l}{\ } \\ 
Simulated SCH		& 1189	& $2.00\ _{\pm0.07}$	& $1.17\ _{\pm0.36}$ & $0.76$ 		& $2.24\ _{\pm0.04}$	& $1.41$ \\
Simulated PL		& 1232	& $2.00\ _{\pm0.04}$	& $24.38\ _{\pm32.13}$ & $0.77$ 		& $2.02\ _{\pm0.02}$	& $0.74$ \\
\hline
\end{tabular}
\label{tab:massfit_appendix_binned}
\end{table*}

\begin{table*}
\centering
\caption{Results of the fit on the cumulative mass functions with the \texttt{mspecfit.pro} code.}
\begin{tabular}{llcccc}
\hline
\multicolumn{1}{l}{Function} & \multicolumn{1}{l}{N$_s$} & \multicolumn{3}{c}{Schechter}  & \multicolumn{1}{c}{Pure PL} \\ 
\multicolumn{1}{l}{\ } & \multicolumn{1}{l}{\ } & \multicolumn{1}{c}{$-\beta$}  & \multicolumn{1}{c}{$M_{0}$ ($10^5$\msun)}  & \multicolumn{1}{c}{$N_0$} & \multicolumn{1}{c}{$-\beta$}   \\ 
\hline
\hline
Simulated SCH		& 1189	& $2.15\ _{\pm0.04}$	& $1.37\ _{\pm0.22}$ & $27\ _{\pm6}$ 	 & $2.30\ _{\pm0.04}$ \\
Simulated PL		& 1232	& $2.00\ _{\pm0.04}$	& $8.05\ _{\pm3.92}$ & $7\ _{\pm6}$ 	 & $2.04\ _{\pm0.03}$ \\
\hline
\end{tabular}
\label{tab:massfit_appendix_cumulative}
\end{table*}

\begin{table*}
\centering
\caption{Results of the Bayesian fit on the mass function.}
\begin{tabular}{llccc}
\hline
\multicolumn{1}{l}{Function} & \multicolumn{1}{l}{N$_s$} & \multicolumn{2}{c}{Schechter}  & \multicolumn{1}{c}{Pure PL} \\ 
\multicolumn{1}{l}{\ } & \multicolumn{1}{l}{\ } & \multicolumn{1}{c}{$-\beta$}  & \multicolumn{1}{c}{Mc ($10^5$\msun)}  & \multicolumn{1}{c}{$-\beta$}   \\ 
\hline
\hline
Simulated SCH		& 1189	& $2.03\ _{\pm0.07}$	& $1.45\ ^{+0.64}_{-0.37}$ &  $2.25\ _{\pm0.04}$ \\
Simulated PL		& 1232	& $1.99\ _{\pm0.03}$	& $74.13\ ^{+257.00}_{-47.22}$ &  $2.01\ _{\pm0.03}$ \\
\hline
\end{tabular}
\label{tab:massfit_appendix_bayesian}
\end{table*}

\begin{figure*}
\centering
\subfigure{\includegraphics[width=\textwidth]{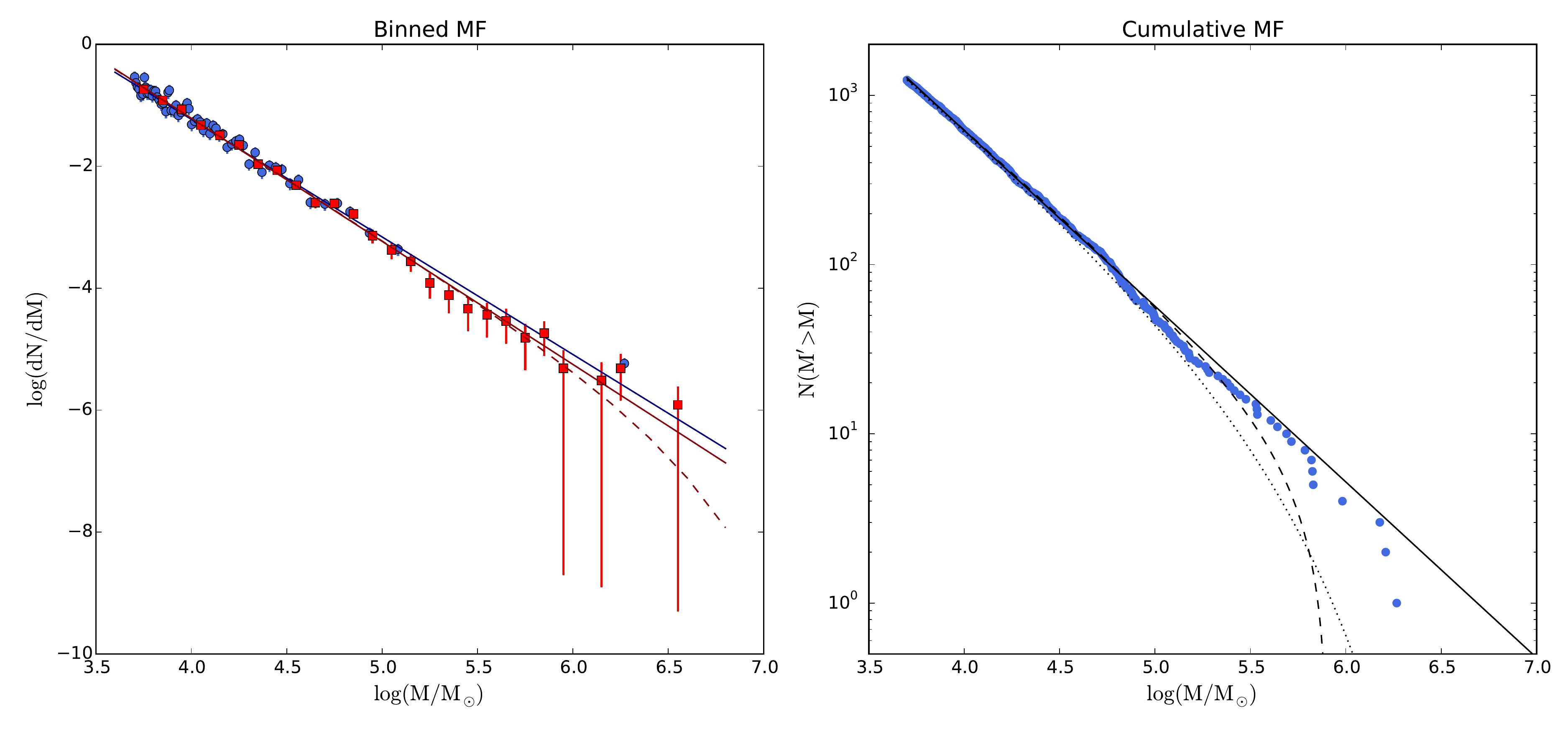}}
\subfigure{\includegraphics[width=0.5\textwidth]{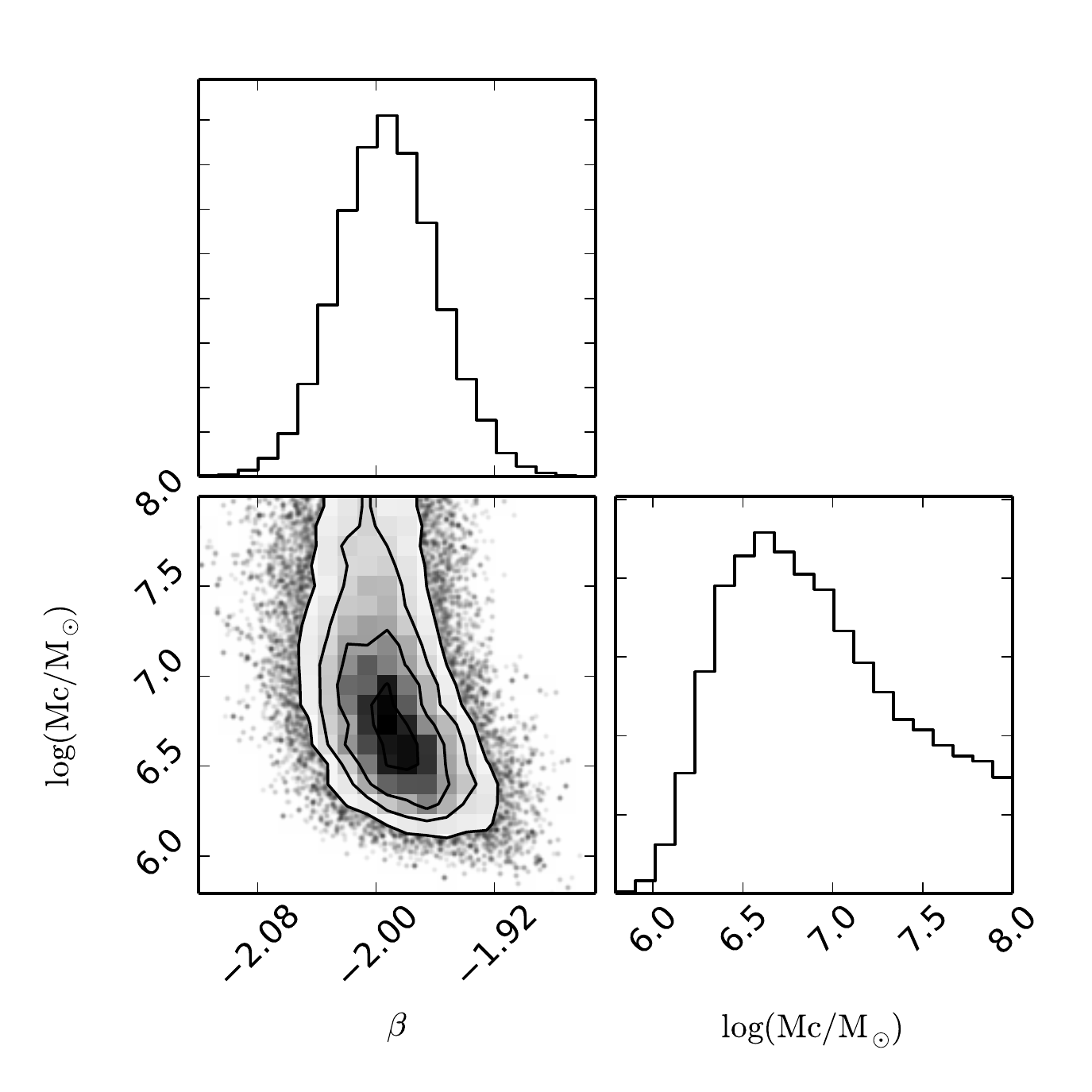}}
\subfigure{\includegraphics[width=0.4\textwidth]{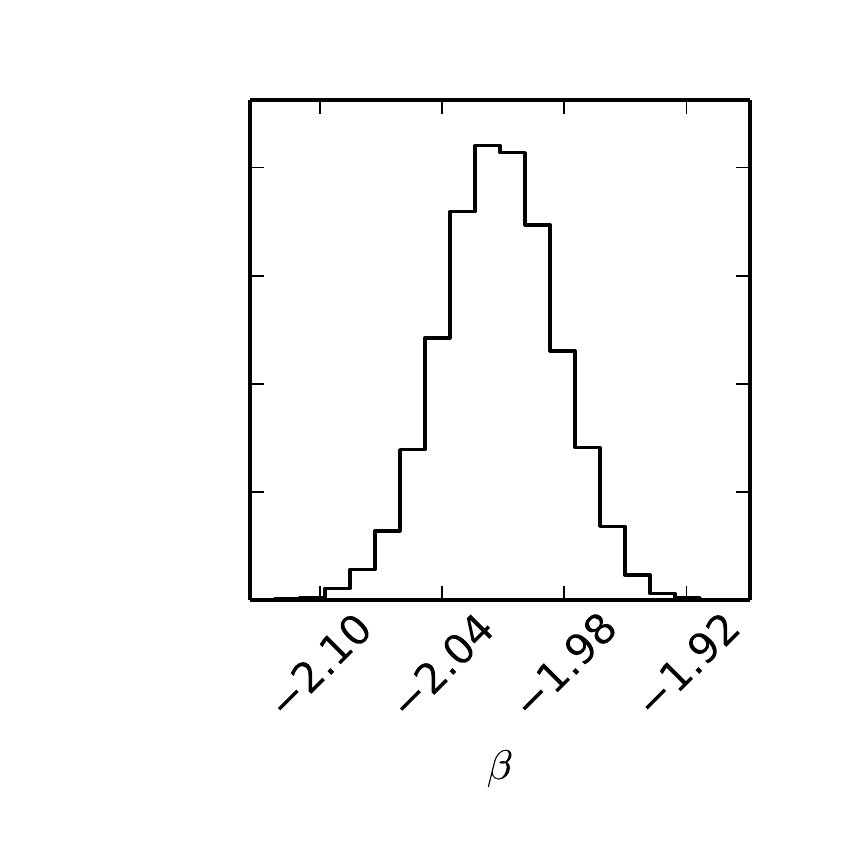}}
\caption{The same as Fig.~\ref{fig:massfit_simsch} but with a mass distribution drawn from a pure power law function with slope $\beta=-2$.}
\label{fig:massfit_simpl}
\end{figure*}



\bsp	
\label{lastpage}
\end{document}